\documentstyle[prl,twocolumn,aps,epsf,eqsecnum,bbox]{revtex}
\begin{document}


\voffset1.5cm


\title{Gluon radiation off hard quarks in a nuclear environment: opacity expansion}
\author{Urs Achim Wiedemann}

\address{Theory Division, CERN, CH-1211 Geneva 23, Switzerland}

\date{\today}
\maketitle

\begin{abstract}
We study the relation between the Baier-Dokshitzer-Mueller-Peign\'e-Schiff
(BDMPS) and Zakharov formalisms for medium-induced gluon radiation off
hard quarks, and the radiation off very few scattering centers. 
Based on the non-abelian Furry approximation for the motion of
hard partons in a spatially extended colour field, we derive a
compact diagrammatic and explicitly colour trivial expression for 
the $N$-th order term of the ${\bf k}_\perp$-differential gluon radiation 
cross section in an expansion in the opacity of the medium. Resumming
this quantity to all orders in opacity, we obtain Zakharov's path-integral 
expression (supplemented with a regularization prescription). This 
provides a new proof of the 
equivalence of the BDMPS and Zakharov formalisms which extends previous 
arguments to the ${\bf k}_\perp$-differential cross section. We give
explicit analytical results up to third order in opacity for both the 
gluon radiation cross section of free incoming and of in-medium 
produced quarks. The $N$-th order term in the 
opacity expansion of the radiation cross section is found to be a 
convolution of the radiation associated to $N$-fold rescattering 
and a readjustment of the probabilities that rescattering occurs
with less than $N$ scattering centers. Both informations can be
disentangled by factorizing out of the radiation cross section a 
term which depends only on the mean free path of the projectile.
This allows to infer analytical expressions for the totally coherent 
and totally incoherent limits of the radiation cross section to arbitrary 
orders in opacity.
\end{abstract}


\section{Introduction}
\label{sec1}

Hard partons, produced in relativistic heavy ion collisions at RHIC 
and LHC, will undergo multiple rescattering inside the nuclear
environment before entering the hadronization process outside the
nuclear environment (or in a very dilute one). This follows from
standard formation time arguments. Prior to hadronization,  
medium-induced radiative energy loss is expected to be 
the main medium modification encountered by hard partons. The
size of this effect depends on the density and nature of the
medium and may be significantly enhanced in a deconfined partonic
medium~\cite{GW94,WGP95}. A corresponding strong medium-modification 
of the high-pt tails of hadronic single particle spectra 
(``jet quenching'')~\cite{WG91,GL98}
is thus a tentative signal for the formation of a deconfined state.
This possibility has motivated many studies of medium-induced 
gluon radiation in recent years~\cite{Z96,BDMPS97a,BDMPS97b,Z98,BDMS98,BDMS-Zak,KST98,WG99,Z99,BDMS99,KST99,BSZ99,GLV99,LS00}.

Most recent studies of the non-abelian energy loss are carried
out within the Gyulassy-Wang (GW) model~\cite{GW94} which mimics the medium by
a set of coloured static scattering centers. Despite its 
simplicity, this model is of phenomenological interest, since the
medium-induced radiative energy loss belongs to those observables
which depend mainly on the average transverse 
colour field strength encountered by the parton rather than 
on the model-specific details with which this colour field
strength is described~\cite{W00}. 

First studies of the GW model~\cite{GW94,WGP95} focussed on the
rescattering of the hard quark and arrived at a radiative energy
loss $dE/dx = {\rm const}$ independent of the path length.
Baier-Dokshitzer-Mueller-Peign\'e-Schiff (BDMPS)~\cite{BDMPS97a} established
later that gluon rescattering diagrams give the dominant contribution.
They found a potentially dramatic linear increase  $dE/dx \propto L $
of the energy loss with the medium thickness $L$. This
can be understood in terms of an uncertainty argument of Brodsky
and Hoyer~\cite{BH93} which relates the average transverse gluon momentum 
$\langle {\bf k}_\perp^2 \rangle$ to the radiative energy loss 
$dE/dx \propto \langle {\bf k}_\perp^2\rangle$. Brownian motion 
of the rescattering gluon then implies $\langle {\bf k}_\perp^2
\rangle \propto L$ and the quadratic $L$-dependence of radiative 
energy loss in the BDMPS formalism. While all these studies
start from the complete set of multiple scattering diagrams,
Zakharov~\cite{Z96,Z98} has advocated a different and very elegant approach
to the same problem. In his path-integral formalism, the radiation 
cross section is determined by a dipole cross section which essentially
measures the difference between elastic scattering amplitudes of
different projectile Fock state components as a function
of impact parameter.
Baier, Dokshitzer, Mueller and Schiff (BDMS) have shown~\cite{BDMS-Zak}  
that the evolution of the rescattering amplitude in the 
BDMPS-formalism is determined by Zakharov's dipole cross section. 

The above mentioned calculations do not have a clear connection to 
experiment since the total radiative energy loss $dE/dx$ is not an 
experimental observable. Unlike the situation in QED where the charged 
projectile can in principle be arbitrarily well separated from its 
radiation, medium-induced QCD bremsstrahlung is only an observable to 
the extent to which it is emitted kinematically well separated outside 
the typical hadronization cone of the hard parton. Realistic energy loss
estimates thus require knowledge about the ${\bf k}_\perp$-differential
gluon radiation spectrum. First calculations of this observable were
published a year ago in the Zakharov~\cite{WG99,Z99} and in the 
BDMPS-formalism~\cite{BDMS99}.
These studies contain essential steps towards a phenomenological
application: especially, Ref.~\cite{WG99} shows how to rewrite
Zakharov's path-integral formalism in a simple, numerically accessible
form, and Ref.~\cite{BDMS99} gives results for the radiation outside
the kinematically unresolvable hadronization cone of the parton.

However, calculating the angular distribution is not the only
problem in making contact with phenomenology.
Another problem is that the concept of a homogeneous medium of finite 
extent, underlying the BDMPS and Zakharov formalisms, may not be 
applicable to heavy ion collisions at RHIC and LHC.
Unlike the situation in QED, where even the thinnest
targets probed in experiments of the Landau-Pomeranchuk-
Migdal effect amount to $\approx 10000-100000$ small-angle scatterings,
one expects for the medium-induced radiation off hard partons in 
relativistic heavy ion collisions a much smaller average number of
rescatterings, say $\approx 2 - 10$~\cite{GLV99}. 
A comparison of the BDMPS- and Zakharov- formalisms with results 
for a small fixed number $N=1, 2, 3,\dots$ of rescatterings is needed
to understand to what extent the concept of a homogeneous
medium can be justified for such an extremely thin medium. The 
only existing studies~\cite{WG99,GLV99} of scenarios with very 
few $N \leq 3$ scattering centers assume implicitly an exclusive 
measurement of jet, gluon and recoil target partons. They are thus 
restricted to the study of a subclass of all available rescattering 
diagrams, and their results do not compare directly to the BDMPS 
and Zakharov formalisms (see below for more details). It is one of 
the results of the present paper to calculate the gluon radiation 
spectrum for a ``medium'' of very few $N \leq 3$ scattering centers, 
including all available rescattering diagrams, thus allowing for
a direct comparison. More generally, we discuss in which sense 
the $N$-th order term in the opacity expansion of Zakharov's result 
is related to radiation off a target of a fixed number $N$ of
scattering centers.

From a technical point of view, the present work is an application
and extension of methods developped in~\cite{W00}. There,
we have derived the non-abelian Furry
approximation for the wavefunction of a hard parton which
undergoes multiple rescattering in a spatially extended colour
field. This Furry wavefunction provides a compact shorthand for
the high-energy limit of the complete class of final state
rescattering diagrams.
Relevant for the present work is that we have developped in 
Ref.~\cite{W00} diagrammatic tools to determine under which 
conditions observables of a multiple partonic rescattering 
process are colour trivial. 
{\it Colour triviality} is the remarkable fact that 
for some rescattering processes, the sum of all contributions to 
the $N$-fold rescattering depends on a unique $N$-th power of the 
$SU(3)$ Casimir operators, rather than to depend on colour 
interference terms associated to more than one colour trace. 
This renders the calculational problem essentially 
abelian and is crucial for going beyond very few rescatterings 
where brute force perturbative calculations are still feasible. 
Colour triviality is thus an important property in the study of 
non-abelian gluon bremsstrahlung.

The present work applies the non-abelian Furry approximation to 
the calculation of the medium-induced gluon radiation cross section.
In section~\ref{sec2}, we derive the radiation cross section in
terms of Furry wavefunctions. From this, we derive in section~\ref{sec3}
a set of diagrammatic identities which automate the proof of colour
triviality. In section~\ref{sec4}, these identities are applied 
to derive an explicitly colour trivial diagrammatic expression 
for the $N$-th order term of the opacity expansion of the 
${\bf k}_\perp$-differential gluon radiation spectrum. Contact
with analytical expressions is made by showing that the sum over
all $N$-th order contributions results in a path-integral expression 
which essentially coincides with Zakharov's result. 
In section~\ref{sec5}, we relate this path-integral
expression to the radiation off a target of a fixed number of
scattering centers.
All calculations in sections~\ref{sec2}-\ref{sec5} are for gluon
radiation off an idealized free {\it in}coming quark which satisfies
plane wave boundary conditions at far backward position. The corresponding
radiation cross sections are denoted by a superscript $(in)$. In 
section~\ref{sec6}, we extend these calculations to the radiation off
{\it nascent} quarks produced in the medium. The gluon radiation
cross section associated with them is characterized by a superscript
$(nas)$ and gives access to the interference pattern between 
hard and medium-dependent radiation. The main results and further
perspectives are finally discussed in the Conclusions.

\section{The formalism}
\label{sec2}

\begin{figure}[h]\epsfxsize=8.7cm 
\centerline{\epsfbox{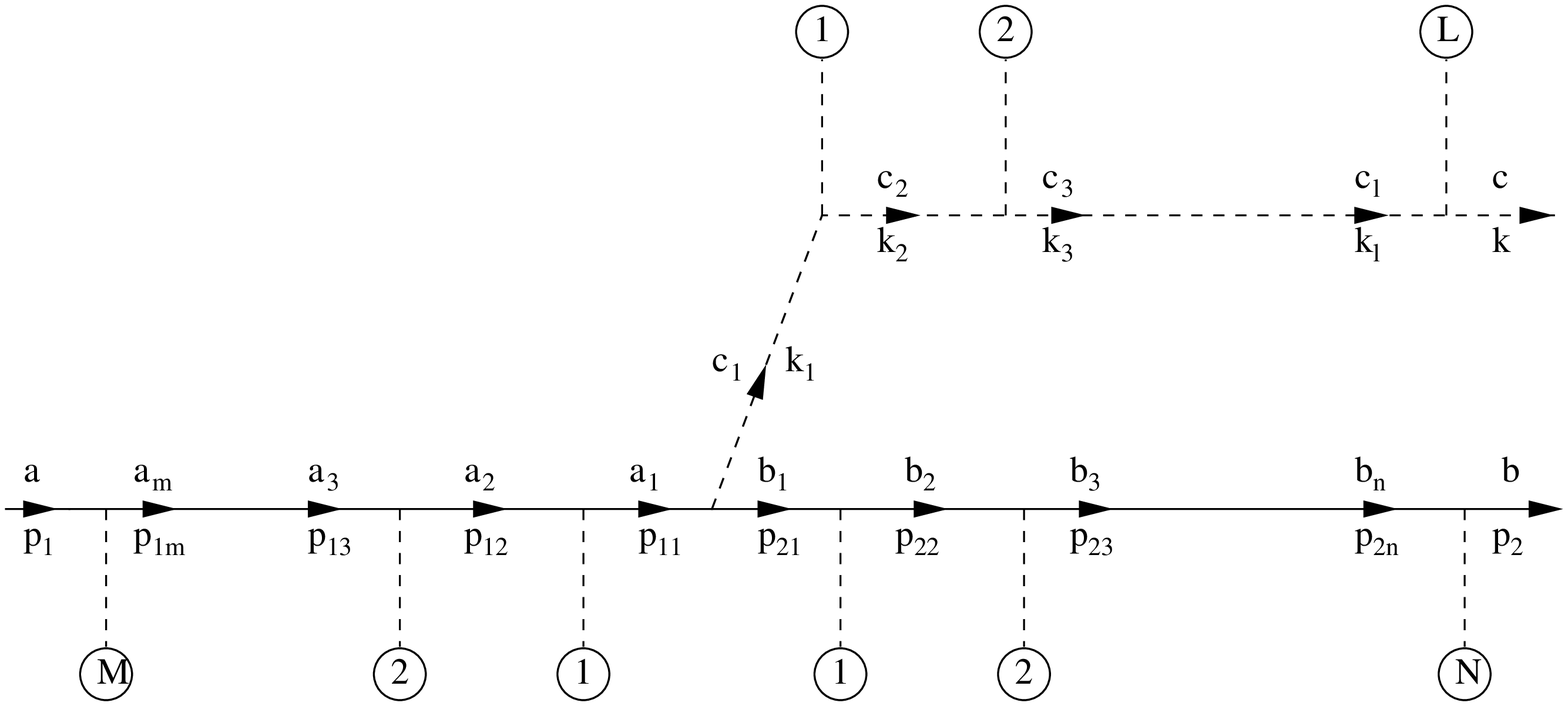}}
\vspace{0.5cm}
\caption{Contribution to the gluon radiation amplitude
(\protect\ref{2.4}), involving $M$-, $L$-, and $N$-fold
rescattering of the in- and outgoing partons respectively.  
}\label{fig1}
\end{figure}

We consider gluon radiation off a hard quark which undergoes multiple
elastic rescattering in a spatially extended colour field. The gluon
of momentum $k = (\omega,{\bf k})$ carries away a fraction $x$ of the
totally available energy of the initial parton, $\omega = x\, E_1$.
A typical contribution to the radiation amplitude is shown in 
Fig.~\ref{fig1}. For the coloured potential describing the medium, 
we choose the ansatz of the Gyulassy-Wang model~\cite{GW94}
\begin{eqnarray}
  A_\mu^{(q)}({\bf x})_{a\, b} &=&
  \delta_{0\mu}\, \sum_{i=1}^\infty\, \varphi^d_i({\bf x})\, 
  \left(T^d\right)_{a\, b}\, ,
  \label{2.1} \\
  A_\mu^{(g)}({\bf x})_{a\, b} &=&
  \delta_{0\mu}\, \sum_{i=1}^\infty\, \varphi^d_i({\bf x})\, 
  \left(i\, f^{adb}\right)\, ,
  \label{2.2} \\
  \varphi^d_i({\bf x}) &=& \varphi({\bf x}-\check{\bf x}_i)\, 
  \delta^{d\, d_i}\, .
  \label{2.3}
\end{eqnarray}
The static scattering potentials $\varphi^d_i({\bf x})$ show a 
sufficiently rapid (e.g. Yukawa-type) spatial fall-off. At the
$i$-th scattering center, a specific colour charge $d=d_i$ is 
exchanged. $T^{d}$, $(d=1, \dots, N_c^2-1)$, denotes the generators 
of the $SU(N_c)$ fundamental representation, and the totally
antisymmetric structure constants $f_{abc}$ denote the adjoint 
representation. The latter appears in the rescattering amplitude
of the emitted gluon.

Summed over arbitrary many rescatterings, the gluon radiation 
amplitude ${\cal M}_{a\to b\, c}$ in this extended colour potential
takes the form
\begin{eqnarray}
  {\cal M}_{a\to b\, c} &=& \int d^3{\bf y}\, I_{a\, a_1}({\bf y},p_1)\, 
  \left( -i\, g_s\, \gamma_{\mu_1} \left(T^{c_1}\right)_{a_1\, b_1}\right)
  \nonumber \\
  && \qquad \times
  e^{-\epsilon |y_l|}\,
  I_{c_1\, c}^{\mu_1}({\bf y},k)\, I_{b_1\, b}({\bf y},p_2)\, .
  \label{2.4}
\end{eqnarray}
Here, the term $e^{-\epsilon |y_l|}$ is the adiabatic switching off
of the interaction term at large distances. In our calculation, the
limit $\epsilon \to 0$ does not commute with the longitudinal
$y_l$-integral, and this regularization has to be carried through
intermediate steps of the calculation~\cite{WG99}.
To explain the notation of (\ref{2.4}), we consider the typical 
diagrammatic contribution given in Fig.~\ref{fig1}.
The $N$-fold rescattering of the quark, emitted from the radiation
vertex with momentum $p_{21}$ and colour $b_1$, and appearing in the
final state with momentum $p_2$ and colour $b$ is determined by
the component $I^{(N)}_{b_1\, b}$ of [see Fig.~\ref{fig1} for details
of notation]
\begin{eqnarray}
  I^{(N)}({\bf y},p_2) &=&
  e^{-i\, {\bf p}_{2,1}\cdot {\bf y}}\,
 {\cal P}\, \left( 
  \prod_{i=1}^N\int {d^3{\bf p}_{2,i}\over (2\pi)^3}\, d^3{\bf x}_i\,
  \right.
                   \nonumber \\
                   && \times
                    {i\, (\not{p}_{2,i} + m)\, \gamma_0\over
                     {p_{2,i}^2 - m^2 + i\,\epsilon}}  
                    \lbrack -i\, A_0^{(q)}({\bf x}_i)\rbrack
                   \nonumber \\
                   && \times
                    \left.
                    \, e^{-i\, {\bf x}_i\cdot 
                      ({\bf p}_{2,i+1} - {\bf p}_{2,i})} \right)\,
                    v({\bf p_2})\, .
  \label{2.5}
\end{eqnarray}
Here, the path-ordering ${\cal P}$ implies that $A_0^{(q)}({\bf x}_{i+1})$
stands to the right of $A_0^{(q)}({\bf x}_i)$, and the momentum transfers 
to the quark line are written as Fourier
transforms of the static scattering potential with respect to the
relative momenta ${\bf p}_{2,i+1} - {\bf p}_{2,i}$. $v({\bf p_2})$
is the spinor of the outgoing quark. In complete
analogy, rescattering effects of the incoming quark are described
by the component $I^{(M)}_{a\, a_1}$ of
\begin{eqnarray}
  I^{(M)}({\bf y},p_1) &=&
 v^{\dagger}({\bf p_1})\, 
 {\cal P}\, \left( 
  \prod_{i=1}^M\int {d^3{\bf p}_{1,i}\over (2\pi)^3}\, d^3{\bf x}_i\,
  \right.
                   \nonumber \\
                   && \times
                    \lbrack -i\, A_0^{(q)}({\bf x}_i)\rbrack
                    {i\, (\not{p}_{1,i} + m)\, \gamma_0\over
                     {p_{1,i}^2 - m^2 + i\,\epsilon}}  
                   \nonumber \\
                   && \times
                    \left.
                    \, e^{-i\, {\bf x}_i\cdot 
                      ({\bf p}_{1,i+1} - {\bf p}_{1,i})} \right)\,
  e^{i\, {\bf p}_{1,1}\cdot {\bf y}}\, \gamma_0\, ,
  \label{2.6}
\end{eqnarray}
and the rescattering effects on the emitted gluon are taken
into account by the component $I^{\mu_1\, (L)}_{c_1\, c}$ of
\begin{eqnarray}
  I^{\mu_1\, (L)}({\bf y},k) &=&
  e^{- i\, {\bf k}_1\cdot {\bf y}}\,
 {\cal P}\, \left( 
  \prod_{i=1}^L\int {d^3{\bf k}_i\over (2\pi)^3}\, d^3{\bf x}_i\,
  \right.
                   \nonumber \\
                   && \times
                    \lbrack -i\, A_0^{(g)}({\bf x}_i)\rbrack
                    {-i\, g^{\mu_i\, \mu_i'}\over
                     {k_i^2 - m^2 + i\,\epsilon}}
                    V_{\mu_i'0\mu_i}
                   \nonumber \\
                   && \times
                    \left.
                    \, e^{-i\, {\bf x}_i\cdot 
                      ({\bf k}_{i+1} - {\bf k}_i)} \right)\, 
                  \epsilon^{\mu_L}
                  \, ,
  \label{2.7}
\end{eqnarray}
where $\epsilon^{\mu_L}$ is the polarization of the gluon.
To calculate the high-energy limit of the radiation spectrum,
one has to keep on the amplitude level the leading order in 
norm and the next-to-leading order in the phase. In this
limit, other scattering contributions (e.g. those involving
the 4-gluon-vertex) are negligible~\cite{BDMPS97a}, and the
expressions (\ref{2.5})-(\ref{2.7}) take the form of
non-abelian Furry wavefunctions~\cite{W00}.
In appendix~\ref{appa}, we show that by approximating
$I^{\mu_1\, (L)}({\bf y},k)$ to leading order $O(1/\omega)$
in the norm and next to leading order in the phase, the
corresponding sum over arbitrary many rescattering centers
takes a particularly simple form
\begin{eqnarray}
 I^{\mu_1}({\bf y},k)
 &=& \sum_{L=0}^\infty I^{\mu_1\, (L)}({\bf y},k)
 \nonumber \\
 &=& \epsilon^{\mu_1}\, e^{- i\, {\bf k}_1\cdot {\bf y}}\,
 \int d{\bf x}_\perp G_{(g)}({\bf y};{\bf x}|\omega)\,
  F({\bf x},{\bf k})\, .
  \label{2.8}
\end{eqnarray}
Here, $F({\bf x},{\bf k})$ denotes the asymptotic plane 
wave
\begin{eqnarray}
  F({\bf x}_\perp,x_l,{\bf p}_2) &=& 
  \exp\left\{- i\,{\bf p}_2^\perp\cdot{\bf x}_\perp
  + i\, { {{\bf p}_2^\perp}^2\over 2\, p_2} x_l\right\}\, ,
  \label{2.9}
\end{eqnarray}
and the Green's function $G_{(g)}$ is an approximate solution of the
non-abelian Dirac equation in the spatially extended colour
potential $A_0^{(q)}$. This Green's function describes the 
leading transverse deviation of the rescattering parton from a 
straight line propagation. It can be represented in terms of a 
path-integral over a path-ordered Wilson line 
$W_{(g)}\bigl([{\bf r}];z,z'\bigr)$,
\begin{eqnarray}
  &&G_{(g)}({\bf r},z;{\bf r}',z'\vert p) 
  =  \nonumber \\
  &&=\int {\cal D}\bbox{r}(\xi)\,
  \exp\left\{ \frac{ip}{2} \int\limits_{z}^{z'}{\it d}\xi\,
  \dot{\bf r}^2(\xi)\right\}\, 
  W\bigl([{\bf r}];z,z'\bigr) \, ,
  \label{2.10}\\
&&  W_{(g)}\bigl([{\bf r}];z,z'\bigr) = {\cal P}\,
  \exp\left\{ -i\, \int\limits_z^{z'}{\it d}\xi\,
              A_0^{(g)}({\bf r}(\xi),\xi) \right\}\, .
  \label{2.11}
\end{eqnarray}
In what follows, we shall explicitly denote or supress the
longitudinal coordinates of these Green's functions, depending
on whether this information can be easily inferred.

In analogy to (\ref{2.8}), we have shown in a previous paper~\cite{W00} 
that the sum over an arbitrary number of final state rescatterings of the 
outgoing quark takes the compact form
\begin{eqnarray}
  &&I({\bf y},p_2) =
  \sum_{N=0}^\infty\, I^{(N)}({\bf y},p_2)
  \nonumber \\
  &&= e^{- i\, p_2\, y_l}\, {\hat D}_2\,
  \int d{\bf x}_\perp\, G_{(q)}({\bf y};{\bf x}|p_2)
  F({\bf x},{\bf p}_2)\, v({\bf p}_2)\, . 
  \label{2.12} 
\end{eqnarray}
As denoted by the subscript $(q)$, the Green's function
contains in this case a Wilson line in the fundamental
quark representation of $SU(N_c)$. The only new ingredient
compared to (\ref{2.8}) is the differential operator 
$\hat D_i$~\cite{KST98,WG99} 
 \begin{eqnarray}
 \hat D_i &=&
  1- i\,\frac{\bbox{\alpha}\cdot\bbox{\nabla}}{2\,E_i} 
   - \frac{\bbox{\alpha}\cdot({\bf p}_i-{\bf n}\,p_i)}{2\,E_i}\, ,
  \nonumber\\
   && \bbox{\alpha} = \gamma_0\, \bbox{\gamma}\quad ;\quad
   z = \bbox{n}\cdot\bbox{x}\quad ;\quad
   p_i = |\bbox{p}_i|\, .
   \label{2.13}
\end{eqnarray}
This operator contains in a configuration space formulation 
the $O(1/E)$ corrections to the propagator of momentum $p_{21}$.
Inclusion of these corrections for the fermion propagators
entering the radiation vertex is necessary, since the leading
$1/E$-contribution cancels due to the Bloch-Nordsieck structure
of the radiation vertex~\cite{W00}.

In terms of the Green's functions $G_{(q)}$ and $G_{(g)}$, the
gluon radiation amplitude (\ref{2.4}) takes the form
\begin{eqnarray}
  &&{\cal M}_{a\to b\, c} = -i \int dy_l\, 
  e^{i\,\bar{q}\, y_l}\, e^{-\epsilon\, |y_l|}
  \int d^2{\bf y}
  \nonumber \\
  &&\qquad \times  \int d^2{\bf x}_1\, d^2{\bf x}_2\, 
  d^2{\bf x}_g\, 
  e^{i{\bf x}_{1\perp}\cdot {\bf p}_{1\perp}
               - i{\bf x}_{2\perp}\cdot {\bf p}_{2\perp}
               - i{\bf x}_{g\perp}\cdot {\bf k}_\perp}
  \nonumber \\
  &&\qquad \times G_{(q)}^{a a_1}({\bf x}_1;{\bf y}|p_1)\, 
  \hat{\Gamma}_{\bf y} \left(T^{c_1}\right)_{a_1b_1}\, 
  G_{(g)}^{c_1c}({\bf y};{\bf x}_g|\omega)
  \nonumber \\
  &&\qquad \times G_{(q)}^{b_1b}({\bf y};{\bf x}_2|p_2)\,
  e^{-i\frac{{\bf p}_{1\perp}^2}{2p_1}x_{l}
      +i\frac{{\bf p}_{2\perp}^2}{2p_2}x_{l}
      -i\frac{{\bf k}_{\perp}^2}{2\omega}x_{l}}\, .
  \label{2.14}
\end{eqnarray}
Here, we have used the fraction $x$ of the incident energy
carried by the emitted gluon, $\omega = x\, E_1$, to write
\begin{equation}
  \bar{q} = p_1-p_2 - \omega = \frac{x\, m_q^2}{2\, (1-x)\, E_1}\, .
  \label{2.15}
\end{equation}
Also, we have introduced the interaction vertex $\hat \Gamma$
as the notational shorthand for the spinor structure of the 
amplitude (\ref{2.14}), 
\begin{equation}
  \hat{\Gamma}_{\bf y} = v^\dagger(p_1)\, \hat D_1 \gamma_0\,
         \epsilon\cdot\gamma\, \hat D_2\, v(p_2)\, .
  \label{2.16}
\end{equation}
This leads to the radiation probability 
\begin{eqnarray}
  &&\langle \vert{\cal M}_{a\to b\, c}\vert^2\rangle = 
    2\, {\rm Re}\, 
    \int_{z_-}^{z_+} dy_l\, \int_{y_l}^{z_+} d{\bar y}_l\, 
    e^{i\bar{q}(y_l-{\bar y}_l)}\, 
    \nonumber \\
    && \quad \times e^{-\epsilon\, |y_l|-\epsilon\, |{\bar y}_l|}
    \int d{\bf y}\, d{\bf \bar y}\,
    d{\bf x}_1\, d{\bf \bar x}_1\,
    d{\bf x}_2\, d{\bf \bar x}_2\,
    d{\bf x}_g\, d{\bf \bar x}_g\, 
     \hat{\Gamma}_{\bf y}\, \hat{\Gamma}^\dagger_{\bf \bar y}
         \nonumber \\
    && \quad \times 
    G_{(q)}^{aa_1}({\bf x}_1;{\bf y}|p_1)\, T^{c_1}_{a_1b_1}\,
    G_{(g)}^{c_1c}({\bf y};{\bf x}_g|\omega)\,
    G_{(q)}^{b_1b}({\bf y};{\bf x}_2|p_2)
    \nonumber \\
    && \quad \times
    G_{(q)}^{b\bar{b}_1}({\bf \bar x}_2;{\bf \bar y}|p_2)\,
    G_{(g)}^{c\bar{c}_1}({\bf \bar x}_g;{\bf \bar y}|\omega)\,
    T^{{\bar c}_1}_{\bar{b}_1\bar{a}_1}\, 
    G_{(q)}^{\bar{a}_1a}({\bf \bar y};{\bf \bar x}_1|p_1)
    \nonumber \\
    && \quad \times    
    e^{i{\bf p}_{1\perp}\cdot ({\bf x}_1 - {\bf \bar x}_1)
      - i{\bf p}_{2\perp}\cdot ({\bf x}_2 - {\bf \bar x}_2)
       - i{\bf k}_{\perp}\cdot ({\bf x}_g - {\bf \bar x}_g)}\, .
    \label{2.17} 
\end{eqnarray}

\begin{figure}[h]\epsfxsize=8.7cm 
\centerline{\epsfbox{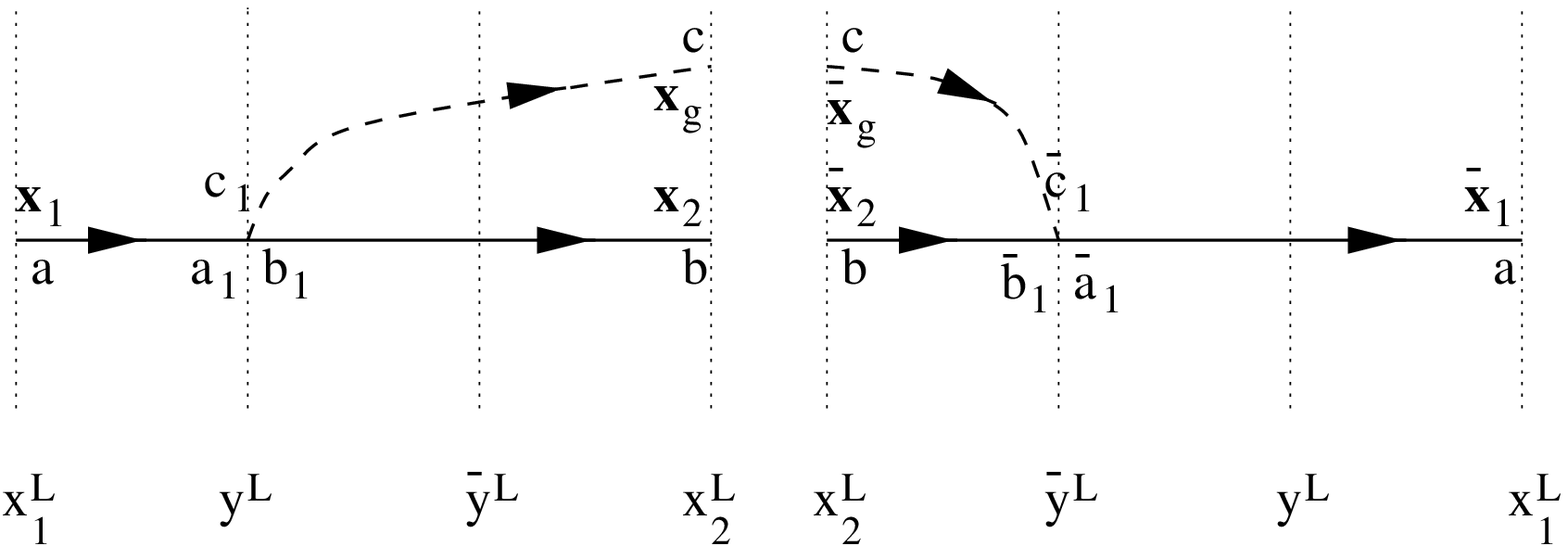}}
\vspace{0.5cm}
\caption{ Diagrammatic representation of the radiation
probability (\protect\ref{2.17}) in configuration space. 
For details, see text.}\label{fig2}
\end{figure}

The radiation probability (\ref{2.17}) has a simple 
diagrammatic representation in configuration space which we give
in Fig.~\ref{fig2}. We denote by solid lines the full quark 
Greens's functions $G_{(q)}$, and by dashed lines the gluon
Green's functions $G_{(g)}$. Contributions to the amplitude are 
on the l.h.s., those to the complex conjugate amplitude on the 
r.h.s. . We consider only the case that the gluon radiation
occurs at larger longitudinal position in the complex conjugate 
amplitude, $\bar{y}^l > y^l$. Taking twice the real part in
(\ref{2.17}) accounts automatically for the other case. Transverse
integration variables and colours are specified at the end points 
of the Green's functions. In this way, the simple diagram of 
Fig.~\ref{fig2} characterizes the rather lengthy expression
(\ref{2.17}) completely.

The following sections are devoted to a study of the 
${\bf k}_\perp$-differential gluon radiation cross section off
a free incoming quark. From (\ref{2.17}), we obtain
\begin{eqnarray}
  &&{d^3\sigma^{(in)}\over d(\ln x)\, d{\bf k}_\perp}
  = {\alpha_s\over (2\pi)^2}\, {1\over 4\, E_1^2\, (1-x)^2}
    \int d{\bf p}_\perp\, 
    \langle \vert{\cal M}_{a\to b\, c}\vert^2\rangle
    \nonumber\\
  &&= C_{\rm pre}\, 
    2\, {\rm Re}\, 
    \int_{z_-}^{z_+} dy_l\, \int_{y_l}^{z_+} d{\bar y}_l\, 
    e^{i\bar{q}(y_l-{\bar y}_l) -\epsilon\, |y_l|-\epsilon\, |{\bar y}_l| }\, 
    \nonumber \\
    && \times 
    \int d{\bf y}\, d{\bf \bar y}\,
    d{\bf x}_1\, d{\bf \bar x}_1\,
    d{\bf x}_g\, d{\bf \bar x}_g\, d{\bf x}_2\,
    e^{- i{\bf k}_{\perp}\cdot ({\bf x}_g - {\bf \bar x}_g)}\, .
         \nonumber \\
    &&\times \Bigg \langle 
    G_{(q)}^{aa_1}({\bf x}_1;{\bf y}|p_1)\, T^{c_1}_{a_1b_1}\,
    \left({\partial\over \partial {\bf y}} 
           G_{(g)}^{c_1c}({\bf y};{\bf x}_g|\omega)\right) 
    \nonumber \\
    &&\quad \times
    G_{(q)}^{b_1b}({\bf y};{\bf x}_2|p_2)\,
    G_{(q)}^{b\bar{b}_1}({\bf x}_2;{\bf \bar y}|p_2)\,
    \nonumber \\
    &&\quad \times
    \left({\partial\over \partial {\bf \bar y}}  
           G_{(g)}^{c\bar{c}_1}({\bf \bar x}_g;{\bf \bar y}|\omega)\right)\,
    T^{{\bar c}_1}_{\bar{b}_1\bar{a}_1}\, 
    G_{(q)}^{\bar{a}_1a}({\bf \bar y};{\bf \bar x}_1|p_1)\Bigg \rangle \, ,
    \label{2.18} 
\end{eqnarray}
where
\begin{equation}
  C_{\rm pre} =  {\alpha_s\over (2\pi)^2}\, 
                {1\over x^2\, E_1^2\, (1-x)^2}
  \label{2.19}
\end{equation}
accounts for a combination of recurring kinematical factors. Expression
(\ref{2.18}) is given in the frame in which the incoming quark has
vanishing transverse momentum ${\bf p}_{1\perp} = 0$. The 
medium average $\langle \dots \rangle$ over the colours and positions 
of the scattering centers is specified in section~\ref{sec2a} below. 
In appendix~\ref{appb},
we give details of why the combination of interaction vertices 
(\ref{2.16}) can be written to leading order in $O(x)$ as derivatives 
acting on the gluon Green's functions. 
  
\subsection{Medium average}
\label{sec2a}

The main tool of our analysis of the radiation cross section
(\ref{2.18}) will be an expansion of the Green's functions
in powers of the scattering potential $A_0$:
\begin{eqnarray}
  &&G({\bf r},z;{\bf r}',z'|p) \equiv G_0({\bf r},z;{\bf r}',z'|p) 
  -i\int\limits_{z}^{z'} d\xi
  \nonumber \\
  && \qquad \times 
  \int d{\bbox \rho}\, G_0({\bf r},z;{\bbox \rho},\xi|p)\, 
  A_0({\bbox \rho},\xi)\, 
  G_0({\bbox \rho},\xi;{\bf r}',z'|p)\,
  \nonumber \\
  && \quad + {\cal P} \int\limits_{z_L}^{x_L} d\xi_1\, 
       \int\limits_{\xi_1}^{x_L} d\xi_2\, 
       \int d{\bbox \rho}_1\, d{\bbox \rho}_2\,
       G_0({\bf r},z;{\bbox \rho}_1,\xi_1|p)
  \nonumber \\
  && \qquad \times  i\, 
  A_0({\bbox \rho}_1,\xi_1)\, 
  G_0({\bbox \rho}_1,\xi_1;{\bbox \rho}_2,\xi_2|p)\, 
  \nonumber \\
  && \qquad \times i\, 
  A_0({\bbox \rho}_2,\xi_2)\, 
  G({\bbox \rho}_2,\xi_2;{\bf r}',z'|p)\, .
  \label{2.20}
\end{eqnarray}
Here, $G_0$ is the free non-interacting Green's function
\begin{equation}
  G_0({\bf r},z;{\bf r}',z'|p) \equiv \frac{p}{2\pi i(z'-z)}
  \exp\left\{ \frac{ip\, \left({\bf r}-{\bf r}'\right)^2}{2(z'-z)} 
              \right\}\, ,
  \label{2.21}
\end{equation}
and the only distinction between the quark and gluon Green's 
function lies in the colour representation of the spatially
extended colour potential $A_0$. An expansion of the radiation
cross section (\ref{2.18}) to order $O(A_0^{2n})$ will turn out 
to be an expansion to $n$-th order in the opacity of the 
medium~\cite{WG99,W00}. 
Accordingly, we shall refer to (\ref{2.20}) as opacity expansion.

We now discuss the calculation of the medium average in (\ref{2.18})
to fixed order in opacity. To this end, we write for a single scattering 
potential, centered at $(\check{\bf r}_i, \check{z}_i)$, 
\begin{eqnarray}
  \varphi_i^a({\bf x}_\perp,\xi) &=& \delta^{a\, a_i}
  \int \frac{d^3\bbox{q}}{(2\pi)^3}\,
  a_0(\bbox{q}_\perp)
  \nonumber \\
  && \qquad \times  
  e^{-i({\bf x}_\perp - \check{\bf r}_i)\cdot \bbox{q}_\perp}
  e^{-i(\xi-\check{z}_i)q_l}\, .
  \label{2.22}
\end{eqnarray}
Here, the high-energy approximation
$a_0(\bbox{q}) \approx a_0(\bbox{q}_\perp)$
implies that the effective momentum transfer occurs at fixed 
longitudinal position $\xi=\check{z}_i$. The medium average 
$\langle \dots \rangle$ is the average over the transverse 
and longitudinal positions $(\check{\bf r}_i, \check{z}_i)$ 
of the scattering potentials and over the colours $a_i$
exchanged with the $i$-th scattering center,
\begin{eqnarray}
  \langle f \rangle &\equiv& \frac{1}{A_\perp}\, 
  \left(\prod_{i=1}^N \sum_{a_i}\, \int d\check{\bf r}_i\, 
    d\check{z}_i\right)\,
  \nonumber \\
  && \qquad \times f(\check{\bf r}_1,\dots,\check{\bf r}_N;
    \check{z}_1,\dots,\check{z}_N; a_1,\dots,a_N)\, .
  \label{2.23}
\end{eqnarray}
$A_\perp$ is a total transverse area which we divide out 
to regain the cross section per unit transverse area.
$N$ is the number of different single scattering potentials
up to which the function $f$ is expanded. 

To simplify notation, we replace in what follows the discrete 
sum over $\check{z}_i$ in (\ref{2.10}) by an integral over the
density $n$ of scattering centers
\begin{equation}
  A_0({\bf x}_\perp,\xi) = \int d\check{z}_i\, n(\check{z}_i)\,
  \varphi_i^a({\bf x}_\perp,\xi)\, T^a\, .
  \label{2.24}
\end{equation}
Since the effective momentum transfer from the single potential
$\varphi_i^a({\bf x}_\perp,\xi)$ occurs at $\xi=\check{z}_i$, we
shall often work with $\xi$ as integration variable. 

\subsection{Leading $O(x)$ approximation}
\label{sec2b}

We restrict our analysis of (\ref{2.18}) to the kinematical region
where the fraction $x$ of the energy carried away by the gluon is
small, $x \ll 1$. This $O(x)$ approximation is commonly used in the 
analysis of the abelian and non-abelian LPM-effect. It focusses on
the kinematical region in which most of the radiation occurs. 
Technically, it allows for two important simplifications:
\begin{enumerate}
\item
  The leading $O(x)$ spin- and helicity-averaged product of the 
  interaction vertices takes in momentum space the simple form
  \begin{equation}
   \hat{\Gamma}_{\bf y}\, \hat{\Gamma}^\dagger_{\bf \bar y}
   \longrightarrow 
   {4\over x^2}\, {\bf k}_{1\perp}\cdot {\bf \bar k}_{1\perp}\, ,
   \label{2.25}
  \end{equation}
  where ${\bf k}_{1\perp}$ and ${\bf \bar k}_{1\perp}$ are the
  transverse momenta of the gluon at the radiation vertex in
  the amplitude (${\bf k}_{1\perp}$) and complex conjugate
  amplitude (${\bf \bar k}_{1\perp}$). In configuration space,
  these momenta are conjugate to the ${\bf y}$-derivatives of the
  gluon Green's functions. Details are explained in
  appendix~\ref{appb}.
\item
  Propagation of the transverse plane waves by Green's
  functions leads in (\ref{2.17}) to the appearance of 
  longitudinal phase factors:
  \begin{eqnarray}
    &&\int d{\bf x}_2\, G_0({\bf x}_1,z_1;{\bf x}_2,z_2|E)\, 
    e^{-i\, {\bf p}_\perp\cdot {\bf x}_2} 
    \nonumber\\
    && = e^{-i\, {\bf p}_\perp\cdot {\bf x}_1}\, 
    \exp\left[-i {{\bf p}_\perp^2\over 2\, E}
      \left(z_2 - z_1\right)\right]\, .
    \label{2.26}
  \end{eqnarray}
  To leading $O(x)$, the longitudinal phase receives only contributions
  from the transverse energy ${\bf p}_\perp^2 / 2\, \omega$ of the gluon.
  Transverse energies ${\bf p}_\perp^2 / 2\, E_1$ of the quarks
  are suppressed by one order in $x$. In this sense, the
  leading $O(x)$ approximation neglects the Brownian motion associated
  with the rescattering of the hard quark. Gluon rescattering is 
  the dominant contribution, as observed by BDMPS.
\end{enumerate}

\section{Opacity expansion}
\label{sec3}

For more than $N=1$ scattering centers, the opacity expansion of the 
radiation cross section 
(\ref{2.18}) involves non-vanishing interference terms between 
amplitudes of different powers in $A_0$. Two examples of non-vanishing 
interference effects between amplitudes of order $O(A_0)$ and $O(A_0^3)$ 
in the $N=2$ opacity expansion are e.g.
\begin{equation}
\epsfxsize=7.7cm 
\centerline{\epsfbox{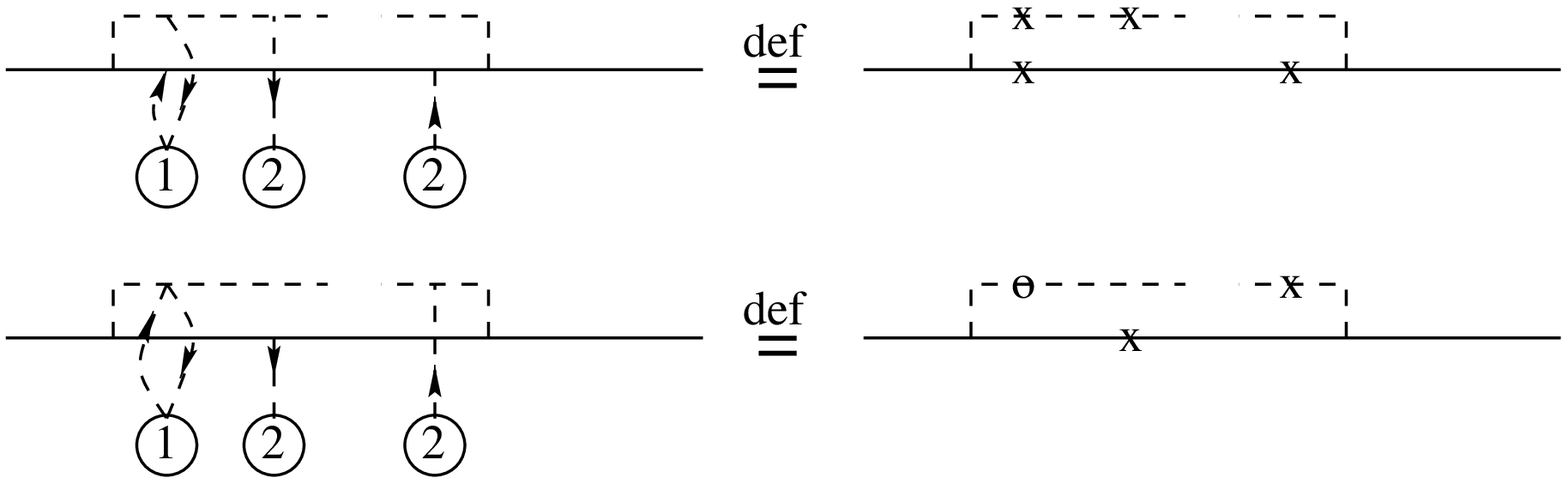}}
\vspace{-1cm}
\label{3.1}
\end{equation}
In what follows, we use the expression {\it contact term} to denote
scattering centers which link with two gluons to one amplitude
[as e.g. the first scattering center in (\ref{3.1})]. The notion
{\it real term} characterizes a scattering center which exchanges
one gluon with the amplitude and one with the complex conjugate
amplitude [as does the second scattering center in (\ref{3.1})].
The structure of the medium average (\ref{2.23}) ensures that all
diagrammatic contributions are combinations of real and contact
terms. For a medium of density $n(\xi)$ and thickness $L$, 
arbitrary combinations of $N$ real and contact terms contribute
to the same order $\left(\alpha_{\rm s} \int_0^\infty n(\xi)\, d\xi\right)^N$
in opacity. Hence, contact term contributions cannot be neglected in
the calculation of (\ref{2.18}). Calculations which do not include
contact terms~\cite{WG99,GLV99} describe an exclusive measurement 
in which the number of recoil target partons is determined.
In section~\ref{sec3a}, we discuss 
the fundamental properties of contact terms as well as a set of 
identities which relate them to real terms. In section~\ref{sec3b},
we use these identities to derive a manifestly colour trivial
expression (\ref{3.13}) for the radiation cross section off $N$
scattering centers. Then, we calculate (\ref{3.13}) in an opacity
expansion for $N=1$ and $N=2$.

\subsection{Notation and identities}
\label{sec3a}

The medium average (\ref{2.23}) ensures that contact terms do not 
transfer a net momentum to the final state. However, they can change 
the transverse momentum between the outgoing particles~\cite{W00}. 
This can be seen, e.g., by writing for the first scattering center in 
(\ref{3.1}) the corresponding contributions (\ref{2.22}) and realizing
that the medium average (\ref{2.23}) results in a $\delta$-function
$\delta^{(2)}\left( \bbox{q}_{1\perp} + \bar{\bbox{q}}_{1\perp}
\right)$ with the transverse momenta of both 
gluons appearing in the same amplitude.

Contact terms satisfy several important identities. All the
following identities can be established by writing out the opacity
expansion (\ref{2.20}) for the last scattering center before the
cut and employing the medium average (\ref{2.23}). Several 
examples of this technique were discussed explicitly in~\cite{W00}.
Here, we consider diagrams to $N$-th order in opacity. We denote by 
a shaded box the longitudinal region in which the $N-1$ first 
interactions take place, and we specify the contribution of the 
last $N$-th scattering center explicitly. We start with the case 
that the last interaction links to the quark line. Irrespective of 
whether it occurs after $\bar{y}_l$,
\begin{equation}
\epsfxsize=7.7cm 
\centerline{\epsfbox{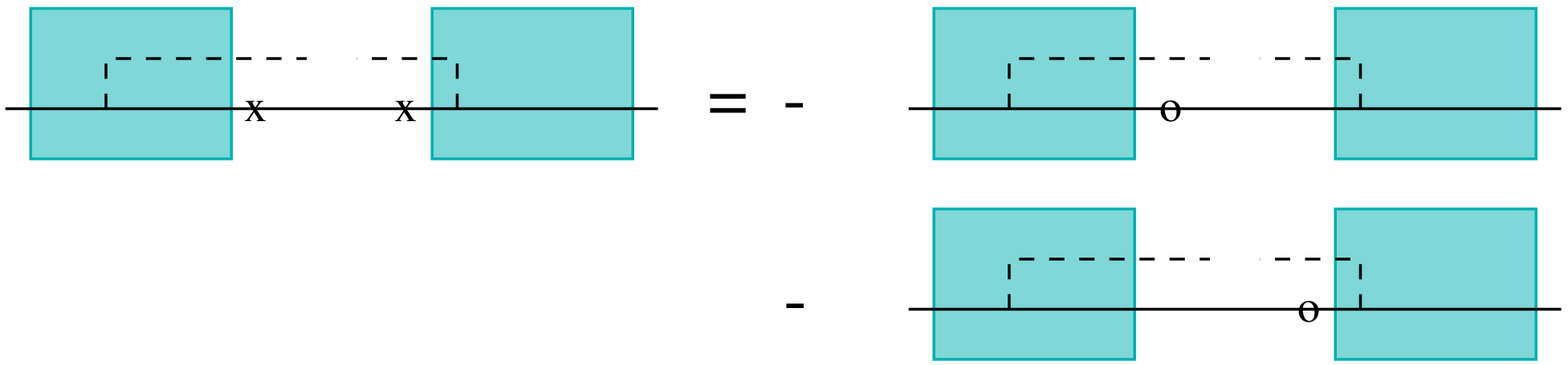}}
\vspace{-1cm}
\label{3.2}
\end{equation}
or before ${y}_l$,
\begin{equation}
\epsfxsize=7.7cm 
\centerline{\epsfbox{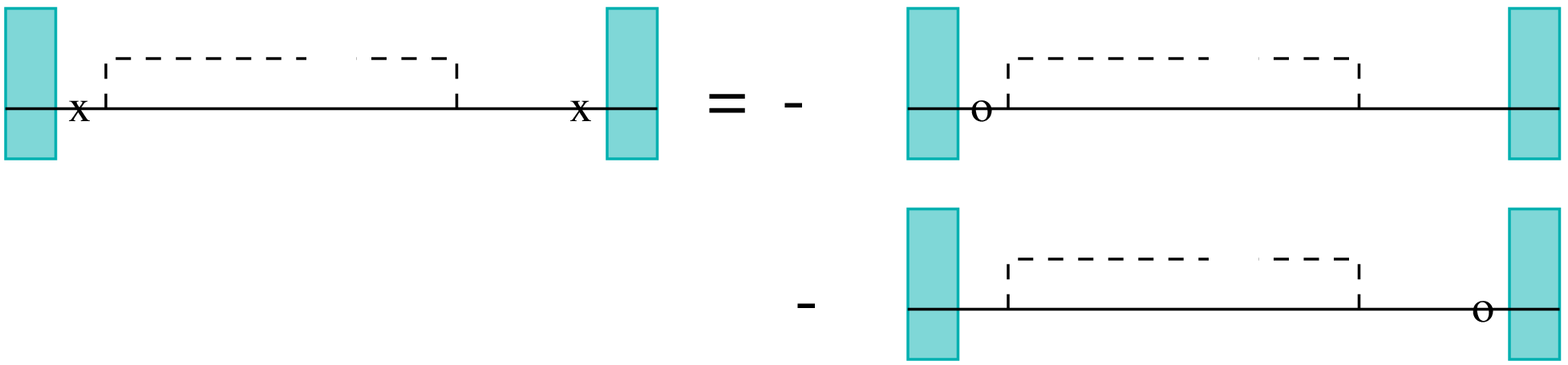}}
\vspace{-1cm}
\label{3.3}
\end{equation}
the real contribution equals in both cases minus the two
corresponding contact terms and thus cancels against them. 
We note that (\ref{3.2}) holds exactly. It is based on the
identity derived in Fig. 7a of Ref.~\cite{W00}. Equation (\ref{3.3}),
in contrast, involves the leading $O(x)$ approximation:
the l.h.s. of (\ref{3.3}) involves a phase $\exp\left[ 
-i \frac{{\bf p}_\perp^2}{2\, p_1}(y_l-\bar{y}_l)\right]$
whose transverse momentum ${\bf p}_\perp^2$ differs by
the amount $\bbox{q}_{N\perp}$ of the last rescattering
from the corresponding phase on the r.h.s. of (\ref{3.3}). 
This phase does not contribute in the leading $O(x)$ approximation
where the momentum transfer to the quark line is neglected and 
only transverse energies of the gluon are kept. The same 
argument allows for an identity concerning the first 
scattering center:
\begin{equation}
\epsfxsize=7.7cm 
\centerline{\epsfbox{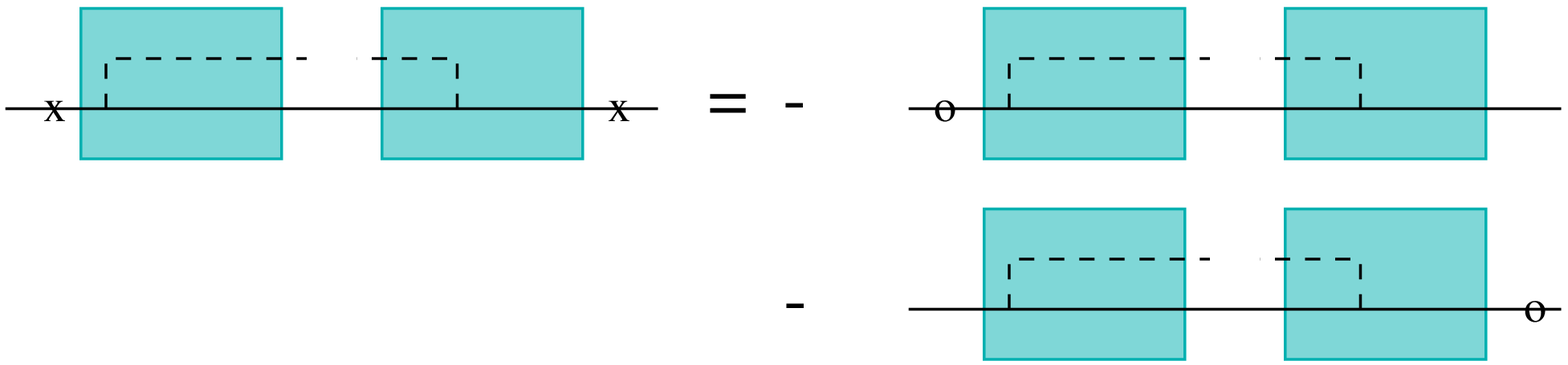}}
\vspace{-1cm}
\label{3.4}
\end{equation}
The important implication of this identity is that the sum of
all rescattering contributions to (\ref{2.18}) which have the
first interaction before $y_l$, vanishes. This eliminates a
large class of diagrams from our calculation.

In addition, there are identities in which one of the 
gluon lines is touched:
\begin{equation}
\epsfxsize=7.7cm 
\centerline{\epsfbox{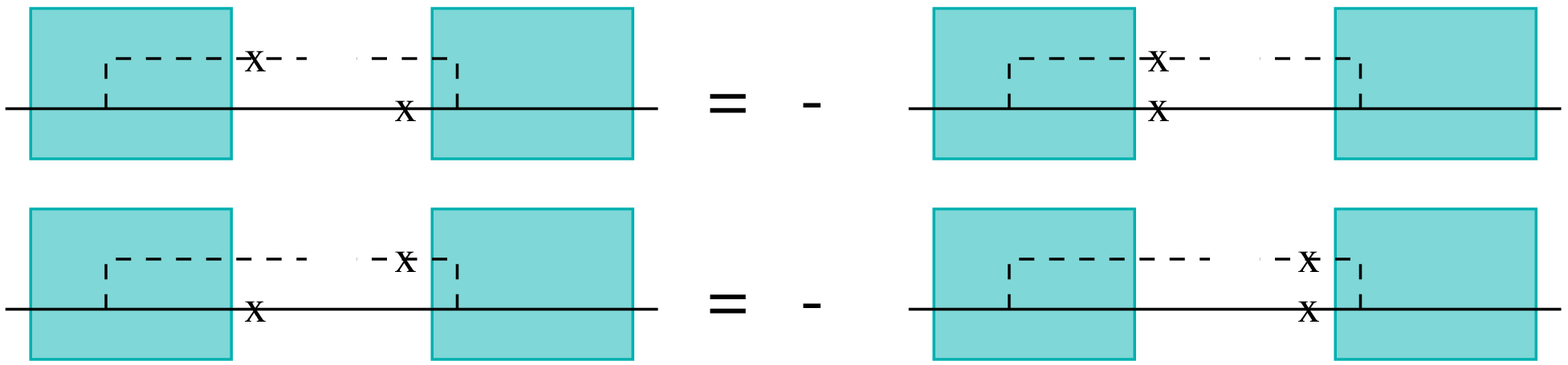}}
\vspace{-1cm}
\label{3.5}
\end{equation}
These identities are exact and were derived already in Fig. 7b 
of Ref.~\cite{W00}.

So far, we have discussed identities for which the last interaction
occured either after $\bar{y}_l$ or before $y_l$. Now we give relations
which hold for a last interaction at longitudinal position
$y_l < \xi < \bar{y}_l$. These identities necessarily involve
knowledge of the colour algebra. The simplest example is
\begin{equation}
\epsfxsize=7.7cm 
\centerline{\epsfbox{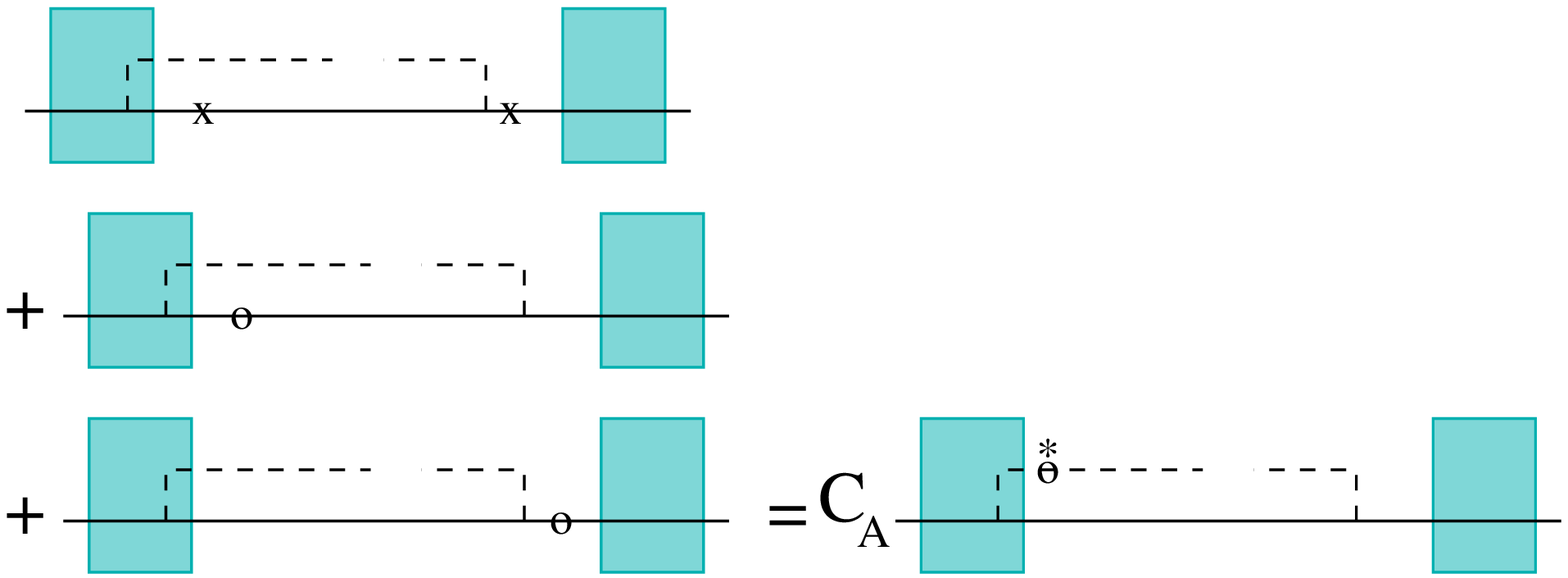}}
\vspace{-1cm}
\label{3.6}
\end{equation}
Here, the r.h.s. shows a star over the interaction term. This 
notation indicates an interaction term without colour factor.
The colour information is absorbed in the prefactor of the diagram. 
To derive (\ref{3.6}), we denote the colour generators of the left 
and right shaded blocks in (\ref{3.6}) by factors $L$ and $R$,
respectively. The colour trace of the first term on the l.h.s. 
then takes the form
\begin{equation}
  {\rm Tr}\left[L\, T_d\, T_c\, T_d\, R\right] = 
  \left(C_F - \frac{C_A}{2}\right)\, 
  {\rm Tr}\left[L\, T_c\, R\right]\, ,
  \label{3.7}
\end{equation}
where $T_d$ is the generator associated with the $N$-th interaction
and $T_c$ that of the quark-gluon emission vertex at $\bar{y}_l$.
The contact terms on the  left and right hand side of (\ref{3.6})
are proportional to ${\rm Tr}\left[L\, T_c\, R\right]$ with prefactors
$- \frac{C_F}{2}$ and $- \frac{C_A}{2}$, respectively. In the 
leading $O(x)$ approximation, there is no complication from the 
momentum structure of the diagrams and equation (\ref{3.7}) thus
ensures the identity (\ref{3.6}). 
  
In a similar way, one can establish an identity for diagrams in 
which one of the interactions links to the gluon line:
\begin{equation}
\epsfxsize=7.7cm 
\centerline{\epsfbox{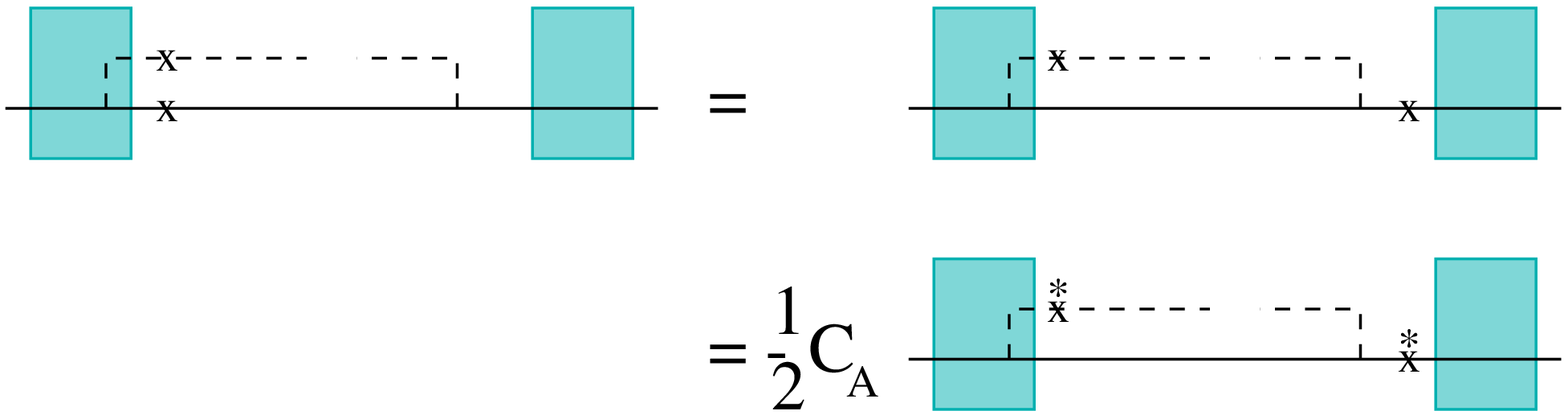}}
\vspace{-1cm}
\label{3.8}
\end{equation}
This is a direct consequence of the relation
\begin{equation}
  i\, f_{\bar{c}dc}\, {\rm Tr}\left[ L\, T_c\, T_d\, R\right] 
  = \frac{1}{2}\, C_A\, {\rm Tr}\left[ L\, T_{\bar{c}}\, R\right]\, .
  \label{3.9}
\end{equation}
%

\subsection{Colour triviality of the gluon radiation cross section}
\label{sec3b}

The identities of the last subsection have two important properties
which we exploit systematically in what follows: (i) they allow
for many diagrammatic cancellations between real and contact terms,
and (ii) they ensure that the remaining terms combine to expressions
which are proportional to $C_A^N$.

\subsubsection{Classification of diagrams}
\label{sec3b1}

In what follows, we consider all diagrammatic contributions which
are $N$-th order in opacity. The $N-1$ first interaction terms
are again summarized diagrammatically in a shaded block on both
sides of the cut. The last interaction is specified explicitly.
We consider three cases\\

\underline{Case I}: The last interaction occurs at longitudinal position
    $\xi_N < y_l$. \\
    In this case, the identity (\ref{3.2}) ensures that the
    sum of all real and contact terms of the last interaction
    vanish.

\underline{Case II}: The last interaction occurs at longitudinal position
    $y_l < \xi_N < \bar{y}_l$.\\
    In this case, we have two real and four contact contributions
    from the last interaction. According to the identities
    (\ref{3.6}) and (\ref{3.8}), they add up in the following
    way:
\begin{equation}
\epsfxsize=7.7cm 
\centerline{\epsfbox{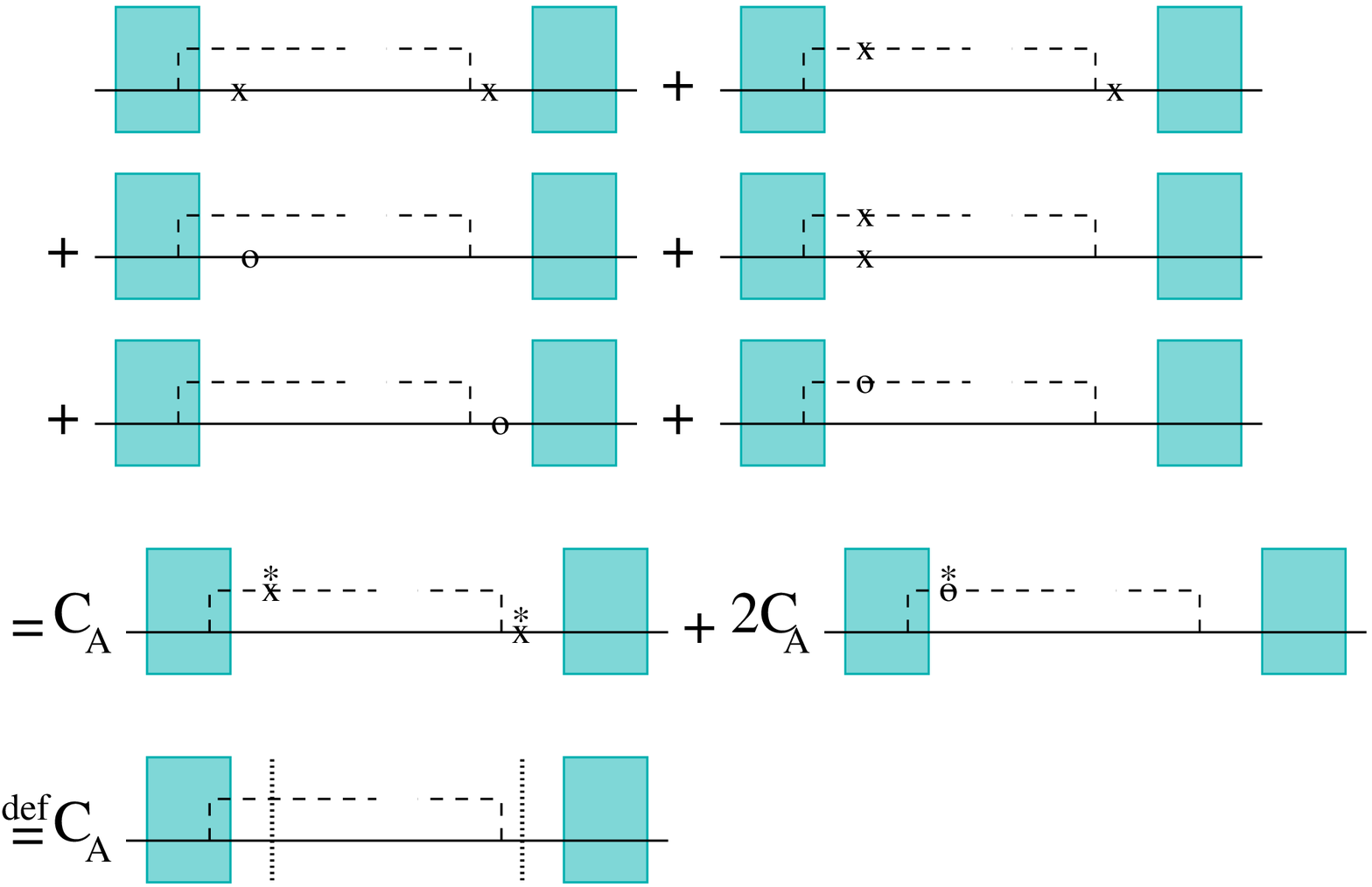}}
\vspace{-1cm}
\label{3.10}
\end{equation}
In the last line, we have introduced a simple shorthand which
summarizes all non-vanishing contributions for $y_l < \xi_N < \bar{y}_l$.
For what follows, it is important that this shorthand comes with a 
unique colour prefactor $C_A$ which absorbs the colour algebra of the
last interaction. 

\underline{Case III}: The last interaction occurs at longitudinal position
    $\bar{y}_l < \xi_N$.\\
    In this case, we have four real and six contact term contributions.
    Identity (\ref{3.2}) ensures that the one real and two contact
    terms shown there cancel each other. Identity (\ref{3.5}) leads
    to the cancellation of two real versus two contact terms. For 
    the remaining one real and two contact terms, we introduce the
    notational shorthand:
\begin{equation}
\epsfxsize=7.7cm 
\centerline{\epsfbox{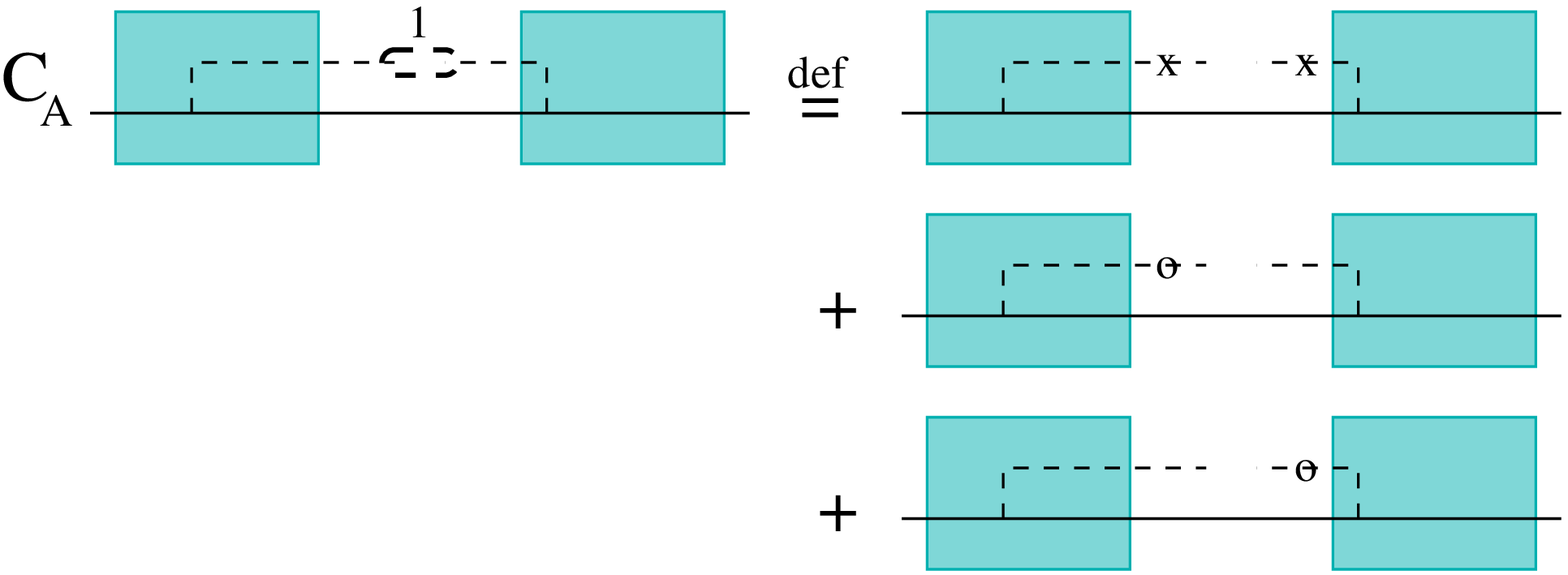}}
\vspace{-1cm}
\label{3.11}
\end{equation}
Again, the sum of all non-vanishing contributions comes with a colour
prefactor $C_A$ which absorbs the colour algebra of the
last interaction.

\subsubsection{Iteration}
\label{sec3b2}
The above classification of diagrams shows that the colour structure
of all surviving contributions to the last $N$-th interaction can
be absorbed in a prefactor $C_A$. This allows for an iteration of
the above procedure to the earlier $(N-1)$-th, $(N-2)$-th, 
$(N-3)$-th etc. interactions. 

We consider first the case that the $m$ last interactions occur
at positions larger $\bar{y}_l$, i.e., $\bar{y}_l < \xi_{N-m} <
\xi_{N-m+1}< \dots < \xi_N$. For the $N$-th interaction, we are
left with the diagrams of (\ref{3.11}). Since the colour structure
is absorbed in the prefactor $C_A$, the arguments leading to
(\ref{3.11}) apply also to the $(N-1)$-th interaction. If the
$(N-(m+1))$-th interaction also occurs after $\bar{y}_l$, then
the sum of all nonvanishing contributions is given by the
recursively defined diagram
\begin{equation}
\epsfxsize=7.7cm 
\centerline{\epsfbox{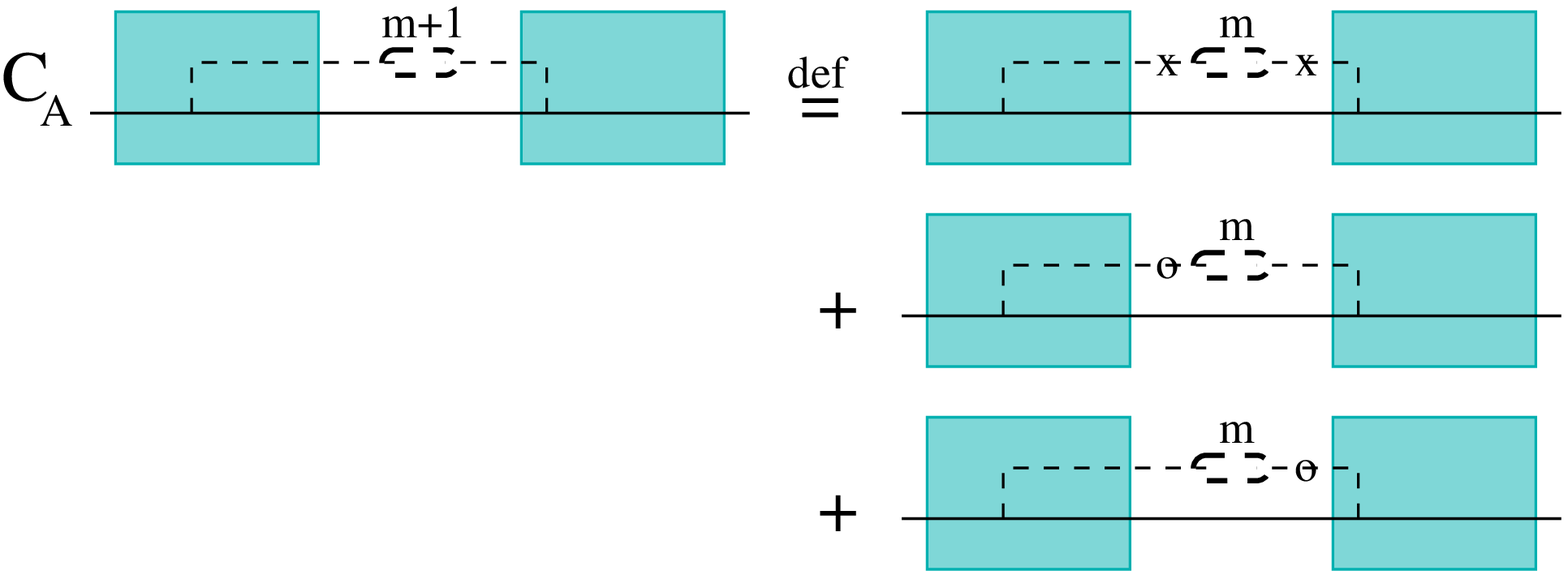}}
\vspace{-1cm}
\label{3.12}
\end{equation}
For $N$-fold rescattering, an arbitrary number $m<N$ interactions 
will occur after $\bar{y}_l$, the remaining $(N-m)$ interactions
will occur at longitudinal positions between $y_l$ and $\bar{y}_l$.
The contributions of all possible diagrams with interactions
before $y_l$ cancels due to the identity (\ref{3.4}).
The radiation cross section (\ref{2.18}) to $N$-th order in opacity
can thus be represented in the compact diagrammatic and explicitly
colour trivial form 
\begin{eqnarray}
  &&{d^3\sigma^{(in)}(N)\over d(\ln x)\, d{\bf k}_\perp}
  = {\alpha_s\over (2\pi)^2}\, {1\over \omega^2}\,
    N_C\, C_F\, C_A^N
    \label{3.13}\\
&&
\epsfxsize=6.0cm 
\centerline{\epsfbox{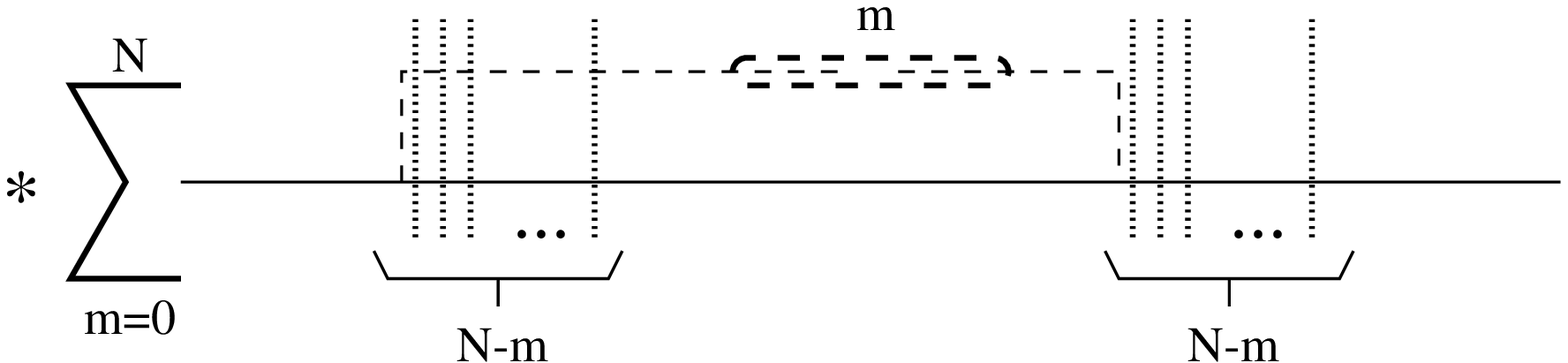}}
\vspace{-1cm}
\nonumber 
\end{eqnarray}
This is the diagrammatic manifestation of the observation of BDMPS
that gluon rescattering alone gives the leading $O(x)$ contribution
to the radiation cross section (\ref{2.18}). As can be seen from 
(\ref{3.10}) and (\ref{3.12}), all $N$ scattering centers involved
transfer momentum to the gluon line.

We introduce for the diagrammatic part
of (\ref{3.13}) an analytical shorthand 

\begin{eqnarray}
&&
\epsfxsize=6.0cm 
\centerline{\epsfbox{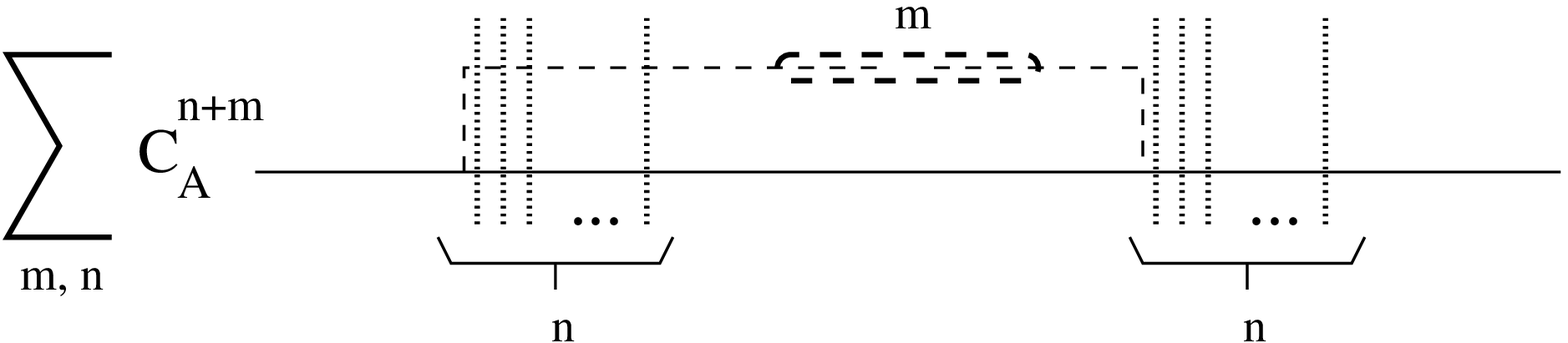}}
\vspace{-1cm}
\nonumber \\
  &&  = \int_Y\, \int_{[q_i,\xi_i]_{i\in [1,N]}}\, 
  \int  d{\bf y}\, d{\bf \bar y}\,
    d{\bf s}\, d{\bf x}_g\, d{\bf \bar x}_g
         \nonumber \\
    &&\qquad \times 
    \left({\partial\over \partial {\bf y}} 
           \tilde{G}({\bf y};{\bf x}_g|\omega)\right)\,
         e^{- i{\bf k}_{\perp}\cdot {\bf x}_g}\, 
    \tilde{G}_0({\bf y};{\bf \bar y}|p_2)
    \nonumber \\
    &&\qquad \times
    e^{i{\bf k}_{\perp}\cdot {\bf \bar x}_g}
    \left({\partial\over \partial {\bf \bar y}}  
           \tilde{G}({\bf \bar x}_g;{\bf \bar y}|\omega)\right)\, 
    \tilde{G}({\bf \bar y},\bar{y}_l;{\bf s},y_l|p_1)\, .
    \label{3.14}
\end{eqnarray}
To specify our notation, we compare (\ref{3.13}) and (\ref{3.14})
to the full radiation cross section (\ref{2.18}). 
The term ${\alpha_s\over (2\pi)^2} {1\over \omega^2}$ in (\ref{3.13}) 
is the leading order $O(x)$ of the prefactor $C_{\rm pre}$. The colour 
algebra of (\ref{2.18}) simplifies to the Casimir factors in (\ref{3.13}). 
Accordingly, the functions $\tilde{G}$ introduced in (\ref{3.14}) are 
abelian. Some quark Green's functions in the diagram of (\ref{3.14}) are 
non-interacting to arbitrary orders in opacity. This allows to  
integrate over ${\bf x}_1$, ${\bf \bar x}_1$ and ${\bf x}_2$ 
in (\ref{2.18}). The new integration variable 
${\bf s}$ in (\ref{3.14}) denotes the transverse coordinate of the quark at
longitudinal position $y_l$ in the complex conjugate amplitude.
The longitudinal $y_l$- and $\bar{y}_l$-integrals in (\ref{2.18}) are 
given by the notational shorthand 
\begin{equation}
  \int_Y f \equiv 2{\rm Re} \int_{z_-}^{z_+} dy_l  
  \int_{y_l}^{z_+} d\bar{y}_l\, e^{-\epsilon |y_l|\, 
  -\epsilon |\bar{y}_l|}\, f \, ,
  \label{3.15}
\end{equation}
where the function $f$ denotes an unspecified integrand.
To $N$-th order in opacity, the medium average $\langle \dots
\rangle$ in (\ref{2.18}) results in $N$ transverse momentum
$\bbox{q}_{i\perp}$-integrals,
\begin{equation}
  \int_{q_i}\, f \equiv 
            \int \frac{d\bbox{q}_{i\perp}}{(2\pi)^2}
                         | a_0(\bbox{q}_{i\perp})|^2\, f\, ,
                         \label{3.16}
\end{equation}
which average over the elastic scattering cross sections 
$| a_0(\bbox{q}_{i\perp})|^2$ for each scattering center.
Also, it leads to $N$ integrals over the allowed longitudinal 
positions $\xi_i\in [\xi_i^{\rm min},\xi_i^{\rm max}]$ of these 
$N$ scattering centers,
\begin{equation}
  \int_{\xi_i} f \equiv \int_{\xi_i^{\rm min}}^{\xi_i^{\rm max}}\, 
  d\xi_i\, n(\xi_i)\, f\, .
  \label{3.17}
\end{equation}
These integrations are combined in (\ref{3.14}) into
\begin{equation}
  \int_{[q_i,\xi_i]_{i\in [1,N]}} f
  = \prod_{i=1}^N \int_{q_i} \int_{\xi_i}\, f\, .
  \label{3.18}
\end{equation}
As a consequence of factorizing in (\ref{2.18}) the colour algebra,
the elastic cross sections $| a_0(\bbox{q}_{i\perp})|^2$ and 
the longitudinal momentum integrals over $\xi_i$, the only remnants
of the opacity expansion (\ref{2.20}) which are left in the integrand 
of (\ref{3.14}) are transverse phase factors. Hence, the functions 
$\tilde{G}$ are defined by replacing in (\ref{2.20}) 
$A_0(\bbox{\rho},\xi_i) \to
e^{-i\bbox{q}_{i\perp}\cdot \bbox{\rho}}$. For each 
$\bbox{q}_{i\perp}$, two phases appear in the integrand of
(\ref{3.14}), one being the complex conjugate of the other. 
To gain familiarity with (\ref{3.14}), one may consider simple
examples like
\begin{eqnarray}
&&
\epsfxsize=6.0cm 
\centerline{\epsfbox{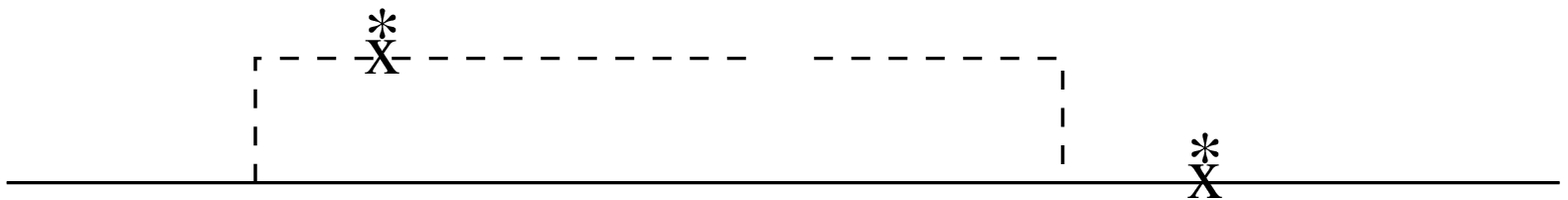}}
\vspace{-1cm}
\nonumber \\
  &&  = \int_Y\, \frac{1}{A_\perp} 
  \int_{q_1} \int_{y_l}^{\bar{y}_l} d\xi_1\, n(\xi_1)\, 
  \int  d{\bf y}\, d{\bf \bar y}\,
    d{\bf s}\, d{\bf x}_g\, d{\bf \bar x}_g\, 
         \nonumber \\
    &&\times 
    \left({\partial\over \partial {\bf y}} \int d\bbox{\rho}_1
            G_0({\bf y};\bbox{\rho}_1,\xi_1|\omega)\, 
            e^{- i \bbox{q}_{1\perp}\cdot \bbox{\rho}_1}\,
            G_0(\bbox{\rho}_1,\xi_1;{\bf x}_g|\omega)\right)
    \nonumber \\
    &&
    \times e^{- i{\bf k}_{\perp}\cdot ({\bf x}_g - {\bf \bar x}_g)}\, 
    G_0({\bf y};{\bf \bar y}|p_2)\, 
    \left({\partial\over \partial {\bf \bar y}}  
           G_0({\bf \bar x}_g;{\bf \bar y}|\omega)\right)\,    
    \nonumber \\
    &&\times
    \int d\bbox{\bar \rho}_1
            G_0(\bar{\bf y};\bbox{\bar \rho}_1,\xi_1|p_1)\, 
            e^{i \bbox{q}_{1\perp}\cdot \bbox{\bar \rho}_1}\,
            G_0(\bbox{\bar \rho}_1,\xi_1;{\bf s},{\bar y}_l|p_1)\, ,
            \nonumber \\
    &&=  \int_Y \int_{q_1}  
    \left({\bf k}_\perp + \bbox{q}_{1\perp}\right)\cdot {\bf k}_\perp\, 
    \int_{y_l}^{\bar{y}_l} d\xi_1\, n(\xi_1)\,     
    \nonumber \\
    && \qquad \times 
    e^{- i\frac{{\bf k}_\perp^2}{2\omega}(\bar{y}_l - \xi_1) 
      - i\frac{\left({\bf k}_\perp+\bbox{q}_{1\perp}\right)^2
               }{2\omega}(\xi_1-y_l)}\, .
    \label{3.19}
\end{eqnarray}
Here, we have used again the leading $O(x)$ approximation to 
neglect the subleading contributions to the longitudinal phase
of the integrand. 

All $N$-th order contributions to the radiation 
spectrum (\ref{3.13}) can be written as a sum of products of the 
form
\begin{eqnarray}
  {d^3\sigma^{(in)}(N)\over d(\ln x)\, d{\bf k}_\perp}
  &=& {\alpha_s\over (2\pi)^2}\, {1\over \omega^2}\,
    N_C\, C_F\, C_A^N
  \nonumber \\
  && \times \int_{q_1} \dots \int_{q_N}
    \sum_{i} \Gamma_{(i)}\, {\cal Z}_{(i)}\, .
    \label{3.20}
\end{eqnarray}
Here, the factor 
\begin{eqnarray}
  {\cal Z}_{(i)} &=& \int_Y \Phi_{(i)}(y_l,\bar{y}_l)\, .
  \label{3.21}  
\end{eqnarray}
denotes an integral over longitudinal phase factors.
To leading $O(x)$, this phase is obtained by iterative use of 
(\ref{2.26}).
The factor $\Phi_{(i)}$ in (\ref{3.21}) is the integral of 
this phase over the allowed longitudinal position
of scattering centers times a factor $-1/2$ for each contact term 
linking to the gluon line. In the example (\ref{3.19}), $\Phi_{(i)}$
is given  by the $\xi_1$-integral. 

The factors $\Gamma_{(i)}$ in (\ref{3.20}) denote the results of
the partial derivatives in (\ref{3.14}) for the $i$-th diagrammatic
contribution. They are easily read off from the diagrams by 
adding up the momentum transfers to the gluon line in both
amplitudes. For the example (\ref{3.19}), this leads to the
factor  $\left({\bf k}_\perp + 
\bbox{q}_{1\perp} \right)\cdot {\bf k}_\perp$.
We now turn to an evaluation of (\ref{3.13})  
in the low opacity expansion for $N=1$ and $N=2$. In section~\ref{sec4},
we derive then an all-order expression (\ref{4.15}) for this radiation 
cross section. As explained in sections~\ref{sec5} and \ref{sec6},
this all order expression gives a much simpler access to the 
opacity expansion since it involves only $N+1$ terms to $N$-th order.

\subsection{$N=1$ Gunion-Bertsch radiation spectrum}
\label{sec3c}

Here, we derive the Gunion-Bertsch radiation spectrum~\cite{GB82} as the 
$N=1$ term in the opacity expansion of (\ref{3.13}). This provides
a first consistency check for our formalism and allows for a
simple illustration of our diagrammatic rules. The $N=1$-term
of (\ref{3.13}) has four contributions
\begin{eqnarray}
  &&{d^3\sigma^{(in)}(N=1)\over d(\ln x)\, d{\bf k}_\perp}
  = {\alpha_s\over (2\pi)^2}\, {1\over \omega^2}\,
    N_C\, C_F\, C_A
    \nonumber\\
&&
\epsfxsize=6.0cm 
\centerline{\epsfbox{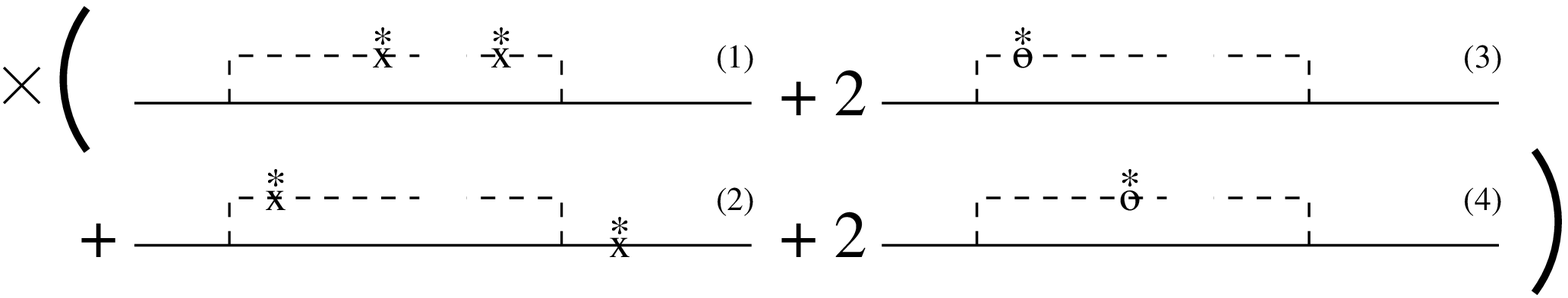}}
\vspace{-1cm}
\nonumber \\
&& = {\alpha_s\over (2\pi)^2}\, {1\over \omega^2}\,
    N_C\, C_F\, C_A \sum_{i=1}^4 \Gamma_{(i)}\, {\cal Z}_{(i)}\, .
    \label{3.22}
\end{eqnarray}
Here, we have labelled the different diagrammatic contributions by 
(1) - (4). Our diagrammatic rules distinguish between interactions
before and after $\bar{y}_l$. For that reason, the diagrams (3) and
(4) are both included in (\ref{3.22}). Contribution (2) is the 
example studied in (\ref{3.19}). The
$\Gamma$-factors are read  off from the $i$-th diagramm by adding 
up the transverse momentum flows into the gluon line on both sides 
of the cut:
\begin{eqnarray}
  \Gamma_{(1)} &=& \left( {\bf k}_\perp + \bbox{q}_{1\perp}\right)
                \cdot \left( {\bf k}_\perp + \bbox{q}_{1\perp}\right)
                \label{3.23}\\
  \Gamma_{(2)} &=& \left( {\bf k}_\perp + \bbox{q}_{1\perp}\right)
                \cdot {\bf k}_\perp 
                \label{3.24}\\
  \Gamma_{(3)} &=&  \Gamma_{(4)} = {\bf k}_\perp \cdot {\bf k}_\perp 
                \label{3.25}
\end{eqnarray}
For the leading $O(x)$ contributions to the factors ${\cal Z}_{(i)}$ 
in (\ref{3.19}), only the transverse momenta of the gluon have to be 
considered. In (\ref{3.22}), the gluon line comes with two possible 
transverse energies
\begin{equation}
  Q = \frac{{\bf k}_\perp^2}{2\omega}\, ,\qquad
   Q_1 = \frac{\left( {\bf k}_\perp 
       + \bbox{q}_{1\perp}\right)^2}{2\omega}\, .
  \label{3.26}
\end{equation}
Including a factor $-1/2$ for each contact term linking twice to the 
gluon line, we find for the phase factors $\Phi_{(i)}$
\begin{eqnarray}
  \Phi_{(1)}(y_l,\bar{y}_l) &=& \int_{\bar{y}_l}^{z_+}
        d\xi\, n(\xi)\, e^{-iQ_1(\bar{y}_l - y_l)}\, ,
        \label{3.27}\\
  \Phi_{(2)}(y_l,\bar{y}_l) &=& \int_{y_l}^{\bar{y}_l}
        d\xi\, n(\xi)\, e^{-iQ(\bar{y}_l - \xi)-iQ_1(\xi - y_l)}\, ,
        \label{3.28}\\ 
  \Phi_{(3)}(y_l,\bar{y}_l) &=& -\int_{\bar{y}_l}^{z_+}
        d\xi\, n(\xi)\, e^{-iQ(\bar{y}_l - y_l)}\, ,
        \label{3.29}\\
  \Phi_{(4)}(y_l,\bar{y}_l) &=& -\int_{y_l}^{\bar{y}_l}
        d\xi\, n(\xi)\, e^{-iQ(\bar{y}_l - y_l)}\, .
        \label{3.30}
\end{eqnarray}
For a homogeneous density distribution $n(\xi)=n_0$ of a medium of 
thickness $L$, the corresponding factors ${\cal Z}_{(i)}$ given in
(\ref{3.21}) are
\begin{eqnarray}
  {\cal Z}_{(1)} &=& \frac{L\, n_0}{Q_1^2}
  \, ,\qquad 
  {\cal Z}_{(2)} = \frac{-2\, L\, n_0}{Q\, Q_1}\, ,
  \nonumber \\
  {\cal Z}_{(3)} &=& \frac{2\, L\, n_0}{Q^2}
  \, ,\qquad 
  {\cal Z}_{(4)} = \frac{- L\, n_0}{Q^2}\, .
  \label{3.31}
\end{eqnarray}
Inserting this into the last line of (\ref{3.22}), we find
\begin{eqnarray}
  {d^3\sigma^{(in)}(N=1)\over d(\ln x)\, d{\bf k}_\perp}
  &=& \frac{\alpha_s}{\pi^2}\, \left(Ln_0\right)
    N_cC_FC_A 
    \nonumber \\
  && \times \int_{q_1} \frac{\bbox{q}_{1\perp}^2}{
      {\bf k}_\perp^2 \left(\bbox{q}_{1\perp} 
        + {\bf k}_\perp\right)^2}\, .
  \label{3.32}
\end{eqnarray}
This is the leading $O(x)$ contribution to the Gunion-Bertsch~\cite{GB82}
radiation cross section times the opacity $L\, n_0$ of a homogeneous
medium of thickness $L$.

\subsection{$N=2$ Opacity expansion}
\label{sec3d}

Before turning in the next section to a systematic all-order
analysis of the radiation cross section (\ref{3.13}), we calculate
here the $N=2$ contribution to the opacity expansion.
%
\begin{figure}[h]\epsfxsize=8.7cm 
\centerline{\epsfbox{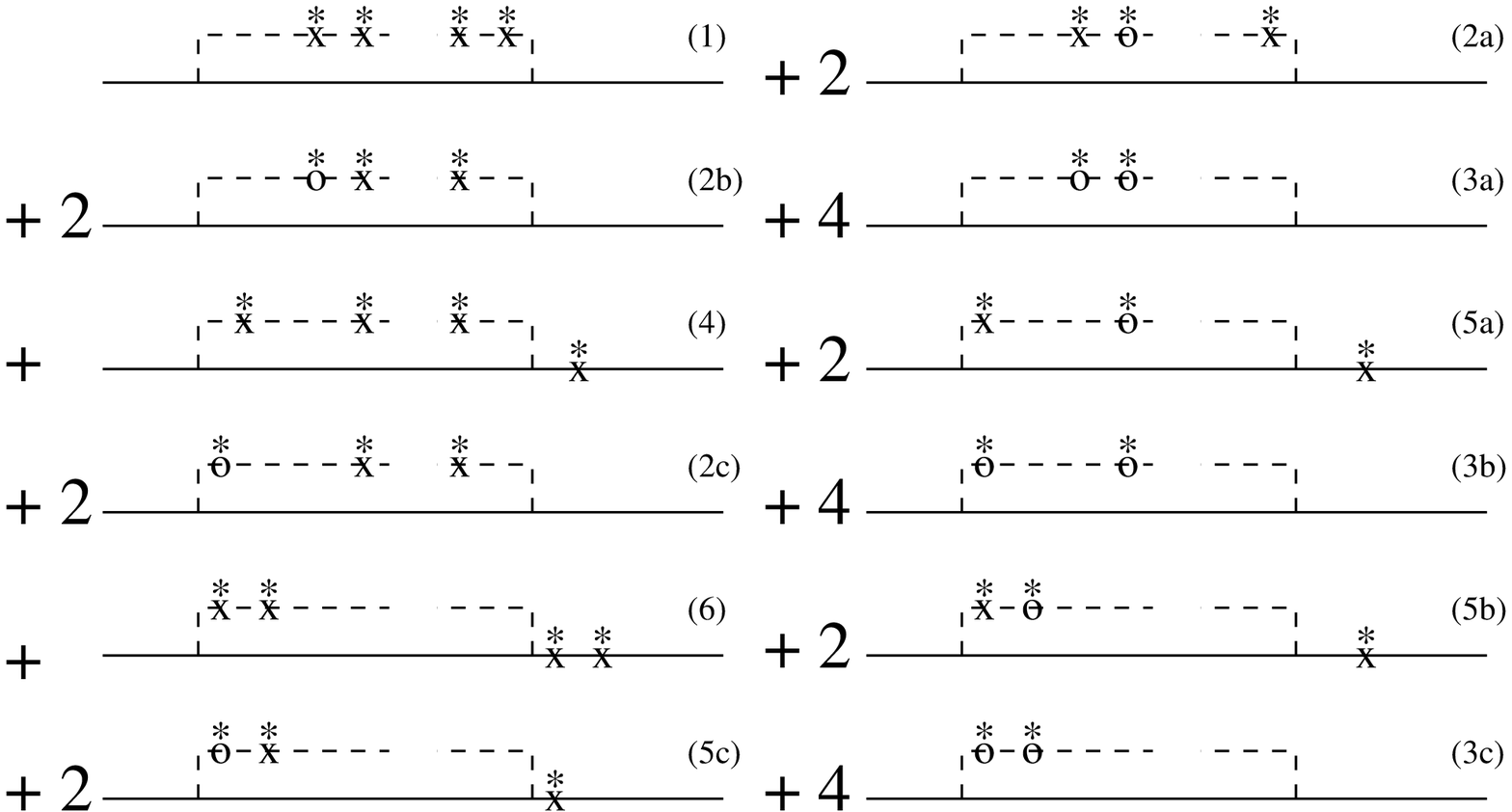}}
\vspace{0.5cm}
\caption{All non-vanishing contributions to the $N=2$ radiation 
cross section as given by (\protect\ref{3.13}). Diagrams are labelled 
with a double index $(ix)$, see text for details.}\label{fig3}
\end{figure}

All terms contributing to the $N=2$ contribution of (\ref{3.13}) are 
listed in Fig.~\ref{fig3}. We have labelled them with a double index  
$(ix)$. The first index $i=1,2,\dots,6$ indicates how the diagram 
contributes to the interaction vertex (\ref{2.25}),
\begin{eqnarray}
  \Gamma_{(1)} &=& \left({\bf k}_\perp + \bbox{q}_{1\perp}
                       + \bbox{q}_{2\perp}\right)^2\, ,
                  \label{3.33}\\   
  \Gamma_{(2)} &=& \left({\bf k}_\perp 
                       + \bbox{q}_{1\perp}\right)^2\, ,
                  \label{3.34}\\   
  \Gamma_{(3)} &=& {\bf k}_\perp^2 \, ,
                  \label{3.35}\\ 
  \Gamma_{(4)} &=& \left({\bf k}_\perp + \bbox{q}_{1\perp}
                       + \bbox{q}_{2\perp}\right)\cdot
                    \left({\bf k}_\perp 
                       + \bbox{q}_{1\perp}\right) \, ,
                  \label{3.36}\\
  \Gamma_{(5)} &=& \left({\bf k}_\perp 
                       + \bbox{q}_{1\perp}\right)\cdot
                    {\bf k}_\perp \, ,
                  \label{3.37}\\  
  \Gamma_{(6)} &=& \left({\bf k}_\perp + \bbox{q}_{1\perp}
                       + \bbox{q}_{2\perp}\right)\cdot
                    {\bf k}_\perp \, .
                  \label{3.38}
\end{eqnarray}
The integration variables $\bbox{q}_{1\perp}$ and 
$\bbox{q}_{2\perp}$ for the first and second momentum
transfer can be relabelled freely. As a consequence, we have to
consider only three different transverse gluon energies
\begin{eqnarray}
  Q &=& \frac{{\bf k}_\perp^2}{2\omega}\, ,\qquad
   Q_1 = \frac{\left( {\bf k}_\perp 
       + \bbox{q}_{1\perp}\right)^2}{2\omega}\, ,
   \nonumber \\
   Q_2 &=& \frac{\left( {\bf k}_\perp + \bbox{q}_{1\perp}
       + \bbox{q}_{2\perp}\right)^2}{2\omega}\, ,
  \label{3.39}
\end{eqnarray}
and no term $\propto \left({\bf k}_\perp + \bbox{q}_{2\perp}\right)$
appears in the interaction vertices (\ref{3.33})-(\ref{3.38}).

The phase factors can be read off from the diagrams of Fig.~\ref{fig3}
as described above. They are now integrals over the allowed positions
$\xi_1$ and $\xi_2$ of both scattering centers, e.g., 
\begin{eqnarray}
  &&\Phi_{(6)}(y_l,\bar{y}_l) = \int_{y_l}^{\bar{y}_l}
        d\xi_1\,\int_{\xi_1}^{\bar{y}_l} d\xi_2\, n(\xi_1)\, n(\xi_2)\,
        \nonumber \\
        && \qquad \qquad \times
        e^{-iQ(\bar{y}_l - \xi_2) - iQ_1(\xi_2-\xi_1) - iQ_2(\xi_1-y_l)}\, .
        \label{3.40}
\end{eqnarray}
The second index $x=a,b,\dots$ labels in Fig.~\ref{fig3} the different
diagrammatic contributions to the same term $\Gamma_{(i)}$. Denoting
by the phases $\Phi_{(i)}$ the sum over all contributions
(e.g. $\Phi_{(3)} = \Phi_{(3a)} + \Phi_{(3b)} + \Phi_{(3c)}$),
we find for the factors ${\cal Z}_{(i)}$ according to (\ref{3.21})
\begin{eqnarray}
  {\cal Z}_{(1)} &=& \frac{L^2}{2\, Q_2^2}\, n_0^2\, ,
                     \label{3.41}\\
  {\cal Z}_{(2)} &=& \frac{-2 + 2\, \cos\left(L\, Q_1\right)
                          }{Q_1^4}\, n_0^2\, ,
                     \label{33.42}\\
  {\cal Z}_{(3)} &=& \frac{-L^2}{2\, Q^2}\, n_0^2\, ,
                     \label{3.43}\\  
  {\cal Z}_{(4)} &=& \frac{2 - 2\, \cos\left(L\, Q_1\right)
                          - L^2\, Q_1^2}{Q_1^3\, Q_2}\, n_0^2\, ,
                     \label{3.44}\\
  {\cal Z}_{(5)} &=& \frac{2 - 2\, \cos\left(L\, Q_1\right)
                          + L^2\, Q_1^2}{Q_1^3\, Q}\, n_0^2\, ,
                     \label{3.45}\\
  {\cal Z}_{(6)} &=& \frac{- 2 + 2\, \cos\left(L\, Q_1\right)
                           }{Q\, Q_1^2\, Q_2}\, n_0^2\, .
                     \label{3.46}
\end{eqnarray}
These terms are multiplied by a factor $-1/2$ for each contact
term linking twice to the gluon line. The corresponding radiation
cross section (\ref{3.20}) shows $L$-dependent oscillations which 
are due to the interference between the gluon radiation of the
two spatially separated scattering centers. We consider especially
for fixed opacity $L\, n_0$ the limits $L\to 0$ and
$L\to \infty$ in which the two scattering centers sit on top
of each other and at arbitrary large separation, respectively.
We find
\begin{eqnarray}
  &&\lim\limits_{L\to\infty}
  {d^3\sigma^{(in)}(N=2)\over d(\ln x)\, d{\bf k}_\perp}
  \Bigg\vert_{L\, n_0 = {\rm fixed}}
  = \frac{\alpha_s}{\pi^2}\, N_c\, C_F\, C_A^2\, 
            \frac{\left(Ln_0\right)^2}{2}\,
  \nonumber \\
  &&\quad \times \int_{q_1}\, \int_{q_2} 
            \left[
   R({\bf k}_\perp+\bbox{q}_{1\perp},\bbox{q}_{2\perp}) 
   - R({\bf k}_\perp,\bbox{q}_{1\perp}) 
  \right]\, ,
  \label{3.47}
\end{eqnarray}
and 
\begin{eqnarray}
  &&\lim\limits_{L\to 0}
  {d^3\sigma^{(in)}(N=2)\over d(\ln x)\, d{\bf k}_\perp}
  \Bigg\vert_{L\, n_0 = {\rm fixed}}
  = \frac{\alpha_s}{\pi^2}\, N_c\, C_F\, C_A^2\, 
            \frac{\left(Ln_0\right)^2}{2}\,
  \nonumber \\
  && \quad \times \int_{q_1}\, 
            \int_{q_2}
  \left[
    R({\bf k}_\perp,\bbox{q}_{1\perp}+\bbox{q}_{2\perp})
        - 2\, R({\bf k}_\perp,\bbox{q}_{1\perp})
      \right]\, .
  \label{3.48}
\end{eqnarray}
We have written these limiting cases in terms of the momentum-dependent
part $R$ of the Gunion-Bertsch QCD radiation spectrum (\ref{3.22}),
\begin{equation}
  R({\bf k}_\perp,\bbox{q}_{\perp}) =  
  \frac{ \bbox{q}_{\perp}^2
          }{{\bf k}_\perp^2\, \left({\bf k}_\perp 
            +\bbox{q}_{\perp}\right)^2}\, ,
        \label{3.49}
\end{equation}
which is associated to a quark of transverse momentum ${\bf k}_\perp$,
receiving a momentum transfer $\bbox{q}_{\perp}$ in the 
scattering.

One may expect that in the coherent factorization limit
$L\to 0$, the radiation spectrum corresponds to one single 
Gunion-Bertsch term $R({\bf k}_\perp,\bbox{q}_{1\perp}+
\bbox{q}_{2\perp})$ with effective momentum transfer
$\left(\bbox{q}_{1\perp}+\bbox{q}_{2\perp}\right)$,
and that in the incoherent limit $L\to \infty$
(which for QCD is of course only of formal interest), the radiation
cross section is the sum of two independent Gunion-Bertsch terms.
The results (\ref{3.48}) and (\ref{3.49}) differ from these
expectations by the term
\begin{eqnarray}
 - \frac{\alpha_s}{\pi^2}\, N_c\, C_F\, C_A^2\, 
            \frac{\left(Ln_0\right)^2}{2}\,
 \int_{q_1}\, \int_{q_2}
    2\, R({\bf k}_\perp,\bbox{q}_{1\perp})\, .
  \label{3.50}
\end{eqnarray}
The origin of this term will be explained in section~\ref{sec5}. 

\section{A path-integral for arbitrary orders in opacity}
\label{sec4}

In this section, we derive a path-integral representation for 
the radiation cross section (\ref{3.13}) which is valid to arbitrary 
orders in opacity. We start from (\ref{3.14}), summed over powers
of opacity:
\begin{eqnarray}
&&
\epsfxsize=6.0cm 
\centerline{\epsfbox{totaln2.eps}}
\vspace{-1cm}
\nonumber \\
  &&  = \frac{1}{A_\perp}\, \int_Y\, 
  \int  d{\bf y}\, d{\bf \bar y}\, d{\bf s}\, d{\bf u}\, 
         \left( {\partial\over \partial {\bf \bar y}} 
                 F({\bf u}-{\bf \bar y})\right)\, 
         \nonumber \\
    &&\qquad \qquad \times 
    {\partial\over \partial {\bf r}_3(y_l)}
    M({\bf r}_1, {\bf r}_2, {\bf r}_3|y_l,\bar{y}_l)\, .
    \label{4.1}
\end{eqnarray}
We have separated in this expression scattering contributions
from longitudinal positions before and after $\bar{y}_l$. The
integration variables ${\bf u}$ and ${\bf \bar y}$ denote the
transverse positions of the gluon at $\bar{y}_l$ in the amplitude
and complex conjugate amplitude, respectively. All
contributions in the region $y_l< \xi < \bar{y}_l $ are given
by
\begin{eqnarray}
  M({\bf r}_1, {\bf r}_2, {\bf r}_3|y_l,\bar{y}_l) &=& 
  \sum_{n=0}^{\infty} C_A^n
  \int_{[q_i,\xi_i]_{i\in [1,n]}}
  \nonumber \\
  && \times   
  \tilde{G}({\bf r}_3(y_l),y_l;{\bf r}_3(\bar{y}_l),\bar{y}_l|\omega) 
  \nonumber \\
  && \times 
  G_0({\bf r}_2(y_l),y_l;{\bf r}_2(\bar{y}_l),\bar{y}_l|p_2)
  \nonumber \\
  && \times   
  \tilde{G}({\bf r}_1(\bar{y}_l),\bar{y}_l;{\bf r}_1(y_l),y_l|p_1)\, ,
  \label{4.2} 
\end{eqnarray}
where the transverse paths ${\bf r}_i(\xi)$, $i\in [1,3]$,
satisfy the boundary conditions
\begin{eqnarray}
  {\bf r}_1(y_l) &=& {\bf s}\, ,\qquad 
  {\bf r}_1(\bar{y}_l) = {\bf \bar y}\, ,
  \label{4.3}\\
  {\bf r}_2(y_l) &=& {\bf y}\, ,\qquad 
  {\bf r}_2(\bar{y}_l) = {\bf \bar y}\, ,
  \label{4.4}\\
  {\bf r}_3(y_l) &=& {\bf y}\, ,\qquad 
  {\bf r}_3(\bar{y}_l) = {\bf u}\, .
  \label{4.5}
\end{eqnarray}
The contributions coming from longitudinal
positions $\xi > \bar{y}_l$ are summarized by the function
\begin{eqnarray}
  &&F({\bf u}-{\bf \bar y}) = \sum_{m=0}^\infty C_A^m 
  \int_{[q_i,\xi_i]_{i\in [1,m]}}\,
   \int d{\bf x}_g\, d{\bf \bar x}_g
   \nonumber \\
   && \qquad \quad 
   \times   
   \tilde{G}({\bf u},\bar{y}_l;{\bf x}_g,z_+|\omega)\, 
   e^{- i{\bf k}_{\perp}\cdot ({\bf x}_g - {\bf \bar x}_g)}\, 
   \nonumber \\
   && \qquad \quad \times 
      \tilde{G}({\bf \bar x}_g,z_+;{\bf \bar y},\bar{y}_l|\omega)\, .
   \label{4.6}
\end{eqnarray}
Here, we have anticipated that the function $F$ depends only on the
difference ${\bf u}-{\bf \bar y}$ of the two integration variables
${\bf u}$ and ${\bf \bar y}$. The consistency of our starting point
(\ref{4.1}) with the rescattering part to the radiation cross 
section (\ref{3.13}) is straightforward to check by inserting
(\ref{4.2})-(\ref{4.6}) into (\ref{4.1}).

In what follows, we discuss how to simplify (\ref{4.2}) and (\ref{4.6}) 
in terms of the dipole cross section
\begin{equation}
  \sigma(\bbox{\rho}) = 2\, C_A\, \int 
  \frac{d\bbox{q}_\perp}{(2\pi)^2}\, |a_0(\bbox{q}_\perp)|^2\, 
  \left( 1 - e^{-i\bbox{q}_\perp\cdot \bbox{\rho}}\right)\, .
  \label{4.7}
\end{equation}
%

\subsubsection{Gluon rescattering for $\xi > \bar{y}_l$}
\label{sec4a1}

We start with the analysis of the function $F$ in (\ref{4.6}) which
describes the gluon rescattering at longitudinal
positions $\xi > \bar{y}_l$. The first order term $m=1$ reads
\begin{eqnarray}
&&
\epsfxsize=6.0cm 
\centerline{\epsfbox{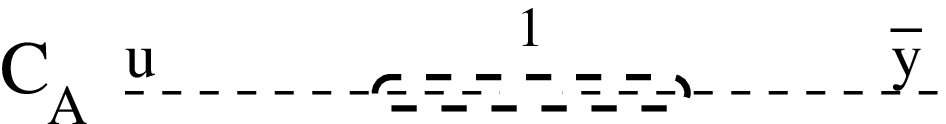}}
\vspace{-1cm}
\nonumber \\
  &&= C_A \int_{\bar{y}_l}^{z_+} d\xi_1\, n(\xi_1) 
  \int   \frac{d\bbox{q}_\perp}{(2\pi)^2}\, 
  |a_0(\bbox{q}_\perp)|^2
  \nonumber \\
  &&  \qquad \times \int  d{\bf x}_g\, d{\bf \bar x}_g\, 
  e^{- i{\bf k}_{\perp}\cdot ({\bf x}_g - {\bf \bar x}_g)}\,
  \int d{\bf r}(\xi_1)\, d{\bf \bar r}(\xi_1)
  \nonumber \\
  && \qquad \qquad \times
  G_0({\bf u},{\bar y}_l;{\bf r}(\xi_1)|\omega)\, 
  G_0({\bf r}(\xi_1);{\bf x}_g,z_+|\omega)\, 
  \nonumber \\
  && \qquad \qquad \times
  G_0({\bf \bar x}_g,z_+;{\bf \bar r}(\xi_1)|\omega)\,  
  G_0({\bf \bar r}(\xi_1);{\bf \bar y},\bar{y}_l|\omega)\,  
  \nonumber \\
  && \qquad \qquad \times
  \left(e^{-i\bbox{q}({\bf r}(\xi_1)- {\bf \bar r}(\xi_1))} 
        -1\right)
    \label{4.8}\\
  &&= -\frac{1}{2} \int_{\bar{y}_l}^{z_+} d\xi_1\, n(\xi_1)\, 
      \sigma\left({\bf u}-{\bf \bar y}\right)\,
      e^{-i{\bf k}_\perp\cdot\left({\bf u}-{\bf \bar y}\right)}\, ,
  \label{4.9}
\end{eqnarray}
The factors in the last line of (\ref{4.8}) correspond to the
three terms on the right hand side of (\ref{3.11}) and can be
combined to a dipole cross section (\ref{4.7}). It is straightforward
to iterate this relation
\begin{eqnarray}
&&
\epsfxsize=6.0cm 
\centerline{\epsfbox{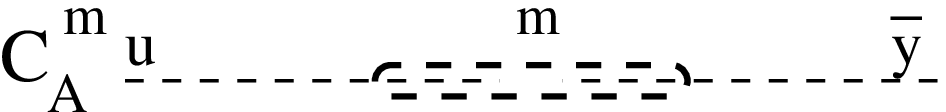}}
\vspace{-1cm}
\nonumber \\
  &&= \frac{1}{m!} \left( -\frac{1}{2} \int_{\bar{y}_l}^{z_+} 
    d\xi\, n(\xi)\, 
      \sigma\left( {\bf u}-{\bf \bar y} \right) \right)^m\,
      e^{-i{\bf k}_\perp\cdot\left( {\bf u}-{\bf \bar y} \right) }\, .
  \label{4.10}
\end{eqnarray}
Here, the prefactor $\frac{1}{m!}$ stems from the longitudinal
position ordering $\xi_i<\xi_{i+1}$ of subsequent interactions.
The sum (\ref{4.6}) over arbitrary $m$-fold rescattering 
contributions can thus be represented by a simple exponential
of the dipole cross section
\begin{eqnarray}
  && F\left({\bf u},{\bf \bar y}\right) =
  e^{-i{\bf k}_\perp\cdot\left( {\bf u}-{\bf \bar{y}} \right) }
  \nonumber \\
  &&\qquad \times
  \exp\left( -\frac{1}{2} \int_{\bar{y}_l}^{z_+} d\xi\, n(\xi)\, 
      \sigma\left({\bf u}-{\bf \bar{y}} \right) \right)\, .
    \label{4.11}
\end{eqnarray}
%
\subsubsection{Rescattering for $y_l< \xi < \bar{y}_l $}
\label{sec4a2}

The rescattering in the region $y_l< \xi < \bar{y}_l $ is
described by the path integral $M$ of (\ref{4.2}). To simplify
this expression, we start from its $n=1$ first order contribution
in opacity
\begin{eqnarray}
  &&M^{(n = 1)}({\bf r}_1, {\bf r}_2, {\bf r}_3|y_l,\bar{y}_l) =
  C_A\, \int \frac{d\bbox{q}_{1\perp}}{(2\pi)^2}\, 
       |a_0(\bbox{q}_{1\perp})|^2 
       \nonumber \\
       && \qquad \times 
       \int_{y_l}^{\bar{y}_l} d\xi_1\, n(\xi_1)\, 
       \int d{\bf r}_1(\xi_1)\, d{\bf r}_2(\xi_1)\, d{\bf r}_3(\xi_1)
       \nonumber \\
       && \qquad \times        
        M_0({\bf r}_1, {\bf r}_2, {\bf r}_3|y_l,\xi_1)\, 
        M_0({\bf r}_1, {\bf r}_2, {\bf r}_3|\xi_1,\bar{y}_l)
       \nonumber \\
       && \qquad \times
       \left(e^{i\bbox{q}_{1\perp}\cdot\left({\bf r}_3(\xi_1)
             - {\bf r}_1(\xi_1)\right)} - 1\right)\, .
       \label{4.12}
\end{eqnarray}
Here, $M_0$ denotes the non-interacting part of $M$,
obtained by replacing $\tilde{G} \to G_0$ in (\ref{4.2}). The
dipole cross section (\ref{4.7}) can be identified with the
$\bbox{q}_{1\perp}$-integral. Equation (\ref{4.12}) allows
for an iteration to arbitrary order $n$. Writing the 
free Green's functions of $M_0$ in a path-integral representation,
using  $\dot{\bf r} = d{\bf r}/d\xi$, we find
\begin{eqnarray}
   &&M({\bf r}_1, {\bf r}_2, {\bf r}_3|y_l,\bar{y}_l) =
   \int {\cal D}{\bf r}_1\,  {\cal D}{\bf r}_2\,   {\cal D}{\bf r}_3
   \nonumber \\
   && \qquad \times 
   \exp\left[ i\frac{E}{2}\int_{y_l}^{\bar{y}_l} d\xi
               \left( x\, \dot{\bf r}^2_3 +  (1-x)\, \dot{\bf r}^2_2 
                      - \dot{\bf r}^2_1\right)\right]
   \nonumber \\
   && \qquad \times 
        \exp\left[ - \frac{1}{2} \int_{y_l}^{\bar{y}_l} d\xi n(\xi)
                \sigma\left({\bf r}_3 - {\bf r}_1\right) \right]\, . 
   \label{4.13}
\end{eqnarray}
%
%
\subsubsection{Radiation cross section in terms of the colour dipole}
\label{sec4a3}

The medium-dependence of the radiation cross section (\ref{4.1})
is contained in the functions $F$ and $M$ of (\ref{4.11}) and (\ref{4.13}).
It is parametrized in terms of the colour dipole cross section 
$\sigma$ and the density of scattering centers $n(\xi)$. Inserting the 
expressions (\ref{4.11}) and (\ref{4.13}) for $F$ and $M$ into (\ref{4.1}), 
a substantial amount of straightforward algebra allows for further 
simplifications. Two of the three path-integrals in (\ref{4.13}) and some
of the transverse integrals in (\ref{4.1}) can be done analytically.
We give these technical details in appendix~\ref{appd}. The final result 
for the radiation cross section (\ref{3.13}) depends to leading
$O(x)$ on one path integral only
\begin{eqnarray}
 &&{\cal K}({\bf r}(y_l),y_l;{\bf r}(\bar{y}_l),\bar{y}_l|\omega)
 \nonumber \\
 &&= \int {\cal D}{\bf r}
   \exp\left[ \int_{y_l}^{\bar{y}_l} d\xi
        \left(i\frac{\omega}{2} \dot{\bf r}^2
          - \frac{1}{2}  n(\xi) \sigma\left({\bf r}\right) \right)
                      \right]\, ,
  \label{4.14}
\end{eqnarray}
and takes the explicit form
\begin{eqnarray}
  &&{d^3\sigma^{(in)}\over d(\ln x)\, d{\bf k}_\perp}
  = {\alpha_s\over (2\pi)^2}\, {1\over \omega^2}\,
    N_C\, C_F\, 
    2{\rm Re} \int_{z_-}^{z_+} dy_l  
  \int_{y_l}^{z_+} d\bar{y}_l\, 
  \nonumber \\
  && \qquad \times 
  e^{-\epsilon |y_l|\, -\epsilon |\bar{y}_l|}
  \int d{\bf u}\,   e^{-i{\bf k}_\perp\cdot{\bf u}}   \, 
  e^{ -\frac{1}{2} \int_{\bar{y}_l}^{z_+} d\xi\, n(\xi)\, 
    \sigma({\bf u}) }\,
  \nonumber \\
  && \qquad \times    {\partial \over \partial {\bf y}}\cdot 
  {\partial \over \partial {\bf u}}\, 
  {\cal K}({\bf y}=0,y_l; {\bf u},\bar{y}_l|\omega) \, .
    \label{4.15}
\end{eqnarray}
This is the main result of the present section. It contains
all the characteristic features of the result of Zakharov~\cite{Z98},
and the extension of Zakharov's result to finite transverse 
momentum~\cite{WG99,Z99}. The dipole cross section $\sigma(\bbox{\rho})$
is the leading $O(x)$ contribution of the $q$-$\bar{q}$-$g$
cross section introduced by Nikolaev and Zakharov~\cite{NZ91}
and used in~\cite{Z98,Z99}. The reason why this cross section 
depends on only one impact parameter $\bbox{\rho}$ is evident
from (\ref{3.13}): in principle, rescattering contributions
can compare the transverse position
between the three combinations $q-q$, $q-g$ or $g-g$ in amplitude 
and complex conjugate  amplitude. In the leading $O(x)$ approximation, 
however, the transverse position of the quark remains unchanged due to 
rescattering. Moreover, the difference between the transverse paths
of the gluon in amplitude and complex conjugate amplitude which is
given by the function $F$ in (\ref{4.11}), is fixed by the measured 
transverse momentum ${\bf k}_\perp$ and shows no time (i.e. $\xi$-)
evolution. Accordingly, there is only one remaining impact parameter 
which measures the transverse separation of the gluon from the 
quark.  

The result (\ref{4.15}) is closely related to the QED radiation
cross section, see Eq. (\ref{5.13}) below. 
The result (\ref{4.15}) differs from the expressions given by
Zakharov~\cite{Z98,Z99} by (i) including the $\epsilon$-regularization 
and (ii) not subtracting from ${\cal K}$ in (\ref{4.15}) the 
zeroth order term ${\cal K}_0$. As discussed elsewhere in detail~\cite{WG99}, 
(i) is indispensable for a quantitative analysis of the radiation
cross section. The subtraction of ${\cal K}_0$ was included in
Ref.~\cite{Z98,Z99} to cancel a potentially dangerous singularity
of the path-integral ${\cal K}$. However, in the presence of 
the $\epsilon$-regularization, this singularity does not exist
and the ${\cal K}_0$-term vanishes. It can thus be included for
technical convenience, but it is not needed.

Our derivation of (\ref{4.15}) provides a new and direct proof
of the equivalence of the BDMPS- and Zakharov-formalism. In
fact, the non-abelian Furry approximation which allowed us
to write the high-energy limit of (\ref{2.4}) is a compact
shorthand of all the rescattering diagrams included in the
analysis of BDMPS~\cite{W00}. Obtaining from this starting point the
path-integral expression (\ref{4.15}) proves the equivalence.
It is an independent confirmation of the corresponding statements 
in Ref.~\cite{BDMS-Zak}, and extends these arguments explicitly to 
the ${\bf k}_\perp$-differential radiation cross section.

%
\section{Probability Conservation to fixed order in opacity}
\label{sec5}

In this section, we introduce a new quantity $\sigma_{\rm cl}$,
obtained by factorizing out of the radiation cross section
(\ref{4.15}) a momentum independent absorption factor,
\begin{equation}
  {d^3\sigma^{(in)}\over d(\ln x)\, d{\bf k}_\perp}
  = e^{-\int d\xi\, n(\xi)\, V_{\rm tot}}\,  
  {d^3\sigma_{\rm cl}\over d(\ln x)\, d{\bf k}_\perp}\, .
  \label{5.1}
\end{equation}
Here, $V_{\rm tot}$ denotes the total elastic cross
section for the projectile scattering off one scattering
center. In what follows, we argue that
the $N$-th order opacity term of $\sigma_{\rm cl}$ reduces
in the coherent and incoherent limiting cases to the classically
expected results for bremsstrahlung associated to $N$-fold
scattering. In the incoherent
limit, it becomes the incoherent sum of $N$ 
Gunion-Bertsch radiation terms 
\begin{eqnarray}
  &&\lim\limits_{L\to\infty}
  {d^3\sigma_{\rm cl}(N)\over d(\ln x)\, d{\bf k}_\perp}
  \Bigg\vert_{L\, n_0 = {\rm fixed}}
  \equiv \frac{\alpha_s}{\pi^2}\, N_c\, C_F\,  
            \frac{\left(L\, n_0\right)^N}{N!}\,
  \nonumber \\
  &&\quad \times \left(\prod_{i=1}^N \int_{q_i} \right)  
            \sum_{i=1}^N
   R\left({\bf k}_\perp+\sum_{j=1}^{i-1}\bbox{q}_{j\perp},
     \bbox{q}_{i\perp}\right)\, .
  \label{5.2}
\end{eqnarray}
In the coherent limit, it reduces to a 
single Gunion-Bertsch radiation term for the combined 
momentum transfer of $N$ single scattering centers,
\begin{eqnarray}
  &&\lim\limits_{L\to 0}
  {d^3\sigma_{\rm cl}(N)\over d(\ln x)\, d{\bf k}_\perp}
  \Bigg\vert_{L\, n_0 = {\rm fixed}}
  \equiv \frac{\alpha_s}{\pi^2}\, N_c\, C_F\,  
            \frac{\left(L\, n_0\right)^N}{N!}\,
  \nonumber \\
  && \quad \times \left(\prod_{i=1}^N \int_{q_i} \right) 
    R\left({\bf k}_\perp,\sum_{i=1}^N\bbox{q}_{i\perp}\right)\, .
  \label{5.3}
\end{eqnarray}
In what follows, all limits $L\to 0$ and $L\to \infty$ are taken
at fixed opacity $L\, n_0 = {\rm fixed}$ without specifying
this explicitly. 
The $N$-th order contribution of (\ref{5.1}) reads
\begin{equation}
  {d^3\sigma^{(in)}(N)\over d(\ln x)\, d{\bf k}_\perp}
  = \sum_{m=0}^N (-1)^m\, w_m\, 
  {d^3\sigma_{\rm cl}(N-m)\over d(\ln x)\, d{\bf k}_\perp}\, ,
  \label{5.4}
\end{equation}
where the weights $w_m$ are the $m$-th order terms of the
absorption factor
\begin{equation}
  w_m = \frac{1}{m!} \left( \int d\xi\, n(\xi)\, 
    V_{\rm tot}\right)^m\, .
  \label{5.5}
\end{equation}
These weights will be seen to appear as a consequence of
probability conservation. 
In section~\ref{sec5a}, we discuss the above relations for QED.
In section~\ref{sec5b}, we extend this discussion to QCD.
%
\subsection{Elastic scattering and radiation in QED}
\label{sec5a}

In the present subsection, we take 
\begin{eqnarray}
  V(\bbox{q}_\perp) &=& |\bar{a}_0(\bbox{q}_\perp)|^2\, ,
  \label{5.6}\\
  V_{\rm tot} &=& \int \frac{d\bbox{q}_\perp}{(2\pi)^2}
  |\bar{a}_0(\bbox{q}_\perp)|^2\, ,
  \label{5.7}
\end{eqnarray}
to be the differential and total elastic scattering cross section for 
an electron scattering off a single electrically charged static potential. 
In general, we distinguish by a bar the QED quantities used in this 
subsection from their QCD counterparts discussed in the rest of this 
paper. The QED dipole cross section is defined by
\begin{equation}  
  \bar{\sigma}(\bbox{\rho}) = 2\, \int 
  \frac{d\bbox{q}_\perp}{(2\pi)^2}\, 
  |\bar{a}_0(\bbox{q}_\perp)|^2\, 
  \left( 1 - e^{-i\bbox{q}_\perp\cdot \bbox{\rho}}\right)\, ,
  \label{5.8}
\end{equation}
and the Fourier transform of $\bar{\sigma}$ takes the form~\cite{WG99}
\begin{eqnarray}
  \bar{\Sigma}({\bf q}_\perp) &=& \frac{1}{2} \int
  \frac{d\bbox{\rho}}{(2\pi)^2}\, 
  \bar{\sigma}(\bbox{\rho})\, e^{i{\bf q}_\perp\cdot \bbox{\rho}}
  \nonumber \\
   &=& - V({\bf q}_\perp) + V_{\rm tot}\, \delta^{(2)}({\bf q}_\perp)\, .
  \label{5.9}
\end{eqnarray}
Furthermore, we denote by
\begin{equation}
  \bar{R}({\bf k}_\perp,{\bf q}_\perp) =
  \frac{x^2{\bf q}_\perp^2}{{\bf k}_\perp^2\, \left({\bf k}_\perp + 
      x{\bf q}_\perp\right)^2}
  \label{5.10}
\end{equation}
the momentum dependence of a Bethe-Heitler radiation term associated
with the incoming electron momentum ${\bf k}_\perp$ and the 
scattering momentum transfer ${\bf q}_\perp$. Finally,
we introduce two shorthands for the average over the transverse 
momenta ${\bf q}_{i\perp}$ of the different scattering centers. 
Depending on whether this average involves
the term $\propto  V_{\rm tot}\, \delta^{(2)}({\bf q}_\perp)$ in
(\ref{5.12}) or not, we write
\begin{eqnarray}
  \int_{{\bar \Sigma}_N} &\equiv&  (-1)^N
  \prod_{i=1}^N \int d{\bf q}_{i\perp}\, 
  \bar{\Sigma}({\bf q}_{i\perp})\, ,
  \label{5.11}\\
  \int_{{\bf q}_i} &\equiv& \int \frac{d{\bf q}_{i\perp}}{(2\pi)^2}\, 
  |\bar{a}_0({\bf q}_{i\perp})|^2\, .
  \label{5.12}
\end{eqnarray}
The QED radiation cross section derived in~\cite{WG99,Z99} can be obtained 
from (\ref{4.15}) by substituting the QED dipole cross section 
$\bar{\sigma}$ of (\ref{5.8}) for the QCD one, and by changing the 
coupling strength $\alpha_{\rm s}\, N_c\, C_F \to \alpha_{\rm em}$,
\begin{eqnarray}
  &&{d^3\bar{\sigma}^{(in)}\over d(\ln x)\, d{\bf k}_\perp}
  = {\alpha_{\rm em}\over (2\pi)^2}\, {1\over \omega^2}\,
    2{\rm Re} \int_{z_-}^{z_+} dy_l  
  \int_{y_l}^{z_+} d\bar{y}_l\, 
  \nonumber \\
  && \qquad \times 
  e^{-\epsilon |y_l|\, -\epsilon |\bar{y}_l|}
  \int d{\bf u}\,   e^{-i{\bf k}_\perp\cdot{\bf u}}   \, 
  e^{ -\frac{1}{2} \int_{z_-}^{{y}_l} d\xi\, n(\xi)\, 
    \bar{\sigma}({\bf u}) }\,
  \nonumber \\
  && \qquad \times    {\partial \over \partial {\bf y}}\cdot 
  {\partial \over \partial {\bf u}}\, 
  {\cal K}({\bf y}=0,y_l; {\bf u},\bar{y}_l|\omega) \, .
    \label{5.13}
\end{eqnarray}
In contrast to the QCD case (\ref{4.15}), the term
$\exp\left( -\frac{1}{2} \int_{z_-}^{{y}_l} d\xi\, n(\xi)\, 
    \bar{\sigma}({\bf u})\right)$ in (\ref{5.13}) ranges
from $z_-$ to $y_l$~\cite{WG99} since it is the rescattering
of the incoming charged projectile (and not the rescattering 
of the outgoing radiated particle as in QCD) which determines 
the QED radiation spectrum.

The following argument is based on the observation that 
the term $\propto V_{\rm tot}\, \delta^{(2)}({\bf q}_\perp)$
in (\ref{5.9}) can be factored out of (\ref{5.13}) in form of
the absorption factor in (\ref{5.1}). The quantity $\sigma_{\rm cl}$
depends only on $V({\bf q}_\perp)$. Moreover, the $N$-th
order term of the radiation cross section (\ref{5.13}) depends only
on the (Fourier transform) of the dipole cross section (\ref{5.9}),
i.e., it is written with the averages (\ref{5.11}). This suggests
that $\sigma_{\rm cl}$ takes a simple form if written in terms 
of the averages (\ref{5.12}). To make more specific statements, we 
now investigate the first few terms in the opacity expansion of
(\ref{5.13}) which were calculated in Ref.~\cite{WG99}.

The $N=1$ term is the Bethe-Heitler radiation
cross section times the opacity $L\, n_0$ of the medium,
\begin{eqnarray}
  \frac{d^3\bar{\sigma}^{(in)}(N=1)}{d(\ln x)\, d{\bf k}_\perp}
  &=& \frac{\alpha_{\rm em}}{\pi^2}\, (L\, n_0)\, \int_{{\bar \Sigma}_1}
  \bar{R}({\bf k}_\perp,{\bf q}_{1\perp})
  \nonumber \\
  &=& \frac{\alpha_{\rm em}}{\pi^2}\, (L\, n_0)\, \int_{{\bf q}_1}
  \bar{R}({\bf k}_\perp,{\bf q}_{1\perp})\, .
  \label{5.14}
\end{eqnarray}
This confirms (\ref{5.4}) up to first order in opacity.
Since the Bethe-Heitler term vanishes for vanishing momentum transfer,
the two medium averages (\ref{5.11}) and (\ref{5.12}) make no
difference for this term. This is different for the higher order
terms for which we discuss here the coherent and incoherent limits.

For two scattering centers, the radiation cross section takes
the coherent limit~\cite{WG99}
\begin{eqnarray}
  &&\lim_{L\to 0} \frac{d^3\bar{\sigma}^{(in)}(N=2)}{d(\ln x)\, 
    d{\bf k}_\perp}
  \nonumber \\
  &&= \frac{\alpha_{\rm em}}{\pi^2} \frac{(L\, n_0)^2}{2}\, 
  \int_{{\bar \Sigma}_2} \, 
  \bar{R}({\bf k}_\perp,{\bf q}_{1\perp}+{\bf q}_{2\perp})
  \nonumber \\
  &&=\frac{\alpha_{\rm em}}{\pi^2} \frac{(L\, n_0)^2}{2}\, 
  \left( \int_{{\bf q}_1} \int_{{\bf q}_2} 
  \bar{R}({\bf k}_\perp,{\bf q}_{1\perp}+{\bf q}_{2\perp})
  \right.
  \nonumber \\
  && \left. \qquad 
  - 2\, V_{\rm tot} \int_{{\bf q}_1} 
  \bar{R}({\bf k}_\perp,{\bf q}_{1\perp})
  \right)\, .
  \label{5.15}
\end{eqnarray}
Here, the first line is the result obtained in Ref.~\cite{WG99},
the second line is obtained by inserting expression (\ref{5.9})
and integrating out the $\delta$-functions of the terms 
$\propto V_{\rm tot}\, \delta^{(2)}({\bf q}_\perp)$. In the
incoherent limit, we find in the same way
\begin{eqnarray}
  &&\lim_{L\to \infty} \frac{d^3\bar{\sigma}^{(in)}(N=2)}{d(\ln x)\, 
    d{\bf k}_\perp}
  \nonumber \\
  &&= \frac{\alpha_{\rm em}}{\pi^2} \frac{(L\, n_0)^2}{2}\, 
  \int_{{\bar \Sigma}_2}\, \left[ 
  \bar{R}({\bf k}_\perp,{\bf q}_{1\perp})
  \right.
  \nonumber \\
  && \left. \qquad \qquad \qquad \qquad +
    \bar{R}({\bf k}_\perp + {\bf q}_{1\perp},{\bf q}_{2\perp})\right]
  \nonumber \\
  &&= \frac{\alpha_{\rm em}}{\pi^2} \frac{(L\, n_0)^2}{2}\, 
   \left( \int_{{\bf q}_1} \int_{{\bf q}_2} 
     \bar{R}({\bf k}_\perp + {\bf q}_{1\perp},{\bf q}_{2\perp})
   \right.
  \nonumber \\
  && \left. \qquad 
  - V_{\rm tot} \int_{{\bf q}_1} 
  \bar{R}({\bf k}_\perp,{\bf q}_{1\perp})
  \right)\, .
  \label{5.16}
\end{eqnarray}
The results (\ref{5.15}), (\ref{5.16}) are the QED analogue of the
expressions (\ref{3.47}) and (\ref{3.48}) obtained for the $N=2$ 
QCD radiation cross section. This confirms (\ref{5.4}) up to
second order in opacity:
\begin{equation} 
   {d^3\bar{\sigma}^{(in)}(N=2)\over d(\ln x)\, d{\bf k}_\perp} = 
  {d^3\bar{\sigma}_{\rm cl}(N=2)\over d(\ln x)\, d{\bf k}_\perp}
  - w_1\, 
  {d^3\bar{\sigma}_{\rm cl}(N=1)\over d(\ln x)\, d{\bf k}_\perp}\, ,
  \label{5.17}
\end{equation}
Here, the weight $w_1 =  (L\, n_0\, V_{\rm tot})$ contains only information
on the mean free path of the projectile. It is the probability of
having one additional interaction with the medium. 
The physical reason for the 
negative terms in (\ref{5.15}) and (\ref{5.16}) is now obvious:

An expansion of (\ref{5.13}) up to second order sums over the two
possibilities that the photon was produced by an interaction of the
electron with either one or two scattering centers. The probability
of this second interaction is a factor $w_1$
smaller than the probability that only one interaction occurs.
Hence, to go consistently from the first ($N=1$) to the second
($N=2$) order approximation of (\ref{5.13}) requires two steps:
(i) the radiation contribution related to two rescatterings of
finite momentum transfers has to be added
[this contribution is denoted by $\sigma_{\rm cl}$].
(ii) the probability for the $N=1$-contribution has to be reduced
by a factor $(1-w_1)$ [in this way, the probabilities
for the $N=1$ and $N=2$-contributions add up to unity]. The second,
negative term in (\ref{5.17}) implements step (ii).
It is this second step which leads to the negative term in the 
coherent limit (\ref{5.15}) and which changes the sign of the second
term in the incoherent limit (\ref{5.16}). To sum up: the opacity 
expansion of (\ref{4.15}) correctly readjusts the probability of
the $\sigma_{\rm cl}(N=1)$ contribution if including the higher 
order $N=2$ in the approximation of (\ref{4.15}). 

Ref.~\cite{WG99} also provides the third order contribution to the
QED radiation cross section. The limiting cases read
\begin{eqnarray}
  &&\lim_{L\to 0} \frac{d^3\bar{\sigma}^{(in)}(N=3)}{d(\ln x)\, 
    d{\bf k}_\perp}
  \nonumber \\
  &&= \frac{\alpha_{\rm em}}{\pi^2} \frac{(L\, n_0)^3}{3!}\, 
  \left( \int_{{\bf q}_1} \int_{{\bf q}_2} \int_{{\bf q}_3} 
  \bar{R}({\bf k}_\perp,{\bf q}_{1\perp}+{\bf q}_{2\perp}+{\bf q}_{3\perp})
  \right.
  \nonumber \\
  && \left. \qquad 
  - 3\, V_{\rm tot} \int_{{\bf q}_1} \int_{{\bf q}_2} 
  \bar{R}({\bf k}_\perp,{\bf q}_{1\perp}+{\bf q}_{2\perp}) \right.
  \nonumber \\
  && \left. \qquad 
  + 3\, V_{\rm tot}^2 \int_{{\bf q}_1} 
  \bar{R}({\bf k}_\perp,{\bf q}_{1\perp}) 
  \right)\, ,
  \label{5.18}
\end{eqnarray}
\begin{eqnarray}
  &&\lim\limits_{L\to\infty}
  {d^3\bar{\sigma}^{(in)}(N=3)\over d(\ln x)\, d{\bf k}_\perp}
  \Bigg\vert_{L\, n_0 = {\rm fixed}}
  = \frac{\alpha_s}{\pi^2}\, N_c\, C_F\, C_A^2\, 
            \frac{\left(Ln_0\right)^3}{3!}\,
  \nonumber \\
  &&\quad \times \int_{q_1}\, \int_{q_2} \, \int_{q_3} 
            \left[
   \bar{R}({\bf k}_\perp+\bbox{q}_{1\perp}+\bbox{q}_{2\perp},
   \bbox{q}_{3\perp})
   \right.
  \nonumber \\
  &&\qquad  \left. 
   -2\, \bar{R}({\bf k}_\perp+\bbox{q}_{1\perp},\bbox{q}_{2\perp})
   + \bar{R}({\bf k}_\perp,\bbox{q}_{1\perp}) 
  \right]\, .
  \label{5.19}
\end{eqnarray}
The relative signs and weights of different Bethe-Heitler terms 
confirm (\ref{5.1}) to third order
\begin{eqnarray}
   {d^3\bar{\sigma}^{(in)}(N=3)\over d(\ln x)\, d{\bf k}_\perp}
   &=& 
  {d^3\bar{\sigma}_{\rm cl}(N=3)\over d(\ln x)\, d{\bf k}_\perp}
  - w_1\, 
  {d^3\bar{\sigma}_{\rm cl}(N=2)\over d(\ln x)\, d{\bf k}_\perp}
  \nonumber \\
  &&
  + w_2
  {d^3\bar{\sigma}_{\rm cl}(N=1)\over d(\ln x)\, d{\bf k}_\perp}\, .
  \label{5.20}
\end{eqnarray}
We note that summed over arbitrary orders of opacity, the factor
$\sum_{j=0}^\infty (-1)^j\, w_j$ in (\ref{5.4}) amounts to a 
momentum-independent normalization. The angular distribution of 
gluon radiation, calculated directly from the all-order expression 
(\ref{5.1}) is not modified by this normalization factor. In an
expansion in opacity, however, the weights $w_i$ do not factorize
from the momentum dependence. The correct angular distribution
for radiation off a target of $N$ scattering centers is given
by the sum over the opacity contributions up to $N$-th order. 
In this way, the results of the BDMPS/Zakharov-formalism can be 
compared directly to results for a finite number of scattering centers.
%
\subsection{Elastic scattering and medium-induced radiation in QCD}
\label{sec5b}

The above arguments confirm the expressions (\ref{5.1})-(\ref{5.5}) 
for QED, and they apply to the QCD radiation spectrum, too. This 
follows from the fact that the QED and QCD radiation spectrum 
are related by simple substitutions which do not affect the form of 
(\ref{4.15}). In QED, however, the only total elastic scattering 
cross section $V_{\rm tot}$ associated to the projectile is that of 
the electron. In contrast, both quark and gluon are charged in QCD
and have a finite elastic scattering cross section. What is 
additionally needed to justify (\ref{5.1})-(\ref{5.5}) in 
QCD is an argument why the total elastic scattering cross section 
$V_{\rm tot}$ in (\ref{5.1})-(\ref{5.5}) should be that of the gluon, 
\begin{equation}
  V_{\rm tot} = C_A\, \int \frac{d\bbox{q}_\perp}{(2\pi)^2}
  |a_0(\bbox{q}_\perp)|^2\, ,
  \label{5.21}
\end{equation}
rather than that of the quark or a combination of both.

The motivation for (\ref{5.21}) comes from the observation of 
section~\ref{sec3a} that the $i$-th scattering center always
contributes with a factor $C_A$ to the radiation cross section,
i.e., the $q$-$g$-projectile scatters effectively with the coupling 
strength of a gluon. This is a consequence of the leading $O(x)$
approximation in which, as observed by BDMPS, only the rescattering 
of the gluon matters. Beyond the leading $O(x)$ approximation, more 
complicated combinations of elastic quark and gluon cross sections 
may be needed, but this is the regime in which colour triviality
breaks down and our formalism does not apply.

In analogy to the QED case, we can argue on the basis of (\ref{5.1})
and (\ref{5.21}) that the $N$-th order opacity term of the QCD radiation 
cross section (\ref{4.15}) is a convolution of the $N$-fold 
scattering contribution and a readjustment of the probabilities
that less than $N$ rescatterings occur. This explains
especially the structure of the $N=2$ QCD contributions calculated
in section~\ref{sec3d}. More details about the opacity expansion
for QCD are given in the next section and in Appendix~\ref{appe}.
%
\section{In-medium production: hard versus medium-induced radiation}
\label{sec6}

The initial state boundary condition in the above calculations is
that of a quark approaching an asymptotic plane wave (\ref{2.9})
at far backward positions. In relativistic heavy ion collisions,
however, the hard (off-shell) parton is produced inside the
collision region and its propagation inside the medium has to 
be followed from a fixed finite time onwards. In what follows, we 
discuss the corresponding gluon radiation cross section $\sigma^{(nas)}$
for a {\it nascent} quark, in contrast to the case for a free {\it in}coming
quark which was discussed in sections~\ref{sec2}-\ref{sec5}.

The radiation cross section $\sigma^{(nas)}$ for nascent quarks is 
obtained by making two simple modifications to our derivation
in sections ~\ref{sec2}-\ref{sec4}:\\
(i) At the beginning of our derivation in section~\ref{sec2}, one has
to replace the incoming plane wave $F({\bf x},p_1)$ by a production
amplitude $J(p_1)$ which specifies the probability that at some
point $(z_0,{\bf 0}_\perp, z_0)$ in the collision region a hard
parton of momentum $p_1$ is produced.
We follow Gyulassy, Levai and Vitev~\cite{GLV99} in factorizing
the production probability $|J(p_1)|^2$ out of the radiation
cross section. This is an assumption on the weak momentum
dependence of this source which is also implicitly present in 
other discussions of radiative energy loss from partons
produced in medium~\cite{Z98}. Since our starting point 
(\ref{2.18}) is written for ${\bf p}_{1\perp}=0$, i.e., 
$F({\bf x},p_1)=1$, this amounts to multiplying (\ref{2.18})
by a factor $|J(p_1)|^2$. To simplify notation, we drop this
factor in what follows: in comparing to experiment, the radiation 
cross section $\sigma^{(nas)}$ given below has to be weighted with 
the production probability of the hard parton. \\
(ii) The parton produced at $(z_0,{\bf 0}_\perp, z_0)$ propagates
in the positive $z$-direction. Accordingly, the gluon emission vertices
in the amplitude and complex conjugate amplitude are constrained to
longitudinal positions $y_l$, $\bar{y}_l$ $>z_0$. This changes the
boundaries of the corresponding integrals in (\ref{2.18}) and survives
all intermediate steps. 

With these two changes in (\ref{2.18}), and choosing $z_0=0$,
the medium-induced gluon radiation cross section off a nascent
quark becomes
\begin{eqnarray}
  &&{d^3\sigma^{(nas)}\over d(\ln x)\, d{\bf k}_\perp}
  = {\alpha_s\over (2\pi)^2}\, {1\over \omega^2}\,
    N_C\, C_F
  \nonumber \\
  && \qquad \times 
    2{\rm Re} \int_{0}^{z_+} dy_l  
  \int_{y_l}^{z_+} d\bar{y}_l\, 
  e^{-\epsilon |y_l|\, -\epsilon |\bar{y}_l|}
  \nonumber \\
  && \qquad \times 
  \int d{\bf u}\,   e^{-i{\bf k}_\perp\cdot{\bf u}}   \, 
  e^{ -\frac{1}{2} \int_{\bar{y}_l}^{z_+} d\xi\, n(\xi)\, 
    \sigma({\bf u}) }\,
  \nonumber \\
  && \qquad \times    {\partial \over \partial {\bf y}}\cdot 
  {\partial \over \partial {\bf u}}\, 
  {\cal K}({\bf y}=0,y_l; {\bf u},\bar{y}_l|\omega) \, .
    \label{6.1}
\end{eqnarray}
This differs from the corresponding result of Zakharov~\cite{Z99} 
by the regularization prescription only. 

In what follows, we shall analyze (\ref{6.1}) in the opacity
expansion. To this end, we expand the path-integral ${\cal K}$
\begin{eqnarray}
 &&{\cal K}({\bf r},y_l;{\bf \bar r},\bar{y}_l) =
    {\cal K}_0({\bf r},y_l;{\bf \bar r},\bar{y}_l)
 \nonumber \\
 && - \int\limits_{z}^{z'}\, {\it d}\xi\, n(\xi)\, 
 \int {\it d}\bbox{\rho}\,
 {\cal K}_0({\bf r},y_l;\bbox{\rho},\xi)\, \frac{1}{2}\,
   \sigma(\bbox{\rho})\, 
   {\cal K}_0(\bbox{\rho},\xi;{\bf \bar r},\bar{y}_l)
 \nonumber \\
 && + \int\limits_{z}^{z'} {\it d}\xi_1\, n(\xi_1)\, 
    \int\limits_{\xi_1}^{z'} {\it d}\xi_2\, n(\xi_2)\, 
    \int {\it d}\bbox{\rho}_1\,{\it d}\bbox{\rho}_2\,
    {\cal K}_0({\bf r},y_l;\bbox{\rho}_1,\xi_1)\,  
    \nonumber \\
 && \times \frac{1}{2}\,
    \sigma(\bbox{\rho}_1)\, 
    {\cal K}(\bbox{\rho}_1,\xi_1;\bbox{\rho}_2,\xi_2)\, \frac{1}{2}\,
    \sigma(\bbox{\rho}_2)\,
    {\cal K}_0(\bbox{\rho}_2,\xi_2;{\bf \bar r},\bar{y}_l)\, .
 \label{6.2} 
\end{eqnarray}
Here, we have suppressed the explicit 
$\omega$-dependence in ${\cal K}$. The corresponding free Green's 
function ${\cal K}_0$ reads
\begin{equation}
  {\cal K}_0({\bf r},y_l;{\bf \bar r},\bar{y}_l)
  = \frac{\omega}{2\pi\, i\, (\bar{y}_l-y_l)}
    \exp\left\{ { {i\omega}
           \left(\bar{\bf r} - {\bf r}\right)^2 \over {2\, (\bar{y}_l-y_l)} }
           \right\}\, .
  \label{6.3}
\end{equation}
The technical steps involved in the opacity expansion are described
in detail in Ref.~\cite{WG99}. We present our result for the 
$N$-th order opacity contribution to (\ref{6.1}) in the form
\begin{eqnarray}
  &&{d^3\sigma^{(*)}(N)\over d(\ln x)\, d{\bf k}_\perp}
  = {\alpha_s\over (2\pi)^2}\, {1\over \omega^2}\,
    N_C\, C_F\, 2\, \int_{\Sigma_N} \sum_{j=0}^N
    \nonumber \\
  && \,\, \times \left[ {\cal Z}_{N,j+1}^{(*)}\, 
                   \left( {\bf k}_\perp + \sum_{i=1}^N {\bf q}_{i\perp}\right)
                   \cdot \left( {\bf k}_\perp + 
                     \sum_{i=1}^{N-j} {\bf q}_{i\perp}\right)
                   \right]\, ,
  \label{6.4}
\end{eqnarray} 
where the integration is defined by (\ref{5.11}) and the factors 
${\cal Z}_{N,j}^{(*)}$ are tabulated in Appendix~\ref{appe}.
The superscript ($* = in$ or $* = nas$) indicates that the result 
differs for the radiation cross sections (\ref{4.15}) and (\ref{6.1})
by the specific form of the factor ${\cal Z}_{N,j}^{(*)}$ only.
We have checked the form of (\ref{6.4}) for $N \leq 3$ by
explicit calculation, but we expect that it holds for arbitrary
$N$. Especially, it is obvious that the $N$-th order term contains
exactly $(N+1)$ different terms: the $j$-th  term in (\ref{6.4}) 
corresponds to the $(N-j)$-th order expansion in ${\cal K}$ and the
$j$-th order expansion of the exponential $\exp \left[-\frac{1}{2} \int d\xi\, 
n(\xi)\, \sigma({\bf u}) \right]$. Moreover, each integrand in
(\ref{6.4}) has to depend on all $N$ transverse momenta, since
the integration (\ref{5.11}) leads to a vanishing contribution
otherwise. 

The $N$-th order opacity expansion of (\ref{6.1}) involves only 
$(N+1)$ different terms. This is a significant simplification in 
comparison to brute force calculations~\cite{GLV99} 
which require already for $N=2$ the calculation of $7^2$ direct
plus $2\times 96$ contact terms~\cite{GLV2}.
In what follows, we discuss the physics contained in this opacity
expansion by studying the coherent and incoherent limits of
(\ref{6.4}).

\subsection{Hard radiation and its $N=1$ modification}
\label{sec6a}

For the gluon radiation spectrum (\ref{4.15}) of a free incoming
quark, the zeroth order in opacity vanishes. A free incoming particle 
does not radiate without interaction. A nascent parton, however, does 
radiate without further interaction. To zeroth order in $L\, n_0$, we find
\begin{equation}
  {d^3\sigma^{(nas)}(N=0)\over d(\ln x)\, d{\bf k}_\perp}
  = {\alpha_s\over \pi^2}\, 
    N_C\, C_F\, \frac{1}{{\bf k}^2_\perp}\, .
  \label{6.5}
\end{equation} 
This is the characteristic gluon radiation spectrum
\begin{equation}
  H({\bf k}_\perp) =  \frac{1}{{\bf k}^2_\perp}\, ,
  \label{6.6}
\end{equation}
associated to the production of a hard parton. In energy
loss calculations which integrate over $\omega$, the above radiation 
cross section (\ref{6.5}) as well as its medium-dependent form
should be multiplied with the quark-gluon splitting function
$\frac{x}{2} \frac{1+(1-x)^2}{x}$ which equals one only in the $O(x)$
approximation~\cite{BDMS99}. Equation (\ref{6.5}) is the QCD analogue 
of the QED radiation accompanying $\beta$-decay. To determine the 
energy loss of hard partons in relativistic heavy ion collisions, we are 
interested in the medium-dependent changes of the radiation
cross section (\ref{6.5}). To first order in opacity, we find
from Appendix~\ref{appe}
\begin{eqnarray}
  {d^3\sigma^{(nas)}(N=1)\over d(\ln x)\, d{\bf k}_\perp}
  &=& {\alpha_s\over \pi^2}\, 
    \frac{N_C\, C_F}{2\,\omega^2}\, \int_{\Sigma_1}
    \frac{-{\bf k}_\perp\cdot {\bf q}_\perp}{Q\, Q_1}\,
    \nonumber \\
    && \qquad \times
    n_0\, \frac{LQ_1 - \sin(LQ_1)}{Q_1}\, .
  \label{6.7}
\end{eqnarray} 
To understand the physical meaning of this expression
it helps to rewrite the coherent and incoherent limiting cases
in terms of the hard and Gunion-Bertsch term:
\begin{eqnarray}
  &&\lim_{L\to 0}{d^3\sigma^{(nas)}(N=1)\over d(\ln x)\, d{\bf k}_\perp}
  = 0\, ,
  \label{6.8}\\
  &&\lim_{L\to\infty}{d^3\sigma^{(nas)}(N=1)\over d(\ln x)\, d{\bf k}_\perp}
  = {\alpha_s\over \pi^2}\, N_C\, C_F\, (n_0\, L)
  \nonumber \\
  && \times  \int_{\Sigma_1}
  \left[ H({\bf k}_\perp + {\bf q}_{1\perp})
         + R({\bf k}_\perp,{\bf q}_{1\perp}) \right]\, .
  \label{6.9}
\end{eqnarray} 
In the incoherent $L\to\infty$ limit, the radiation cross section
(\ref{6.1}) expanded up to first order in opacity, takes the form
\begin{eqnarray}
  &&\lim_{L\to\infty}\sum_{m=0}^{N=1}
  {d^3\sigma^{(nas)}(m)\over d(\ln x)\, d{\bf k}_\perp}
  = {\alpha_s\over \pi^2}\, N_C\, C_F
  \nonumber \\
  &&\qquad \times \left[ (1-w_1)H({\bf k}_\perp)
                  + n_0\, L\, \int_{{\bf q}_1} 
                  H({\bf k}_\perp + {\bf q}_{1\perp})
                  \right.
  \nonumber \\
  && \qquad \left. \quad + n_0\, L\, \int_{{\bf q}_1} 
                  R({\bf k}_\perp,{\bf q}_{1\perp})
                  \right]\, .
  \label{6.10}
\end{eqnarray}
The three terms on the r.h.s. of (\ref{6.10}) have a simple physical 
meaning: the first is the hard, medium-independent radiation (\ref{6.6})
reduced by the probability $w_1 =  n_0\, L\, V_{\rm tot}$ that
one interaction of the projectile occurs in the medium. The
second term describes the hard radiation component which rescatters
once in the medium. The third term is the medium-induced Gunion-Bertsch
contribution associated with the rescattering. In general, the
$L$-dependence of (\ref{6.7}) reflects the interference
pattern between the different contributions and can be quite 
complicated. For the case $N=1$, we have seen that this interference
pattern interpolates between simple and physically intuitive limiting 
cases.

\subsection{Coherent and Incoherent Limits for $N=2$ and $N=3$}
\label{sec6b}

The general $L$-dependent expressions for the second and third orders
in the opacity expansion of (\ref{6.1}) can easily be written down 
by using (\ref{6.4}) and the corresponding results tabulated in 
Appendix~\ref{appe}. In the present section, we
focus entirely on the coherent and incoherent limiting cases. For
$N=2$, they read
\begin{eqnarray}
  &&\lim_{L\to 0}{d^3\sigma^{(nas)}(N=2)\over d(\ln x)\, d{\bf k}_\perp}
  = 0\, ,
  \label{6.11}\\
  &&\lim_{L\to\infty}{d^3\sigma^{(nas)}(N=2)\over d(\ln x)\, d{\bf k}_\perp}
  = {\alpha_s\over \pi^2}\, N_C\, C_F\, \frac{(n_0\, L)^2}{2}
  \nonumber \\
  && \times  \int_{\Sigma_2}
  \left[ H({\bf k}_\perp + {\bf q}_{1\perp}+ {\bf q}_{2\perp})
         + R({\bf k}_\perp +{\bf q}_{1\perp},{\bf q}_{2\perp}) \right]\, .
  \label{6.12}
\end{eqnarray} 
Adding this second order contribution to (\ref{6.10}), a simple
systematic starts to emerge:
\begin{eqnarray}
  &&\lim_{L\to\infty}\sum_{m=0}^{N=2}
  {d^3\sigma^{(nas)}(m)\over d(\ln x)\, d{\bf k}_\perp}
  = {\alpha_s\over \pi^2}\, N_C\, C_F
  \nonumber \\
  && \qquad \times \left[ (1-w_1+w_2)\, H({\bf k}_\perp) \right.
  \nonumber \\
  &&\qquad \quad                  
  + (1-w_1)\, n_0\, L\, \int_{{\bf q}_1} 
                  H({\bf k}_\perp + {\bf q}_{1\perp})
  \nonumber \\
  && \qquad \quad + \left. \frac{(n_0\, L)^2}{2}\, \int_{{\bf q}_1} 
                  \int_{{\bf q}_2}
                  H({\bf k}_\perp + {\bf q}_{1\perp} + {\bf q}_{2\perp})
                  \right]
  \nonumber \\
  && + \lim_{L\to\infty} \left[ 
    {d^3\sigma_{\rm cl}(2)\over d(\ln x)\, d{\bf k}_\perp}
    + (1-w_1)\, 
    {d^3\sigma_{\rm cl}(1)\over d(\ln x)\, d{\bf k}_\perp} \right]\, .
  \label{6.13}
\end{eqnarray}
Both the $m$-fold rescattered hard term and the Gunion Bertsch
contribution for $m$-fold rescattering are weighted with the 
expansion of the absorption factor up to order $(N-m)$.
For $N=3$, we find from the results in Appendix~\ref{appe}
\begin{eqnarray}
  &&\lim_{L\to 0}{d^3\sigma^{(nas)}(N=3)\over d(\ln x)\, d{\bf k}_\perp}
  = 0\, ,
  \label{6.14}\\
  &&\lim_{L\to\infty}{d^3\sigma^{(nas)}(N=3)\over d(\ln x)\, d{\bf k}_\perp}
  = {\alpha_s\over \pi^2}\, N_C\, C_F\, \frac{(n_0\, L)^3}{3!}
  \nonumber \\
  && \times  \int_{\Sigma_3}
  \left[ H({\bf k}_\perp + {\bf q}_{1\perp}+ {\bf q}_{2\perp}
         + {\bf q}_{3\perp})\right.
  \nonumber \\
  && \qquad \qquad \left.
  + R({\bf k}_\perp +{\bf q}_{1\perp}+{\bf q}_{2\perp},
  {\bf q}_{3\perp}) \right]\, .
  \label{6.15}
\end{eqnarray} 
Adding this contribution to (\ref{6.13}) confirms the indicated 
systematics. Rather than spelling out the result, we extrapolate
in the following section our analytical findings to arbitrary $N$.

\subsection{Extrapolation of limiting cases to arbitrary $N$}
\label{sec6c}

The results of our explicit calculations up to third order,
combined with the physical arguments given above, suggest
results valid to arbitrary orders $N$. For the case of a free
incoming quark, these results are given
in equations (\ref{5.2}) - (\ref{5.4}). For the case of an
in-medium produced quark, we write them in terms of the shorthand
\begin{eqnarray}
 H^{(m)}({\bf k}_\perp) &\equiv& \frac{\alpha_{\rm s}}{\pi^2}
 N_c\, C_F\, \frac{(n_0\, L)^m}{m!}\, 
 \left( \prod_{i=1}^m \int_{{\bf q}_i}\right)
 \nonumber \\
 && \quad \times
 H\left({\bf k}_\perp+ \sum_{i=1}^m {\bf q}_{i\perp}\right)\, .
  \label{6.16}
\end{eqnarray}
This describes the hard radiation component which rescatters on
$m$ well-separated scattering centers. For the incoherent limit
of (\ref{6.1}), expanded up to $N$-th order, we obtain by 
extrapolating the systematics observed up to $N=3$:
\begin{eqnarray}
  &&\lim_{L\to\infty}\sum_{m=0}^{N}
  {d^3\sigma^{(nas)}(m)\over d(\ln x)\, d{\bf k}_\perp}
  = \sum_{m=0}^N (-1)^{N-m}\, w_{N-m}\, H^{(m)}({\bf k}_\perp)
  \nonumber \\
  && \qquad + 
  \sum_{m=0}^N (-1)^{N-m}\, w_{N-m}\,
  \lim_{L\to\infty}
    {d^3\sigma_{\rm cl}(m)\over d(\ln x)\, d{\bf k}_\perp}\, .
  \label{6.17}
\end{eqnarray}
In the coherent limit, we infer from (\ref{6.11}) and (\ref{6.14}):
\begin{equation}
  \lim_{L\to 0}\sum_{m=0}^{N}
  {d^3\sigma^{(nas)}(m)\over d(\ln x)\, d{\bf k}_\perp}
  =  {\alpha_s\over \pi^2}\, N_C\, C_F\, \frac{1}{{\bf k}_\perp^2}\, .
  \label{6.18}
\end{equation}
In this coherent limit, there is no medium modification to the 
hard radiation (\ref{6.5}), irrespective of the opacity of the target:
locating the medium at (and only at) the creation point of the parton, 
the projectile looses the possibility of interacting with the medium. 

What matters physically is, how medium-effects switch on with increasing
thickness $L$ of the medium. This $L$-dependence will interpolate 
smoothly between the totally coherent and totally incoherent limits
discussed here. 

\section{Conclusion}
\label{sec7}

Colour triviality is an important tool in the study of the nuclear
dependence of hard QCD processes. For colour trivial observables,
the $N$-fold gluon exchange with the nuclear environment reduces
to the $N$-th power of a colour Casimir operator. This renders
the calculational problem abelian. 

In the present work, we have given a simple diagrammatic proof
of the colour triviality for the $N$-th order opacity contribution
to the ${\bf k}_\perp$-differential gluon radiation cross section 
off a hard quark. The diagrammatic identities developped to this end
are a direct consequence of the non-abelian Furry approximation for 
hard partons in a nuclear environment. Summing up the contributions
to the radiation cross sections to all orders in opacity, we have
derived Zakharov's path-integral formalism and we have given a new 
proof of its equivalence to the BDMPS-formalism. 

For practical applications, it is important that our approach gives
an easy access to the low opacity expansion of the radiation cross 
sections (\ref{4.15}) and (\ref{6.1}). 
The $N$-th order term involves only $(N+1)$ terms which are relatively
easy to calculate. In contrast to previous brute force 
calculations~\cite{GLV99} which have to deal with an exponentially 
increasing number of terms, our approach is thus well-suited for
the calculation of a realistic number of scattering centers. This
is illustrated by the simple form of our explicit results up to third
order in opacity. 

As discussed in section~\ref{sec5}, the information contained in the 
$N$-th order opacity term to the gluon radiation cross section is 
a convolution of (i) the radiation associated to the rescattering 
off $N$ scattering centers and (ii) a readjustment
of the probabilities that rescattering occurs with $m<N$
scattering centers only. We have shown how to disentangle this
information by factorizing in (\ref{5.1}) the radiation cross section 
into two terms. In this way, at least the coherent and incoherent
limits of the involved $L$-dependent interference pattern became
accessible analytically to arbitrary order. 

Most important for applications to relativistic heavy ion collisions 
at RHIC and LHC, we have extended these results in section~\ref{sec6} 
to the gluon radiation off quarks produced in the medium. The 
corresponding radiation cross section (\ref{6.1}) gives access to the 
interference pattern between the hard radiation (associated to the 
production of the hard parton) and the medium-induced radiation. Its
coherent limit shows no medium dependence. The incoherent
limit confirms the classical picture of an incoherent sum of 
rescattering contributions of the hard radiation and Gunion-Bertsch
radiation terms. The dependence of gluon radiation on the thickness 
$L$ of the medium is expected to interpolate smoothly between these 
analytically accessible limiting cases. The detailed study of this 
$L$-dependence awaits further work, not only in the opacity expansion
discussed here, but also in the dipole approximation discussed
previously~\cite{WG99}.

\acknowledgements
I thank S. Catani and Y. Dokshitzer for helpful discussions 
about section~\ref{sec5}.\\
{\it Note added in proof:} M. Gyulassy, P. Levai and I. Vitev  
informed me of an opacity expansion derived independently in~\cite{GLV2}.

\appendix
\section{Non-abelian Furry approximation for rescattering gluons}
\label{appa}

In this appendix, we simplify the gluon rescattering contribution
(\ref{2.7}). We consider on-shell gluons 
$k_\mu = (\omega, {\bf k}_\perp, k_l)$
with a physical transverse polarization
\begin{equation}
 \epsilon_\mu(k) = (\epsilon_0, \bbox{\epsilon}_\perp, -\epsilon_0)\, ,
 \qquad \epsilon_0 = \frac{\bbox{\epsilon}_\perp\cdot {\bf k}_\perp}
                          {\omega + k_l}\, .
 \label{a.1}
\end{equation}
To leading order in energy, the contraction of this polarization
vector with the momentum dependent part of the three-gluon vertices
occuring in (\ref{2.7}) takes the form
\begin{eqnarray}
  &&V_{\mu_{i-1}0\mu_i}(k_{i-1},k_i-k_{i-1},k_i)\, 
        \epsilon^{\mu_i}(k) 
        \nonumber \\
  && \qquad
        = -2\, \omega\, \epsilon^{\mu_{i-1}}(k)
        + O(\omega^0)\, .
  \label{a.2}
\end{eqnarray}
This allows to rewrite (\ref{2.7}) 
\begin{eqnarray}
  I^{\mu_1\, (L)}({\bf y},k) &=& \epsilon^{\mu_1}(k)
 {\cal P}\, \left( 
  \prod_{i=1}^L\int {d^3{\bf k}_i\over (2\pi)^3}\, d^3{\bf x}_i\,
  \right.
                   \nonumber \\
                   && \times
                    \lbrack -i\, A_0^{(g)}({\bf x}_i)\rbrack
                    {-i\, (-2\, \omega)\over
                     {k_i^2 + i\,\epsilon}}
                   \nonumber \\
                   && \times
                    \left.
                    \, e^{i\, {\bf k}_i\cdot 
                      ({\bf x}_i - {\bf x}_{i-1})} \right)\, 
                    e^{- i\, {\bf k}\cdot {\bf x}_L}\, ,
  \label{a.3}
\end{eqnarray}
where we have rearranged the phases, using ${\bf x}_0 = {\bf y}$.
The longitudinal momentum integrals can be done by contour integration
\begin{eqnarray}
  &&\int {dk_{i,l}\over (2\pi)}\,
   {i \over {k_i^2 + i\,\epsilon}}\,
   e^{i\,k_{i,l}\, (x_{i,l} - x_{i-1,l})}\nonumber \\
   && \qquad
   = {1\over 2\, k_{i,l}}
    \Theta(x_{i,l} - x_{i-1,l})\, e^{i\,k_{i,l}\, (x_{i,l} - x_{i-1,l})}\, .
  \label{a.4}
\end{eqnarray}
On the r.h.s. of this equation, $k_{i,l}$ is determined by the pole
value to order $O(1/\omega)$, $k_{i,l} = \omega\, - 
\textstyle{k_{i\perp}^2\over 2\, \omega}$. Keeping in (a.3) the
norm to leading order in $1/\omega$ but the phase to next to leading
order, we write
\begin{eqnarray}
  I^{\mu_1\, (L)}({\bf y},k) &=& \epsilon^{\mu_1}\,
  e^{-\, i\, \omega\, y_l}
 {\cal P}\, \left( 
  \prod_{i=1}^L\int d^3{\bf x}_i\, \Theta(x_{i,l} - x_{i-1,l})
  \right.
                   \nonumber \\
                   && \times \left.
                   G_0({\bf x}_{i-1},{\bf x}_i|\omega)\, 
                    \lbrack -i\, A_0^{(g)}({\bf x}_i)\rbrack \right)\,
                   \nonumber \\
                   && \times
                    e^{- i\, {\bf k}_\perp\cdot {\bf x}_{L,\perp}
                       + i \frac{{\bf k}_\perp^2}{2\, \omega} x_{L,l}}\, ,
  \label{a.5}
\end{eqnarray}
where the transverse momentum integrals are identified with the
free Green's functions
\begin{eqnarray}
  G_0({\bf x}_{i-1};{\bf x}_i\vert \omega)
     &=& \int {d^2{\bf k}_{i,\perp}\over (2\pi)^2}\, 
                    e^{i\, {\bf k}_{i,\perp}\cdot 
                      ({\bf x}_{i,\perp} - {\bf x}_{i-1,\perp})}\,
    \nonumber \\
  &&\qquad \quad \times
                    e^{- i{k_{i,\perp}^2\over 2\, \omega}
                       (x_{i,l} - x_{i-1,l})}\, .
  \label{a.6}
\end{eqnarray}
To rewrite the path-ordered product in (\ref{a.5}), we introduce 
the shorthand
\begin{eqnarray}
  &&G^{(L)}({\bf y}_\perp,y_l;{\bf x}_\perp,x_l\vert \omega) =
  \nonumber \\
  && \quad {\cal P}\, \left( \prod_{i=1}^L\int d^3{\bf x}_i\, 
            \Theta(x_{1,l} - x_{i-1,l})\right)\,
      G_0({\bf y}_\perp,y_l;{\bf x}_{1,\perp},x_{1,l}\vert p)
  \nonumber \\
  && \times \left( \prod_{i=1}^L
     \lbrack -i\, A_0({\bf x}_i)\rbrack\, 
            G_0({\bf x}_{i,\perp},x_{i,l};{\bf x}_{i+1,\perp},
                      x_{i+1,l}\vert p)\, \right)\, ,
  \label{a.7}
\end{eqnarray}
where ${\bf x}_{L+1}={\bf x}$. This allows us to write
\begin{eqnarray}
  I^{\mu_1\, (L)}({\bf y},k) &=& \epsilon^{\mu_1}\,
  e^{-\, i\, \omega\, y_l}
  \nonumber \\
  && \times
  \int d{\bf x}_\perp G^{(L)}({\bf y};{\bf x}|\omega)\,
  F({\bf x},{\bf k})\, .
  \label{a.8}
\end{eqnarray}
%
\section{Simplifying the interaction vertex}
\label{appb}

Here, we discuss details of how to simplify the interaction
vertex (\ref{2.16}) in the leading $O(x)$ approximation. 
The ${\bf y}$-derivatives in the differential operators $\hat D_1$ 
and $\hat D_2$ are conjugate to the transverse momenta $p_{11}^\perp$ 
and $p_{21}^\perp$ of the quark entering and leaving the radiation 
vertex, see Fig.~\ref{fig1} for notation. As a consequence, the 
interaction vertex (\ref{2.15})
takes a particularly simple form in a frame in which 
$p_{11}^\perp =  p_{21}^\perp$. Depending on the spin of the
ingoing ($\lambda_q = \pm \textstyle{1\over 2}$) and outgoing
($\lambda_{q'} = \pm \textstyle{1\over 2}$) quark
and the gluon helicity ($\lambda_g = \pm 1$), it reads in this
frame (see Refs.~\cite{KST98,WG99} for a derivation)
\begin{eqnarray}
  \hat{\Gamma}_y\bigl(\lambda_q=\lambda_{q'}, \lambda_g\bigr)
  &=& -i\,\lambda_g\, 
      \left[ \lambda_g\, (2-x) + 2\,\lambda_q\, x\right]
  \nonumber \\
   && \times \left( {\partial\over \partial y_1} 
             -i\, \lambda_g\, {\partial\over \partial y_2}
             \right)\, ,
  \label{b.1} \\
  \widehat\Gamma_r\bigl(\lambda_q=-\lambda_{q'}, \lambda_g\bigr)
  &=& 2\, m_q\, x\, \lambda_g\, 
  \delta_{\lambda_g\, , 2\lambda_q} \, ,
  \label{b.2}
\end{eqnarray}
where the differential operators in (\ref{b.1}) 
act on the two different transverse components ${\bf y} = (y_1,y_2)$
of the Green's function for the outgoing quark. The important
property of this representation of the spinor structure is that
the spin- and helicity-averaged combination $\hat{\Gamma}_{\bf y}\, 
\hat{\Gamma}^\dagger_{\bf y'}$ takes the simple form
\begin{equation}
  \hat{\Gamma}_{\bf y}\, \hat{\Gamma}^\dagger_{\bf y'}
  = g_{nf}
    {\partial\over \partial {\bf y_*}}\cdot
    {\partial\over \partial {\bf y'_*}}
    + g_{sf}\, ,
  \label{b.3}
\end{equation}
where the prefactors for the spin-flip $g_{sf}$ and non-flip
$g_{nf}$ term read
\begin{equation}
  g_{nf} = \lbrack 4 - 4x + 2x^2\rbrack \, ,
  \qquad g_{sf} = 2\, m_q^2\, x^2\, .
  \label{b.4}
\end{equation}
For the case of photon emission, one can choose to redefine 
the longitudinal $z$-axis to be parallel to the emitted photon 
${\bf k}$~\cite{KST98,WG99}. This ensures $p_{11}^\perp =  p_{21}^\perp$ 
and allows to eliminate all spinor dependence from the radiation 
probability with the help of (\ref{b.3}) and $y_*=y$. For the 
gluon radiation amplitude, this trick cannot be used. The reason 
is that due to the rescattering of the gluon, the direction 
${\bf k}$ of the final state gluon is not aligned to the 
direction ${\bf k}_1$ with which the gluon leaves the radiation 
vertex. To use the simple expression (\ref{b.3}) nevertheless,
we introduce the derivative $-i \partial/ \partial {\bf y}_g$, where the
subscript denotes that this partial derivative is acting on the
${\bf y}$-dependence of the gluon Green's function $G_{(g)}$.
$-i \partial/ \partial {\bf y}_g$ is the conjugate of the 
direction ${\bf k}_{1\perp}$. In the frame parallel to 
${\bf k}_1$, the conjugate of the momentum $p_{21}^\perp$
thus takes the form
\begin{equation}
  {\partial\over \partial {\bf y}} \longrightarrow
   {\partial\over \partial {\bf y}_*} =
    {\partial\over \partial {\bf y}} - \frac{1-x}{x}
  {\partial\over \partial {\bf y}_g}\, .
  \label{b.5}
\end{equation}
With this replacement, the vertex functions in (\ref{b.1}) and 
(\ref{b.3}) can be used for the discussion of the gluon radiation
spectrum in a frame in which the incoming quark momentum ${\bf p}_1$
defines the longitudinal direction.

In the leading $O(x)$ approximation, only the partial
derivative w.r.t. ${\bf y}_g$ survives in (\ref{b.5}), the
spin-flip contribution can be neglected, and the
spin- and helicity averaged combination (\ref{b.3}) 
takes in momentum space the simple form (\ref{2.25}).
%
\section{Cancelation of a class of contact terms}
\label{appc}

To arrive at the expression (\ref{3.13}) for the gluon radiation
cross section in the presence of $N$ scattering centers, we
have included arbitrary combinations of real and contact terms
to fixed order in opacity. However, contributions to the radiation
cross section have to involve at least one (real or contact)
interaction in both the amplitude and complex amplitude. The
reason for that is that due to energy-momentum conservation,
an amplitude which is touched neither by real nor by contact
terms cannot contribute to a final state. According to this
counting, diagrams which have $N$ contact terms in one amplitude
should not be included in the calculation of the $O(N)$ 
contribution. Here, we show that their inclusion in the
derivation of (\ref{3.13}) is allowed, since this set of 
diagrams adds up to zero at arbitrary order in opacity.

For our proof, we consider the set of $O(N)$ diagrams with $N$
contact terms in the amplitude, and $m$ of these contact terms
occuring at longitudinal positions $> y_l$. These contributions
have the form
\begin{eqnarray}
\epsfxsize=6.5cm 
\epsfbox{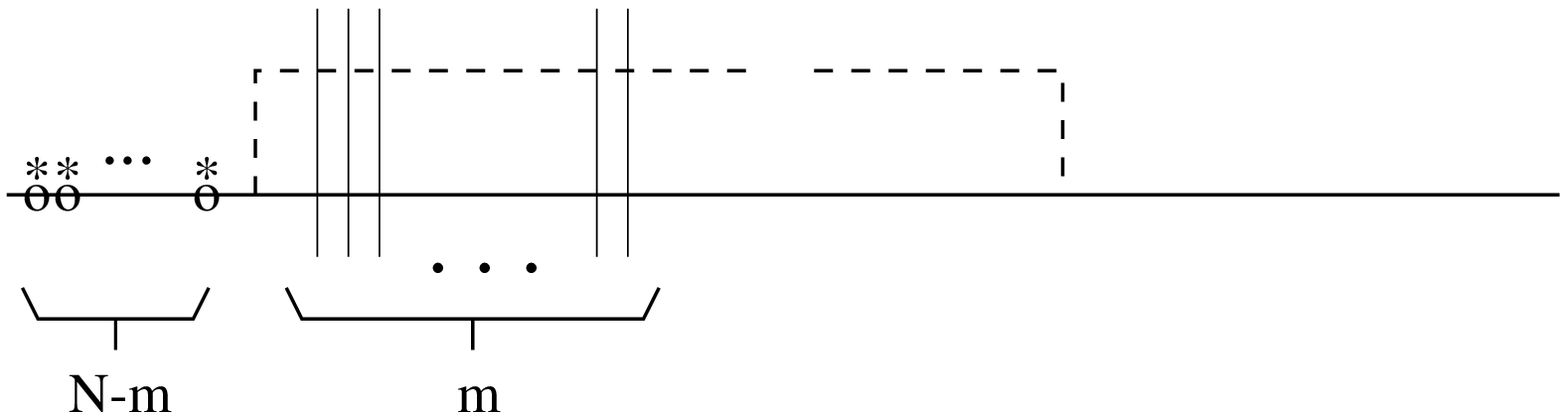}
\vspace{-1cm}
\label{c.1} 
\end{eqnarray}
where each thin vertical line denotes a contact term linking
either twice to the quark line or twice to the gluon line
or once to the quark and once to the gluon line. In what
follows, we consider the subset of these diagrams which contributes
to an arbitrary but fixed interaction vertex $\Gamma_{(i)}$. The
longitudinal phase factor $\Phi$ associated with this set of
diagrams takes the form
\begin{equation}
 \Phi(y_l,\bar{y}_l) = \varphi(y_l)\, e^{-i\, Q\, \bar{y}_l}\, .
 \label{c.2}
\end{equation}
The reason for this factorization of the $\bar{y}_l$-dependence
is that the longitudinal positions $\xi_i$, $i\in [1,N]$,
of the $N$ scattering centers (which are integration variables
in the calculation of $\Phi$) do not have $\bar{y}_l$ as 
integration boundary. As a consequence, the only possible
$\bar{y}_l$-dependence of $\Phi$ comes from the phase in the
conjugate interaction free amplitude where the gluon has 
transverse energy $Q$.

 For each class of diagrams (\ref{c.1}), we find the 
corresponding class of diagrams which has $N$ contact terms
on the opposite side of the cut
\begin{eqnarray}
\epsfxsize=6.5cm 
\epsfbox{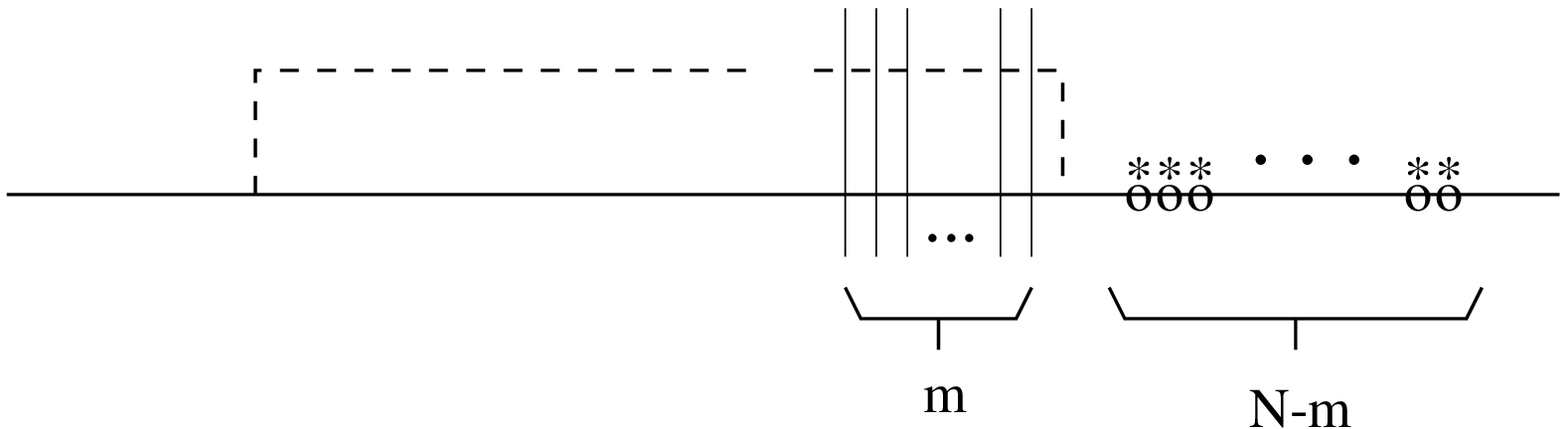}
\vspace{-1cm}
\label{c.3} 
\end{eqnarray}
For the class of diagrams to fixed $\Gamma_{(i)}$, the
phase corresponding to (\ref{c.3}) reads
\begin{equation}
 \Phi^*(\bar{y}_l,y_l) = \varphi^*(\bar{y}_l)\, e^{i\, Q\, y_l}\, .
 \label{c.4}
\end{equation}
It is related to (\ref{c.2}) by complex conjugation and interchange
$\bar{y}_l \leftrightarrow y_l$. The combined ${\cal Z}$-factor for the
sum of both phases (\ref{c.2}) and (\ref{c.4}) reads
according to (\ref{3.19})
\begin{eqnarray}
  {\cal Z}_{\rm comb} &=& 2{\rm Re}
  \int_{z_-}^{z_+} dy_l \int_{y_l}^{z_+} d\bar{y}_l\, 
  e^{-\epsilon |\bar{y}_l| - \epsilon |y_l|}
  \nonumber \\
  && \times \left(\Phi(y_l,\bar{y}_l) + \Phi(\bar{y}_l,y_l)\right)\, .
  \label{c.5}
\end{eqnarray}
The second term in (\ref{c.5}) can be rewritten in the form
\begin{eqnarray}
  &&\int_{z_-}^{z_+} dy_l \int_{z_-}^{z_+} d\bar{y}_l\, 
  \Theta\left(y_l - \bar{y}_l \right)\, \Phi(\bar{y}_l,y_l)
  \nonumber \\
  && = \int_{z_-}^{z_+} dy_l \int_{z_-}^{z_+} d\bar{y}_l\, 
  \Theta\left(\bar{y}_l - y_l\right)\, \Phi(y_l,\bar{y}_l)\, .  
  \label{c.6}
\end{eqnarray}
This allows to write for (\ref{c.5})
\begin{eqnarray}
  {\cal Z}_{\rm comb} &=& 2{\rm Re}
  \int_{z_-}^{z_+} dy_l \int_{z_-}^{z_+} d\bar{y}_l
  \nonumber \\
  && \qquad \times e^{-\epsilon |\bar{y}_l| - \epsilon |y_l|}\, 
  \Phi(y_l,\bar{y}_l)\, .
  \label{c.7}
\end{eqnarray}
The $\bar{y}_l$-integration of this expression can be done
explicitly and results in ${\cal Z}_{\rm comb} = 0$. This 
proves our claim that the sum of all $O(N)$ contributions
which have $N$ contact terms on one side of the cut, vanishes.

%
\section{Derivation of the radiation cross section (\ref{4.15})}
\label{appd}

The radiation cross section (\ref{4.15}) is obtained by inserting
the path-integral $M$ of (\ref{4.13}) and the function $F$ of
(\ref{4.11}) into (\ref{4.11}). To simplify the path-integral $M$, 
we change coordinates
\begin{eqnarray}
  {\bf r}_a(\xi) &=& x\, {\bf r}_3(\xi) + (1-x)\, {\bf r}_2(\xi)\, ,
  \label{d.1} \\
  {\bf r}_b(\xi) &=& {\bf r}_3(\xi) - {\bf r}_2(\xi)\, ,
  \label{d.2}
\end{eqnarray}
and we introduce relative and average pair coordinates
\begin{eqnarray}
  \bbox{\hat{\rho}}(\xi) &=& {\bf r}_a(\xi) - {\bf r}_1(\xi)\, ,
  \label{d.3}\\
  \bbox{\bar \rho}(\xi) &=& {\bf r}_a(\xi) + {\bf r}_1(\xi)\, .
  \label{d.4}
\end{eqnarray}
In terms of these coordinates, the path-integral $M$ of (\ref{4.13}) 
reads 
\begin{eqnarray}
   &&M({\bf r}_1, {\bf r}_2, {\bf r}_3|y_l,\bar{y}_l) =
   \int {\cal D}\bbox{\hat{\rho}}\,  {\cal D}\bbox{\bar \rho}\,
   {\cal D}{\bf r}_b
   \nonumber \\
   && \qquad \times 
   \exp\left[ i\frac{E}{2}\int_{y_l}^{\bar{y}_l} d\xi
               \left(\dot{\bbox{\hat{\rho}}}\cdot\dot{\bbox{\bar \rho}}
                 +  (1-x)\,x\, \dot{\bf r}^2_b 
                      \right)\right]
   \nonumber \\
   && \qquad \times 
        \exp\left[ - \frac{1}{2} \int_{y_l}^{\bar{y}_l} d\xi n(\xi)
                \sigma\left(\bbox{\hat{\rho}}
                + (1-x)\, {\bf r}_b\right) \right]\, . 
   \label{d.5}
\end{eqnarray}
The path-integral over ${\cal D}\bbox{\bar \rho}$ can be done
analytically (see Appendix B of Ref.~\cite{WG99} for more 
details on this step)
\begin{eqnarray}
   &&M({\bf r}_1, {\bf r}_2, {\bf r}_3|y_l,\bar{y}_l) =
   - \left(\frac{E}{2\pi\, i(\bar{y}_l-y_l)}\right)^2
   \nonumber \\
   && \qquad \times 
   e^{\frac{iE}{2(\bar{y}_l-y_l)}\left[
       2{\bf \bar y}\cdot\left({\bf v} + {\bf s}\right) + {\bf v}^2
       - {\bf s}^2\right]}
   \nonumber \\
   && \qquad \times    \int {\cal D}{\bf r}_b
   \exp\left[ \int_{y_l}^{\bar{y}_l} d\xi
        \left(i\frac{E\,(1-x)\,x}{2} \dot{\bf r}^2_b 
          \right. \right.
   \nonumber \\
   && \qquad \qquad \quad \left. \left.
          - \frac{1}{2}  n(\xi) \sigma\left(\bbox{\hat{\rho}_s}
                + (1-x)\, {\bf r}_b\right) \right)
                      \right]\, ,
   \label{d.6}
\end{eqnarray}
where
\begin{equation}
  {\bf v} = x({\bf u}-{\bf \bar y}) - x{\bf r}_3(y_l) - (1-x){\bf y}\, ,
  \label{d.7}
\end{equation}
and $\bbox{\hat{\rho}}_s$ denotes the straight line path
\begin{equation}
  \bbox{\hat{\rho}}_s(\xi) = x({\bf u}-{\bf \bar y})
  - ({\bf v}+{\bf s})
  \frac{\bar{y}_l - \xi}{\bar{y}_l-y_l}\, .
  \label{d.8}
\end{equation}
Here, we have kept the dependence on ${\bf r}_3(y_l)$ explicitly,
since we need the corresponding derivative in (\ref{4.1}). We
now shift the intergration variable ${\bf u} \to {\bf u} + {\bf \bar y}$
in (\ref{4.1}), replacing the partial derivative w.r.t. ${\bf \bar y}$
by the partial derivative w.r.t. ${\bf u}$ and doing a partial
integration:
\begin{eqnarray}
&&
\epsfxsize=6.0cm 
\centerline{\epsfbox{totaln2.eps}}
\vspace{-1cm}
\nonumber \\
  &&  = \frac{1}{A_\perp}\, \int_Y\, 
  \int  d{\bf y}\, d{\bf \bar y}\, d{\bf s}\, d{\bf u}\, 
                 F({\bf u})\, {\partial\over \partial {\bf u}}
                 \cdot 
         \nonumber \\
    &&\qquad \qquad \times 
    {\partial\over \partial {\bf r}_3(y_l)}
    M^{({\bf u} \to {\bf u} + {\bf \bar y})}
    ({\bf r}_1, {\bf r}_2, {\bf r}_3|y_l,\bar{y}_l)\, .
    \label{d.9}
\end{eqnarray}
Here, the superscript $M^{({\bf u} \to {\bf u} + {\bf \bar y})}$ on
$M$ indicates that the ${\bf u}$ variable is shifted. Starting from
(\ref{d.6}), the ${\bf \bar y}$- and ${\bf s}$-integration is now
trivial:
\begin{eqnarray}
  &&\int d{\bf \bar y}\, d{\bf s}\,
    M^{({\bf u} \to {\bf u} + {\bf \bar y})}
    ({\bf r}_1, {\bf r}_2, {\bf r}_3|y_l,\bar{y}_l)
    \nonumber \\
   && \qquad =
   \int {\cal D}{\bf r}_b
   \exp\left[ \int_{y_l}^{\bar{y}_l} d\xi
        \left(i\frac{E\,(1-x)\,x}{2} \dot{\bf r}^2_b 
          \right. \right.
   \nonumber \\
   && \qquad \qquad \quad \left. \left.
          - \frac{1}{2}  n(\xi) \sigma\left(x{\bf u}
                + (1-x)\, {\bf r}_b\right) \right)
                      \right]\, .
   \label{d.10}
\end{eqnarray}
The approximations
used to derive (\ref{4.1}) are valid to leading order $O(x)$. To this
order, the path-integral in (\ref{d.10}) coincides with the path-integral
${\cal K}$ given in (\ref{4.14}). To obtain our final result (\ref{4.15})
for the radiation cross section, we note that 
the boundary condition on this path-integral is
${\bf r}_b(y_l) = {\bf r}_3(y_l) - {\bf y} = 0$ and
${\bf r}_b(\bar{y}_l) = {\bf u}$. We can thus replace in (\ref{d.9}) 
$\frac{\partial}{\partial {\bf r}_3(y_l)} \to
\frac{\partial}{\partial {\bf r}_b(y_l)}$. The integrand of 
(\ref{d.9}) is then ${\bf y}$-independent, and the ${\bf y}$-integral
cancels against the total transverse area $A_\perp$. This leads
to the result (\ref{4.15}).

%
\section{Gluon radiation up to third order in opacity}
\label{appe}

Here, we give explicit results for the low opacity expansion of the 
gluon radiation cross sections (\ref{4.15}) and (\ref{6.1}) up to
third order. We have assumed a medium of homogeneous density $n_0$
and finite thickness $L$. By explicit calculation, we have found
for $N \leq 3$ that the $N$-th order terms are of the form (\ref{6.4}).
We denote by $Q_j$ the transverse energy of the gluon after $j$
momentum transfers from the medium, see (\ref{3.39}) for definition.
The ${\cal Z}$-factors entering (\ref{6.4}) are:

\underline{Results for $\frac{d^3\sigma^{(in)}(N)}{d(\ln x)\, 
d{\bf k}_\perp}$}\\
\noindent
A. For $N=1$:
\begin{eqnarray}
  Z^{(in)}_{1,1} &=& \frac{L\, n_0}{2\ Q_1^2}\, ,
  \label{e.1}\\
  Z^{(in)}_{1,2} &=& - \frac{L\, n_0}{Q\, Q_1}\, .
  \label{e.2}
\end{eqnarray}
\noindent
B. For $N=2$:
\begin{eqnarray}
  Z^{(in)}_{2,1} &=& n_0^2\, \frac{L^2}{4\ Q_2^2}\, ,
  \label{e.3}\\
  Z^{(in)}_{2,2} &=& n_0^2\, \frac{2 -2\cos\left(LQ_1\right) - L^2Q_1^2
                            }{2\, Q_1^3\, Q_2}\, ,
  \label{e.4}\\
  Z^{(in)}_{2,3} &=& n_0^2\, \frac{-1 +\cos\left(LQ_1\right)
                            }{Q\, Q_1^2\, Q_2}\, .
  \label{e.5}
\end{eqnarray}
\noindent
C. For $N=3$:
\begin{eqnarray}
  Z^{(in)}_{3,1} &=& n_0^3\, \frac{L^3}{12\ Q_3^2}\, ,
  \label{e.6}\\
  Z^{(in)}_{3,2} &=& - n_0^3\, \frac{6\sin\left(LQ_2\right) - 6LQ_2 + L^3Q_2^3
                            }{6\, Q_2^4\, Q_3}\, ,
  \label{e.7}\\
  Z^{(in)}_{3,3} &=& n_0^3\,\frac{Q_1^3\left[\sin\left(LQ_2\right)-LQ_2\right]
                            }{Q_1^3\, (Q_1-Q_2)\, Q_2^3\, Q_3}
                          \nonumber \\
          && - n_0^3\, \frac{Q_2^3\left[\sin\left(LQ_1\right)-LQ_1\right]
                            }{Q_1^3\, (Q_1-Q_2)\, Q_2^3\, Q_3}\, ,
  \label{e.8}\\
  Z^{(in)}_{3,4} &=& n_0^3\,\frac{Q_1^2\left[LQ_2-\sin\left(LQ_2\right)\right]
                            }{Q\, Q_1^2\, (Q_1-Q_2)\, Q_2^2\, Q_3} 
                           \nonumber \\
          && - n_0^3\, \frac{Q_2^2\left[LQ_1-\sin\left(LQ_1\right)\right]
                            }{Q\, Q_1^2\, (Q_1-Q_2)\, Q_2^2\, Q_3}\, .
  \label{e.9}
\end{eqnarray}

\underline{Results for $\frac{d^3\sigma^{(nas)}(N)}{
d(\ln x)\, d{\bf k}_\perp}$}\\
\noindent
A. For $N=1$:
\begin{eqnarray}
  Z^{(nas)}_{1,1} &=& n_0\, \frac{LQ_1-\sin\left(LQ_1\right)}{Q_1^3}\, ,
  \label{e.10}\\
  Z^{(nas)}_{1,2} &=& - n_0\, \frac{LQ_1-\sin\left(LQ_1\right)}{Q\, Q_1^2}\, .
  \label{e.11}
\end{eqnarray}
\noindent
B. For $N=2$:
\begin{eqnarray}
  Z^{(nas)}_{2,1} &=& n_0^2\, \frac{2\cos\left(LQ_2\right) - 2 + L^2Q_2^2
                            }{2\, Q_2^4}\, ,
  \label{e.12}\\
  Z^{(nas)}_{2,2} &=&  n_0^2\, \frac{Q_1^3\left[2-2\cos\left(LQ_2\right)
                             -L^2Q_2^2\right]
                            }{2\, Q_1^3\, (Q_1-Q_2)\, Q_2^3}
                          \nonumber \\
          && - n_0^3\, \frac{Q_1^3\left[2-2\cos\left(LQ_1\right)
                             -L^2Q_1^2\right]
                            }{2\, Q_1^3\, (Q_1-Q_2)\, Q_2^3}\, ,
  \label{e.13}\\
  Z^{(nas)}_{2,3} &=&  n_0^2\, \frac{Q_1^2\left[-1+\cos\left(LQ_2\right)\right]
                            }{Q\, Q_1^2\, (Q_1-Q_2)\, Q_2^2}
                          \nonumber \\
          && - n_0^3\, \frac{Q_2^2\left[-1+\cos\left(LQ_1\right)\right]
                            }{Q\, Q_1^2\, (Q_1-Q_2)\, Q_2^2}\, .
  \label{e.14}
\end{eqnarray}
\noindent
C. For $N=3$:
\begin{eqnarray}
  Z^{(nas)}_{3,1} &=& n_0^3\, \frac{6\sin\left(LQ_3\right) - 6LQ_3 + L^3Q_3^3
                            }{6\, Q_3^5}\, ,
  \label{e.15}\\
  Z^{(nas)}_{3,2} &=&  n_0^3\, \frac{Q_3^4\left[6\sin\left(LQ_2\right) 
                                         - 6LQ_2 + L^3Q_2^3\right]
                            }{6\, Q_2^4\, (Q_2-Q_3)\, Q_3^4}
                          \nonumber \\
          && - n_0^3\, \frac{Q_2^4\left[6\sin\left(LQ_3\right) 
                                         - 6LQ_3 + L^3Q_3^3\right]
                            }{6\, Q_2^4\, (Q_2-Q_3)\, Q_3^4}
  \label{e.16}\\
  Z^{(nas)}_{3,3} &=& \frac{n_0^3\, \sin\left(LQ_1\right)
                            }{Q_1^3\, (Q_1-Q_2)\, (Q_1-Q_3)}
                          \nonumber \\
          && + \frac{n_0^3\, \sin\left(LQ_2\right)
                            }{Q_2^3\, (Q_2-Q_1)\, (Q_2-Q_3)}  
              \nonumber \\
          && + \frac{n_0^3\, \sin\left(LQ_3\right)
                            }{Q_3^3\, (Q_3-Q_1)\, (Q_3-Q_2)} 
              \nonumber \\
          && - L\frac{Q_2Q_3 + Q_1Q_2+Q_1Q_3}{Q_1^2\, Q_2^2\, Q_3^2}\, ,
  \label{e.17}\\
  Z^{(nas)}_{3,4} &=& - \frac{n_0^3\, \sin\left(LQ_1\right)
                            }{Q\, Q_1^2\, (Q_1-Q_2)\, (Q_1-Q_3)}
                          \nonumber \\
          && - \frac{n_0^3\, \sin\left(LQ_2\right)
                            }{Q_2^2\, (Q_2-Q_1)\, (Q_2-Q_3)\, Q}  
              \nonumber \\
          && - \frac{n_0^3\, \sin\left(LQ_3\right)
                            }{Q_3^2\, (Q_3-Q_1)\, (Q_3-Q_2)\, Q} 
              \nonumber \\
          && + L\frac{n_0^3}{Q\, Q_1\, Q_2\, Q_3}\, ,
  \label{e.18}
\end{eqnarray}

%


\begin{references}
\bibitem{GW94} 
 M. Gyulassy and X.-N. Wang, Nucl. Phys. {\bf B420}
 (1994) 583.
\bibitem{WGP95}
 X.-N. Wang, M. Gyulassy and M. Pl\"umer, Phys. Rev. {\bf D51}
 (1995) 3436.
\bibitem{WG91}
 X.-N. Wang and M. Gyulassy,  Phys. Rev. {\bf D44} (1991) 3501. 
\bibitem{GL98}
 M. Gyulassy and P. Levai, Phys. Lett. {\bf B442} (1998) 1.
\bibitem{Z96} 
 B.G. Zakharov, JETP Letters {\bf 63} (1996) 952, 
 {\bf 65} (1997) 615.
\bibitem{BDMPS97a}
 R. Baier, Y.L. Dokshitzer, A.H. Mueller, S. Peign\'e and D. Schiff,
 Nucl. Phys. {\bf B483} (1997) 291.
\bibitem{BDMPS97b}
 R. Baier, Y.L. Dokshitzer, A.H. Mueller, S. Peign\'e and D. Schiff,
 Nucl. Phys. {\bf B484} (1997) 265.
\bibitem{Z98} 
 B.G. Zakharov, Phys. Atom. Nucl. {\bf 61} (1998) 838
 [Yad. Fiz. {\bf 61} (1998) 924], hep-ph/9807540.
\bibitem{BDMS98}
 R. Baier, Y.L. Dokshitzer, A.H. Mueller and D. Schiff,
 Phys. Rev. {\bf C58} (1998) 1706.
\bibitem{BDMS-Zak}
 R. Baier, Y.L. Dokshitzer, A.H. Mueller and D. Schiff,
 Nucl. Phys. {\bf B531} (1998) 403.
\bibitem{KST98} B.Z. Kopeliovich, A. Sch\"afer and A.V. Tarasov,
  Phys. Rev. {\bf C59} (1999) 1609.
\bibitem{WG99} U.A. Wiedemann and M. Gyulassy, Nucl. Phys.
 {\bf B560} (1999) 345-382.
\bibitem{Z99}
 B.G. Zakharov, JETP Lett. {\bf 70} (1999) 176. 
\bibitem{BDMS99}
 R. Baier, Y.L. Dokshitzer, A.H. Mueller and D. Schiff,
 Phys. Rev. {\bf  C60} (1999) 064902.
\bibitem{KST99} B.Z. Kopeliovich, A. Sch\"afer and A.V. Tarasov,
  Phys. Rev. {\bf D62} (2000) 054022.
\bibitem{BSZ99}
 R. Baier, D. Schiff, and B.G. Zakharov, hep-ph/0002198,
 submitted to Ann. Rev. Nucl. Part. Sci. 
\bibitem{GLV99} M. Gyulassy, P. Levai and I. Vitev, 
 Nucl. Phys. {\bf B571} (2000) 197. 
\bibitem{LS00}
 I.P. Lokhtin and A.M. Snigirev, Phys. Lett. {\bf B440} (1998) 163.
 and hep-ph/0004176.
\bibitem{W00}
 U.A. Wiedemann, Nucl. Phys. {\bf B582} (2000) 409.
\bibitem{BH93}
 S.J. Brodsky and P. Hoyer, Phys. Lett. {\bf B 298} (1993) 165.
\bibitem{NZ91}
 N.N. Nikolaev and B.G. Zakharov, Z. Phys. {\bf C49} (1991) 607.
\bibitem{GB82}
 J.F. Gunion and G. Bertsch, Phys. Rev. {\bf D 25} (1982) 746.
\bibitem{GLV2}
 M. Gyulassy, P. Levai and I. Vitev, nucl-th/0005032 and
 nucl-th/0006010. 
\end{references}
\end{document}